\documentclass[a4paper,11pt]{article}
\pdfoutput=1 

\usepackage{jcappub} 

\usepackage[T1]{fontenc} 
\usepackage{graphicx}
\usepackage{epsfig}
\usepackage{rotate}
\usepackage{amsmath}
\usepackage{amssymb}
\usepackage{amsfonts}
\usepackage{bm}
\usepackage{verbatim}
\usepackage{tablefootnote}
\usepackage{enumerate}
\usepackage{afterpage}
\usepackage{xcolor}
\usepackage{natbib, hyperref}

\UseRawInputEncoding

\newcommand{\ud}{\mathrm{d}}
\newcommand{\p}{\partial}
\newcommand{\cH}{\mathcal{H}}

\def\be{\begin{equation}}
\def\ee{\end{equation}}
\def\bea{\begin{eqnarray}}
\def\eea{\end{eqnarray}}

\newcommand{\red}[1]{{\color{black}{#1}}}

\title{Relativistic and wide-angle corrections to galaxy power spectra}

\author{Sheean Jolicoeur$^{1,a}$, S\^ecloka L. Guedezounme$^{2,b}$, \\ Roy Maartens$^{2,3,4}$, Pritha Paul$^{5}$, Chris Clarkson$^{5,2}$, \\ Stefano Camera$^{6,7,8,2}$}
\affiliation{\small $^{1}$Department of Physics, Stellenbosch University, Matieland 7602, South Africa \\ 
$^{2}$Department of Physics \& Astronomy, University of Western Cape, Cape Town 7535, South Africa \\
$^{3}$Institute of Cosmology \& Gravitation, University of Portsmouth, Portsmouth PO1 3FX, UK\\
$^{4}$National Institute for Theoretical \& Computational Sciences, Cape Town 7535, South Africa\\
$^{5}$School of Physics \& Astronomy, Queen Mary University of London, London E1 4NS, UK\\
$^{6}$Dipartimento di Fisica, Universit\`a degli Studi di Torino, 
Torino, 10125, Italy\\
$^{7}$Istituto Nazionale di Fisica Nucleare, Sezione di Torino,  
Torino, 10125, Italy\\
$^{8}$Istituto Nazionale di Astrofisica, Osservatorio Astrofisico di Torino, 
Pino Torinese, 10025, Italy}

\emailAdd{$^{a}$jolicoeursheean@gmail.com}
\emailAdd{$^{b}$seclokaguedezounme@gmail.com}
\usepackage[toc]{appendix}

\abstract{Galaxy surveys contain  information on the largest scales via wide-angle and relativistic contributions. By combining two different galaxy populations, we can suppress the strong cosmic variance on ultra-large scales and thus enhance the detectability of the signals. 
The relativistic  Doppler and Sachs-Wolfe effects are of a similar magnitude to the leading wide-angle corrections, so that it is important to treat them together, especially since they can partially cancel. The power spectra depend on the choice of line of sight for each galaxy pair and we present results for {a general} 
line of sight.  Then we estimate the detection significance of the auto- and cross-power spectra for a variety of cases. We use two futuristic galaxy samples based on a `beyond-DESI' survey and a  SKA Phase 2 survey, covering 15,000\,deg$^2$ up to $z=1$. We find a detection significance for the total relativistic wide-angle effects that ranges from $\sim 5\sigma$ to $>15\sigma$, depending on the line-of-sight configuration.}

\begin{document}
\maketitle
\date{\today}
%\flushbottom

\section{Introduction}

The dark matter density contrast can be traced by galaxies, allowing us to probe the Universe and thus test cosmological models and gravity itself. The underlying dark matter distribution imparts peculiar velocities to galaxies, which distort the galaxy positions along the line of sight. These redshift-space  distortions are dominated by the Kaiser effect but also include further relativistic effects \cite{McDonald:2009ud,Yoo:2010ni,Bonvin:2011bg,Challinor:2011bk}.
In a Fourier space analysis of redshift-space distortions, it is common to use the {plane-parallel or flat-sky approximation} in which the line-of-sight direction from an observer to different galaxy pairs is fixed. 

For next-generation surveys with wide sky coverage, the flat-sky limit cannot be expected to deliver the necessary theoretical accuracy. 
In fact it is necessary for consistency to include the relativistic corrections with the wide-angle corrections -- since they are of the same order of magnitude on ultra-large scales. Furthermore, these corrections can reinforce each other and can also partially cancel each other, depending on the properties of the tracers.
A theoretically complete solution requires  a full-sky analysis, using the 2-point correlation function or its angular harmonic (or spherical Fourier-Bessel) transform and including all wide-angle and relativistic contributions \cite{Challinor:2011bk,Bonvin:2011bg,Bertacca:2012tp,Yoo:2013zga,Tansella:2018sld,Gebhardt:2021mds}.\footnote{See e.g. \cite{Szalay:1997cc,Bharadwaj:1998bq,Matsubara:1999du,Szapudi:2004gh,Papai:2008bd,Raccanelli:2010hk} for earlier work on general wide-angle correlations, without including  relativistic effects.} However this is computationally very intensive and a simpler approach is the approximate inclusion of leading-order {wide-angle} corrections to the plane-parallel limit (see e.g.
\cite{Reimberg:2015jma,Castorina:2017inr,Benabou:2024tmn}).  Recent works have included relativistic effects in the wide-angle corrections of the galaxy power spectrum \cite{Grimm:2020ays} and its multipoles \cite{Beutler:2018vpe,Castorina:2021xzs,Elkhashab:2021lsk,Noorikuhani:2022bwc,Paul:2022xfx}.

%A common approach to derive the wide-angle corrections to the Fourier power spectrum is to perturbatively compute the expansions of appropriate wide-angle terms in configuration space and then take the Fourier transform of the result \cite{Reimberg:2015jma,Castorina:2017inr,Beutler:2018vpe,Paul:2022xfx}. 
\red{In this paper, we follow 
%an alternative 
the perturbative approach adopted by \cite{Noorikuhani:2022bwc}.} We generalise their result from the midpoint line of sight to all possible lines of sight. We include the standard redshift-space distortions and the non-integrated relativistic corrections from Doppler and Sachs-Wolfe effects, thereby generalising \cite{Paul:2022xfx} which only considered the Doppler contribution. The integrated relativistic corrections from  lensing magnification and time delay effects are included in the midpoint results of \cite{Noorikuhani:2022bwc} but we omit these effects in our generalisation, leaving them for future work. We investigate the detectability of the relativistic, wide-angle and combined corrections by computing the $\chi^{2}$ for two futuristic galaxy surveys and their cross power.
\red{If these corrections  to the standard power spectra are detectable, then constraints on cosmological parameters could be affected when neglecting the corrections. In particular, we expect that measurements of the local primordial non-Gaussianity parameter, $f_{\rm NL}$, whose signal is strongest on very large scales, could be affected by relativistic wide-angle corrections.}

\red{We consider linear scalar perturbations on a late-time flat $\Lambda$CDM background, in the Newtonian gauge. The perturbed metric is
\begin{align}
    \ud s^2=a^2(\eta)\,\Big[-(1+2\Phi)\,\ud \eta^2 + (1-2\,\Phi)\,\ud \bm{x}^2 \Big],
\end{align}
where $a$ is the scale factor, $\eta$ is conformal time and $\Phi$ is the gravitational potential.
The matter 4-velocity is
\begin{align}
    u^\mu = \frac{1}{a}\,\Big(1-\Phi, v^i \Big),
\end{align}
where the  peculiar velocity $v^i=\ud x^i/\ud\eta=a\,u^i$ is  irrotational, so that  $v_i=\partial_i V$, where $V$ is the velocity potential.}

\section{Relativistic and wide-angle corrections in the number counts}
 
For a given source population $a$, 
the  dominant terms in the redshift-space number density contrast at linear order {(see \cite{Saga:2021jrh} for beyond-linear effects)}  are 
\begin{equation}
\Delta_{a}(z,\bm{x}_{a}) = b_{a}(z)\,\delta(z,\bm{x}_{a})- \frac{1}{\cH(z)}\,\p_{\|}^{2}V(z,\bm{x}_{a})\,,
\label{e2}
\end{equation}
where $|\bm{x}_a|$ is the \red{background} comoving line-of-sight distance to the \red{background} {redshift $z$}, $b_{a}$ is the bias of tracer $a$, $\cH = (1+z)^{-1}\,H= H_{0}\,(1+z)^{-1}\,[(1+z)^3\,\Omega_{m0}+1-\Omega_{m 0}]^{1/2}$ is the conformal Hubble parameter, 
$\delta$ is the \red{comoving} matter density contrast, 
%$v_i=\partial_iV$ is the peculiar velocity, 
{and} $\p_{\|} =\hat{\bm{x}}_{a}\cdot\bm\nabla$ is the  derivative along the line of sight $\hat{\bm{x}}_{a}$. 
In Fourier space \eqref{e2} is 
\begin{equation}
\Delta_{a}(z,\bm{k},{\hat{\bm{x}}_a}) =  \Big[b_{a}(z)+f(z)\,\big(\hat{\bm{k}}\cdot \hat{\bm{x}}_{a}\big)^{2}\Big]\,\delta(z,\bm{k}) \;,\label{e6}
\end{equation}
where $f=-\ud\ln D/\ud\ln(1+z)$ is the linear matter growth rate,  $D$ is the growth factor (normalized to 1 today), and we used the continuity equation $V = -(\mathcal{H}/k^{2})\,f\,\delta$. 

First, we consider the non-integrated relativistic corrections to the density contrast in \eqref{e2}, which are the Doppler (D) and gravitational potential  ($\Phi$) terms. In Newtonian gauge \cite{Challinor:2011bk}:
\begin{eqnarray}
\Delta^{\mathrm{D}}_{a}(z,\bm{x}_{a}) &=& \bigg[\mathcal{E}_a(z) - 2\,\mathcal{Q}_{a}(z) + \frac{2\,\big[\mathcal{Q}_{a}(z)-1\big]}{x_{a}\,\mathcal{H}(z)} - \frac{\mathcal{H}(z)^{\prime}}{\mathcal{H}(z)^{2}}\bigg]\,\p_{\|}V(z,\bm{x}_{a})\;, \label{eDop} \\ %\nonumber \\
\Delta^{\Phi}_{a}(z,\bm{x}_{a}) &=& \bigg[-1-\mathcal{E}_a(z) + 4\mathcal{Q}_{a}(z) - \frac{2\,\big[\mathcal{Q}_{a}(z)-1\big]}{x_{a}\,\mathcal{H}(z)} + \frac{\mathcal{H}(z)^{\prime}}{\mathcal{H}(z)^{2}}\bigg]\Phi(z,\bm{x}_{a}) \nonumber \\
&&{}+\frac{1}{\mathcal{H}(z)}\Phi^{\prime}(z,\bm{x}_{a}) + (3-\mathcal{E}_a)\,\mathcal{H}(z)\,{V}(z,\bm{x}_{a}) \label{eSW}\;,
\end{eqnarray}
where $\Phi=-(3/2)\Omega_{m}(\mathcal{H}/k)^{2} \delta$ by the Poisson equation. Here $\mathcal{E}$ (which is often denoted $b_{\rm e}$ or $f_{\rm evo}$) and $\mathcal{Q}$ ($=5s/2$) are the evolution and magnification biases (see e.g. \cite{Maartens:2021dqy}): 
\begin{align}
\mathcal{E}_{a} = -\frac{\partial \ln n_a}{\partial \ln (1+z)}\,,\qquad
\mathcal{Q}_{a} = -\frac{\partial \ln n_a}{\partial \ln L_{{\rm c},a}}\,,
\end{align}
where $n_a$ is the background comoving number density of sample $a$ and
$L_{{\rm c},a}$ is the luminosity cut of the sources detected, given the specifications of the survey under consideration.

It is important to highlight the point that here $\delta$ is the matter density contrast in comoving gauge. When we include ultra-large scale effects from relativistic or wide-angle corrections, it is advisable to use the comoving density contrast $\delta$ \cite{Challinor:2011bk,Bruni:2011ta}. If this is not done, then two important relations above break down on ultra-large scales and need to be modified:\\
(a)~the scale-independent bias relation $\delta_a(z,\bm k)=b_a(z)\,\delta(z,\bm k)$, and \\
(b)~the simple form of the Poisson equation, $\nabla^2\Phi\propto\delta$.

In Fourier space, \eqref{eDop} and \eqref{eSW} may be written as \begin{equation}
\Delta^{\mathrm{D}}_{a}
= \mathrm{i}\,\frac{1}{k}\gamma^{\mathrm{D}}_{a}
\big(\hat{\bm{k}}\cdot \hat{\bm{x}}_{a}\big) \delta
\quad \text{and} \quad  \Delta^{\Phi}_{a} 
= \frac{1}{k^{2}}\gamma^{\Phi}_{a}\,
\delta
\;, \label{e30_1} 
\end{equation}
where
\begin{eqnarray}
\gamma^{\mathrm{D}}_{a} &=& \mathcal{H}f\bigg[\mathcal{E}_a - 2\mathcal{Q}_{a} + \frac{2\big(\mathcal{Q}_{a}-1\big)}{x_{a}\,\mathcal{H}} - \frac{\mathcal{H}^{\prime}}{\mathcal{H}^{2}}\bigg]\;,\label{gamma1} \\
\gamma^{\Phi}_{a} &=& \frac{3}{2}\Omega_{m}\mathcal{H}^2\bigg[2 + \mathcal{E}_a- f -4\mathcal{Q}_{a} + \frac{2\big(\mathcal{Q}_{a}-1\big)}{x_{a}\,\mathcal{H}} - \frac{\mathcal{H}^{\prime}}{\mathcal{H}^{2}}\bigg]+\mathcal{H}^2 f(3-\mathcal{E}_a)  \;.\label{gamma2}
\end{eqnarray}
From now on, we neglect terms
\red{in the 2-point correlations $\langle \Delta_a\,\Delta_b \rangle$, i.e.\ quadratic  in the density contrast,}
that are of order $\delta(\mathcal{H}/k)^n$, where $n>2$. For consistency, all terms with $n\leq 2$ should be included \red{in the  2-point correlations and power spectra}.

The relativistic corrections are added to the density contrast in \eqref{e6} to give
\begin{align}
\Delta_{a}(z,\bm{k},{\hat{\bm{x}}_a}) &=  \Big[b_{a}(z)+f(z)\big(\hat{\bm{k}}\cdot \hat{\bm{x}}_{a}\big)^{2} + \mathrm{i}\,\frac{1}{k}\gamma^{\mathrm{D}}_{a}{(z)} \big(\hat{\bm{k}}\cdot \hat{\bm{x}}_{a}\big) +  \frac{1}{k^{2}}\gamma^{\Phi}_{a}{(z)} \Big]\delta(z,\bm{k}) \nonumber \\
&= \mathcal{K}_{a}(z,\bm{k},{\hat{\bm{x}}_a})\,\delta(z,\bm{k})\;,\label{etot}
\end{align}
where $\mathcal{K}_{a}$ is the total Fourier kernel for tracer $a$. 
Typically \eqref{etot} is simplified using the plane-parallel approximation. However, the inclusion of relativistic corrections naturally requires the additional inclusion of wide-angle corrections, which are of the same parametric order, i.e.\ $O[\delta(\mathcal{H}/k)]$. In order to implement wide-angle corrections, we need to specify the geometric set-up.

\begin{figure}[! ht]
\centering
\includegraphics[width=8.0cm]{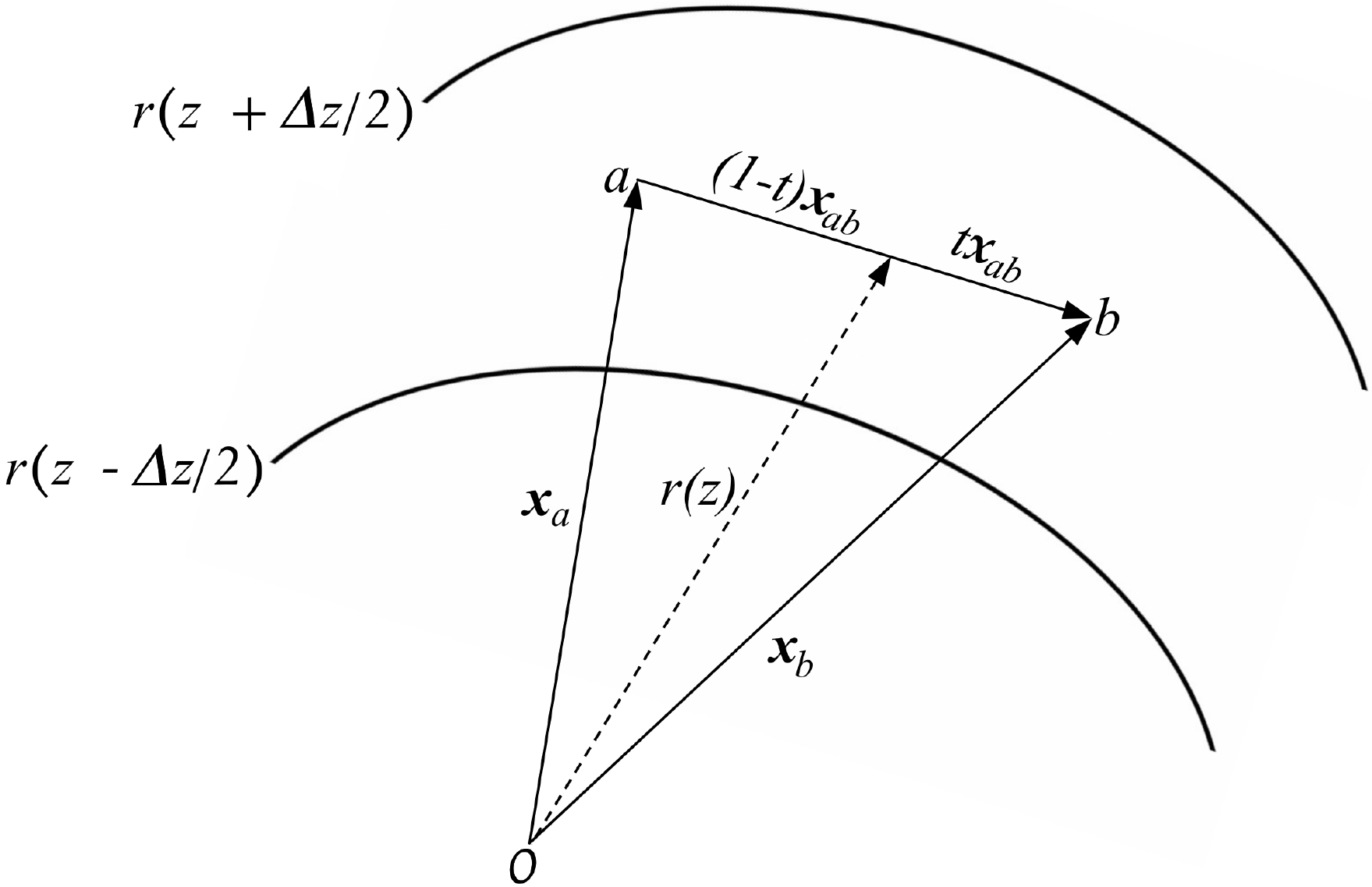}
\caption{Geometry of the two-point correlation function for a pair of fluctuations.} \label{fig1}
\end{figure}

We choose a line-of-sight vector $\bm{r}$ from the observer to the separation vector $\bm{x}_{ab}=\bm x_a-\bm x_b$ (see \autoref{fig1}):
\begin{equation}
\bm{x}_{a} = \bm{r} - (1-t)\,\bm{x}_{ab} \quad \text{and} \quad \bm{x}_{b} = \bm{r} +t\,\bm{x}_{ab}\;,\label{e11}
\end{equation}
where $0 \le t \le 1$. The endpoint ($t=0$ or 1) and midpoint ($t=1/2$) are typical choices.

Then we define the perturbation parameter vector \cite{Noorikuhani:2022bwc}
\begin{equation}
\bm{\epsilon} = \frac{\bm{x}_{ab}}{r}\;,\label{e12}
\end{equation}
which requires that the galaxy separation is less than the line-of-sight distance, i.e., $x_{ab}<r$ (see \cite{Benabou:2024tmn} for a detailed discussion of the validity of the perturbative wide-angle expansion). In Fourier space, $x_{ab}<r$ imposes a minimum $k_\epsilon(z)$
for the validity of the perturbative expansion:
\begin{align}\label{keps}
    k> k_\epsilon(z)= \frac{1}{r(z)}\,.
\end{align}
We follow \cite{Noorikuhani:2022bwc} in neglecting the radial variations in a redshift bin, effectively treating all quantities in a bin as at the mid-redshift.
\red{Although it is straightforward to include these corrections (see \cite{Paul:2022xfx})
they significantly complicate the expressions and we leave 
their inclusion for future work. An example of the effect of 
radial corrections is shown in \autoref{fig6x} below.}
%See \cite{Paul:2022xfx} for the inclusion of the radial corrections.

We rewrite
\eqref{e11} as
\begin{equation}
\bm{x}_{a} = \big[\hat{\bm{r}} - (1-t)\,\bm{\epsilon}\big]\,r \quad \text{and} \quad \bm{x}_{b} = \big(\hat{\bm{r}} + t\,\bm{\epsilon}\big)\,r\;.\label{e13}
\end{equation} 
Taking the products $\bm{x}_{a} \cdot \bm{x}_{a}$, it follows that
\begin{eqnarray}
\frac{1}{x_{a}} &=& \frac{1}{r}\Big[1 + (1-t)\big(\bm{\epsilon} \cdot \hat{\bm{r}}\big) + \frac{3}{2}(1-t)^{2}\big(\bm{\epsilon} \cdot \hat{\bm{r}}\big)^{2} - \frac{1}{2}(1-t)^{2}\epsilon^{2}\Big]\;,\label{e14} \\
\frac{1}{x_{b}} &=& \frac{1}{r}\Big[1 - t\big(\bm{\epsilon} \cdot \hat{\bm{r}}\big) + \frac{3}{2}t^{2}\big(\bm{\epsilon} \cdot \hat{\bm{r}}\big)^{2} - \frac{1}{2}t^{2}\epsilon^{2}\Big] \;.\label{e15}
\end{eqnarray}
\red{Here, and in the remainder of the paper, we neglect terms of $\mathcal{O}({\epsilon}^3)$. Consequently, from now on all relevant expressions are implicitly understood to hold up to $\mathcal{O}\big({\epsilon}^2, \delta (\cH/k)^2\big)$.}
%where we keep terms up to $O({\epsilon}^2)$. 
Using the above results in \eqref{e13}, we obtain
\begin{align}
\hat{\bm{x}}_{a} &= \Big[1 + (1-t)\big(\bm{\epsilon} \cdot \hat{\bm{r}}\big) + \frac{3}{2}(1-t)^{2}\big(\bm{\epsilon} \cdot \hat{\bm{r}}\big)^{2} - \frac{1}{2}(1-t)^{2}\epsilon^{2}\Big] \hat{\bm{r}} - (1-t)\big[1 + (1-t)\big(\bm{\epsilon} \cdot \hat{\bm{r}}\big)\big]\bm{\epsilon}\;,\label{e16} \\
\hat{\bm{x}}_{b} &= \Big[1 - t\big(\bm{\epsilon} \cdot \hat{\bm{r}}\big) + \frac{3}{2}t^{2}\big(\bm{\epsilon} \cdot \hat{\bm{r}}\big)^{2} - \frac{1}{2}t^{2}\epsilon^{2}\Big]\hat{\bm{r}} + t\big[1 - t\big(\bm{\epsilon} \cdot \hat{\bm{r}}\big)\big]\bm{\epsilon}\;.\label{e17}
\end{align}

We  decompose $\bm{k}$ and $\bm{\epsilon}$ into their Cartesian components $(k_x,k_y,k_z)$, taking the $z$-axis to be along the 
line of sight $\bm{r}$ to redshift $z$. Then using \eqref{e16} and \eqref{e17}, we can write the total Fourier kernel given in \eqref{etot} as
\begin{equation}
\mathcal{K}_{a} = \mathcal{K}^{\mathrm{S}}_{a} + \mathcal{K}^{\mathrm{R}}_{a} + \mathcal{K}^{\mathrm{W}}_{a}+ \mathcal{K}^{\mathrm{RW}}_{a}\,,
\label{e17_1}
\end{equation}
where the $\mathcal{K}^{\mathrm{I}}_{a}$ are as follows.
\begin{itemize}
\item S: The standard Newtonian plane-parallel Kaiser part:
\begin{align}
\mathcal{K}^{\mathrm{S}}_{a} &=  b_a + f\left(\frac{k_{z}}{k}\right)^{2}\quad \mbox{where} \quad
k_{z} = \bm{k} \cdot \hat{\bm{r}} \equiv \mu k\;. \label{ee17_1}
\end{align}
\item R: The plane-parallel relativistic corrections to the standard kernel:
\begin{align}
\mathcal{K}^{\mathrm{R}}_{a} &= \mathcal{K}^{\mathrm{D}}_{a} + \mathcal{K}^{\Phi}_{a}\;, \quad \mbox{where}\quad\mathcal{K}^{\mathrm{D}}_{a} = \mathrm{i}\,\frac{k_{z}}{k}\,\gamma^{\mathrm{D}}_{a}\;, \qquad \mathcal{K}^{\Phi}_{a} = \frac{1}{k^{2}}\,\gamma^{\Phi}_{a}\;. \label{e17_2}  
\end{align}
\item W: The Newtonian wide-angle corrections  to the standard kernel:\\  \hspace*{2cm}~{$\mathcal{K}^{\mathrm{W}}_{a}$, given in \autoref{AppA}.}
\item RW: The wide-angle corrections to the plane-parallel relativistic kernel:\\  \hspace*{2cm} $\mathcal{K}^{\mathrm{RW}}_{a}$, given in \autoref{AppA}. 
\end{itemize}
Then the full kernel, i.e.\ standard + combined relativistic and wide-angle corrections,  is
\begin{align}
\mathcal{K}_{a} = \mathcal{K}^{\mathrm{S}}_{a} +  
\mathcal{K}^{\mathrm{W}}_{a} + \mathcal{K}^{\mathrm{R}}_{a}  + \mathcal{K}^{\mathrm{RW}}_{a} \,.
\end{align}

\section{Auto- and cross-power spectra}

The Fourier power spectra are
\begin{equation}
P_{ab}(\bm{k},{\bm r}) = \int \frac{\ud^{3}\bm k^{\prime}}{(2\pi)^{3}} \int \ud^{3}\bm{\epsilon}\;r\,\mathrm{e}^{-\mathrm{i}\,r(\bm{k}-\bm{k}^{\prime})\cdot \bm{\epsilon}}\,\mathcal{K}_{a}(\bm{k}^{\prime},\bm{\epsilon},{\bm r})\,\mathcal{K}_{b}^{*}(\bm{k}^{\prime},\bm{\epsilon},{\bm r})\,P(k^{\prime})\;, \label{eWPower}
\end{equation}
where $P$ is the \red{linear power spectrum of the comoving matter density contrast $\delta$.} Here and below, we omit the dependencies on $z$ for brevity. 
These are the {\em`local'} power spectra, i.e.\ the Fourier power spectra at line-of-sight position $\bm{r}(z)$. The {\em `full'} power spectra are then given by an integral over all lines of sight \cite{Noorikuhani:2022bwc},
\begin{align}
P_{ab}^{\rm full}(\bm{k})= \frac{1}{V_{\rm s}}\int \ud^3\bm r\, P_{ab}(\bm{k},\bm r)  \,,  
\end{align}
where $V_{\rm s}(z)$ is the survey volume. In this paper we will work with the local power spectra.

The product of Fourier kernels in  \eqref{eWPower} can be written in the form \cite{Noorikuhani:2022bwc}:
\begin{equation}
\mathcal{K}_{a}(\bm{k}',\bm{\epsilon},{\bm r})\,\mathcal{K}_{b}^{*}(\bm{k}',\bm{\epsilon},{\bm r}) = \sum_{l,m,n}\,\mathcal{C}_{lmn}^{(ab)}(\bm{k}',{\bm r})\,\epsilon_{x}^{l}\,\epsilon_{y}^{m}\,\epsilon_{z}^{n}
\red{\quad \mbox{where}\quad l+m+n \le 2}
\;.\label{e18}
\end{equation}
Here the coefficients $\mathcal{C}_{lmn}^{(ab)}$ are independent of $\bm{\epsilon}$ and \red{the condition on $l,m,n$ excludes terms of $\mathcal{O}(\epsilon^3)$}. Then the power spectrum  \eqref{eWPower} becomes
\begin{equation}
P_{ab}(\bm{k},{\bm r}) = \sum_{l,m,n}\, \int \frac{\ud^{3}\bm{k}^{\prime}}{(2\pi)^{3}} \int \ud^{3}\bm{x}_{ab}\;\mathrm{e}^{-\mathrm{i}(\bm{k}-\bm{k}^{\prime})\cdot \bm{x}_{ab}}\,\epsilon_{x}^{l}\,\epsilon_{y}^{m}\,\epsilon_{z}^{n}\,\mathcal{C}_{lmn}^{(ab)}(\bm{k}^{\prime},{\bm r}) \,P(k^{\prime})\;,\label{e19}
\end{equation}
where $r\,\ud^{3}\bm\epsilon = \ud^{3}\bm{x}_{ab}$ from \eqref{e12}. The useful relation 
\begin{equation}
\mathrm{e}^{-\mathrm{i}(\bm{k}-\bm{k}^{\prime})\cdot \bm{x}_{ab}}\,\epsilon_{x}^{l}\,\epsilon_{y}^{m}\,\epsilon_{z}^{n}=
\left(\frac{\mathrm{-i}}{r}\right)^{\!l+m+n}\!\left(\frac{\p}{\p k_{x}^{\prime}}\right)^l \! \left(\frac{\p}{\p k_{y}^{\prime}}\right)^m \! \left(\frac{\p}{\p k_{z}^{\prime}}\right)^n \! \big[\mathrm{e}^{-\mathrm{i}(\bm{k}-\bm{k}^{\prime})\cdot \bm{x}_{ab}}\big] \;,\label{e20}
\end{equation}
leads to
\begin{align}
P_{ab} = \sum_{l,m,n}\, 
\left(\frac{-\mathrm{i}}{r}\right)^{l+m+n}\, \int \frac{\ud^{3}\bm{k}^{\prime}}{(2\pi)^{3}}\;I_{lmn}^{(ab)}\;,\label{e21} 
\end{align}
{where}
\begin{align}
I_{lmn}^{(ab)} = \mathcal{C}_{lmn}^{(ab)}\,P
\left(\frac{\p}{\p k_{x}^{\prime}}\right)^l \! \left(\frac{\p}{\p k_{y}^{\prime}}\right)^m \! \left(\frac{\p}{\p k_{z}^{\prime}}\right)^n
\int \ud^{3}\bm{x}_{ab}\,\mathrm{e}^{-\mathrm{i}(\bm{k}-\bm{k}^{\prime})\cdot \bm{x}_{ab}}\;.\label{e22}
\end{align}

Here we moved the $\bm{\epsilon}$-independent coefficients and derivative operators out of the $\bm{x}_{ab}$ integral, which gives the 3D Dirac-delta function: 
\begin{equation}
\int \ud^{3}\bm{x}_{ab}\,\mathrm{e}^{\mathrm{i}(\bm{k}^{\prime}-\bm{k})\cdot \bm{x}_{ab}} = (2\pi)^{3}\delta^{\mathrm{Dirac}}(\bm{k}^{\prime}-\bm{k}). \label{e23}
\end{equation}
Using the  identity 
\begin{equation}
\int \ud x\,f(x)\frac{\p^{q}}{\p x^{q}}\big[\delta^{\mathrm{Dirac}}(x-x_{0})\big] = (-1)^{q}\frac{\p^{q}}{\p x^{q}}f(x)\bigg|_{x=x_{0}}\;,\label{e24}
\end{equation} 
we find that
\begin{align}
P_{ab} &= \sum_{l,m,n}\, \left(\frac{\mathrm{i}}{r}\right)^{l+m+n} \!\left(\frac{\p}{\p k_{x}^{\prime}}\right)^l \! \left(\frac{\p}{\p k_{y}^{\prime}}\right)^m \! \left(\frac{\p}{\p k_{z}^{\prime}}\right)^n\Big[\mathcal{C}_{lmn}^{(ab)}\,P\Big]
\notag \\ &
= P^{\mathrm{S}}_{ab} + P^{\mathrm{R}}_{ab} + P^{\mathrm{W}}_{ab} + P^{\mathrm{RW}}_{ab}
%O\big[\epsilon^3,\delta({\cal H}/k)^3 \big]
\;.\label{ePtotal}
\end{align}
The non-vanishing coefficients $\mathcal{C}_{lmn}^{(ab)}$ in \eqref{ePtotal} are given in \autoref{AppB} and 
the individual spectra on the right of \eqref{ePtotal} are as follows:
\begingroup
\allowdisplaybreaks
\begin{eqnarray}
P^{\mathrm{S}}_{ab} &=& \mathcal{K}^{\rm S}_{a} \, \mathcal{K}^{\rm S}_{b}\, P  \;, \label{P1}\\
P^{\mathrm{R}}_{ab} &=& \Big[ \mathcal{K}^{\rm S}_{a} \, \mathcal{K}^{\rm D*}_{b} + \mathcal{K}^{\rm S}_{b} \, \mathcal{K}^{\rm D}_{a}  + \mathcal{K}^{\rm S}_{a} \, \mathcal{K}^{\rm \Phi}_{b}  + \mathcal{K}^{\rm S}_{b} \, \mathcal{K}^{\rm \Phi}_{a}  + \mathcal{K}^{\rm D}_{a} \, \mathcal{K}^{\rm D*}_{b} \Big] P \;, \label{P2} \\
%\end{align}
%together with
%\begingroup
%\allowdisplaybreaks
%\begin{align}
%\begin{eqnarray}
P^{\mathrm{W}}_{ab} &=& \frac{1}{(k r)^{2}} f  \Bigg\{ 8 t (1 - t)  f \mu^{2}  [1 + 2 \mu^{2} ( - 5 + 6 \mu^{2})] 
 \label{P3} \notag
\\\notag
&& {} \hspace{1.5cm} + 2  \mu^2  (- 7 + 10 \mu^{2}) \Big[t^{2} \,  k \,  \partial_{k} \mathcal{K}^{\rm S}_{a} + (1 - t)^{2} \,  k \,  \partial_{k} \mathcal{K}^{\rm S}_{b}\Big] 
\\\notag
&& {} \hspace{1.5cm} +  4 \mu^3  \Big[ t^{2} \,  k \,  \partial_{k_z} \mathcal{K}^{\rm S}_{a} + (1 - t)^{2} \,  k \,  \partial_{k_z} \mathcal{K}^{\rm S}_{b}\Big]
\\\notag
&& {} \hspace{1.5cm} + (1 - \mu^2)  ( - 1 + 4 \mu^2) \Big[ t^{2} \,  k^2 \, \partial_{k}^{2}  \mathcal{K}^{\rm S}_{a} + (1 - t)^{2} \,  k^2 \, \partial_{k}^{2}  \mathcal{K}^{\rm S}_{b}\Big] 
\\\notag
&& {} \hspace{1.5cm} + 2 (1 - \mu^2) \mu  \Big[t^{2} \,  k^2 \, \partial_{k_z} \partial_{k} \mathcal{K}^{\rm S}_{a} + (1 - t)^{2} \,  k^2 \, \partial_{k_z} \partial_{k}  \mathcal{K}^{\rm S}_{b}\Big]   \Bigg\} P
\\\notag
&& {} + \mathrm{i} \, \frac{2}{k r} f \mu \Big\{ 2 \mu^{2} \Big[t \mathcal{K}^{\rm S}_{a} - (1 - t) \mathcal{K}^{\rm S}_{b}\Big]
\\\notag
&& {} \hspace{1.8cm} + (1 - \mu^{2}) \Big[ t \,  k \,  \partial_{k} \mathcal{K}^{\rm S}_{a}  -  (1 - t) \,  k \,  \partial_{k} \mathcal{K}^{\rm S}_{b}\Big]
\Big\} P
\\\notag
&& {} + \frac{2}{(k r)^{2}} f \Bigg\{ 2 t (1 - t)  f  \mu^{2}   (1 - \mu^2) (- 2 + 9 \mu^{2}) 
\\\notag
&& {} \hspace{2.0cm} +  \mu^{2}  (- 7 + 10 \mu^{2})  \Big[t^{2} \mathcal{K}^{\rm S}_{a} + (1 - t)^{2} \mathcal{K}^{\rm S}_{b}\Big]  
\\\notag
&& {} \hspace{2.0cm} + (1 - \mu^2) (- 1 + 4 \mu^2)  \Big[t^{2}   \,  k \,  \partial_{k} \mathcal{K}^{\rm S}_{a} + (1 - t)^{2}   \,  k \,  \partial_{k} \mathcal{K}^{\rm S}_{b}\Big] 
\\\notag
&& {} \hspace{2.0cm} + (1 - \mu^2)  \mu  \Big[t^{2}   \,  k \,  \partial_{k_{z}} \mathcal{K}^{\rm S}_{a} + (1 - t)^{2}   \,  k \,  \partial_{k_{z}} \mathcal{K}^{\rm S}_{b}\Big] 
\Bigg\} \, k \, \partial_{k} P 
\\\notag
&& {} + \mathrm{i} \, \frac{2}{k r}  f  \mu  (1 - \mu^{2}) \Big[t \mathcal{K}^{\rm S}_{a} - (1 - t) \mathcal{K}^{\rm S}_{b}\Big]  \, k \, \partial_{k} P 
\\\notag
&& {} + \frac{1}{(k r)^{2}} f  (1 - \mu^{2})  \Bigg\{ 4 t  (1 - t) f \mu^{2} (1 - \mu^{2}) 
\\ && {} \hspace{3.0cm} 
+ (- 1 + 4 \mu^2)  \Big[t^{2}  \mathcal{K}^{\rm S}_{a} + (1 -  t)^{2}  \mathcal{K}^{\rm S}_{b}\Big]  \Bigg\} \, k^{2} \, \partial_{k}^{2} P \;,  \label{P3}\\
%and
%\begin{eqnarray}
P^{\mathrm{RW}}_{ab} &=& f \Bigg\{ -\frac{1}{k r} \frac{\mathcal{H}}{k}  \bigg\{2 \mu^{2} \bigg[t \bigg(2 \mathcal{Q}_{b} - \mathcal{E}_{b} +  \frac{\mathcal{H}^{'}}{\mathcal{H}^2}\bigg) \mathcal{K}^{\rm S}_{a}  + (1 - t) \bigg(2 \mathcal{Q}_{a} - \mathcal{E}_a +  \frac{\mathcal{H}^{'}}{\mathcal{H}^2}\bigg) \mathcal{K}^{\rm S}_{b}\bigg] \notag 
\\ \notag
&& {} \hspace{0.5cm} +  (1 - \mu^{2}) \bigg[t \bigg(2 \mathcal{Q}_{b} - \mathcal{E}_{b}  +  \frac{\mathcal{H}^{'}}{\mathcal{H}^2}\bigg) \,  k \, \partial_{k} \mathcal{K}^{\rm S}_{a}
 + (1 - t) \bigg(2 \mathcal{Q}_{a} - \mathcal{E}_a +  \frac{\mathcal{H}^{'}}{\mathcal{H}^2}\bigg) \,  k \,  \partial_{k} \mathcal{K}^{\rm S}_{b}\bigg]   \bigg\}
\\ \notag
&& {} \hspace{0.5cm} + \frac{2}{(k r)^{2}}  \Big\{ (1 - 4 \mu^{2}) \Big[t (1 - \mathcal{Q}_{b}) \mathcal{K}^{\rm S}_{a} + (1 - t) (1 - \mathcal{Q}_{a}) \mathcal{K}^{\rm S}_{b}\Big] 
\\ \notag
&& {} \hspace{2.0cm} + ( 2 \mu^2-1) \Big[t (1 - \mathcal{Q}_{b})  \,  k \, \partial_{k} \mathcal{K}^{\rm S}_{a} + (1 - t) (1 - \mathcal{Q}_{a}) \,  k \, \partial_{k} \mathcal{K}^{\rm S}_{b}\Big]
\\ \notag
&& {} \hspace{2.0cm} + \mu \Big[t (1 - \mathcal{Q}_{b}) \,  k \, \partial_{k_z} \mathcal{K}^{\rm S}_{a} + (1 - t) (1 - \mathcal{Q}_{a}) \,  k \, \partial_{k_z} \mathcal{K}^{\rm S}_{b}\Big]  \Big\} 
\\ \notag
&& {} \hspace{0.5cm} + \mathrm{i} \,  \frac{2}{k r} \mu \Big\{2 \mu^{2} \Big[ t \mathcal{K}^{\rm D}_{a} - (1 - t) \mathcal{K}^{\rm D*}_{b}\Big]  
\\ \notag
&& {} \hspace{0.5cm} +  (1 - \mu^{2}) \Big[ t \,  k \,  \partial_{k} \mathcal{K}^{\rm D}_{a} - (1 - t) \,  k \,  \partial_{k} \mathcal{K}^{\rm D*}_{b}\Big] \Big\} 
\Bigg\} P
\\ \notag
&& {} +  f  \Bigg\{  \frac{1}{k r} \frac{\mathcal{H}}{k}  ( \mu^{2}-1)  \bigg[t  \bigg(2 \mathcal{Q}_{b} -  \mathcal{E}_{b} +  \frac{\mathcal{H}^{'}}{\mathcal{H}^2}\bigg) \mathcal{K}^{\rm S}_{a}
 + (1 - t)  \bigg( 2 \mathcal{Q}_{a} -  \mathcal{E}_a +  \frac{\mathcal{H}^{'}}{\mathcal{H}^2} \bigg) \mathcal{K}^{\rm S}_{b}\bigg] 
\\ \notag
&& {} \hspace{1.0cm} + \frac{2}{(k r)^2} ( 2 \mu^{2}-1) \Big[t  (1 - \mathcal{Q}_{b}) \mathcal{K}^{\rm S}_{a}  + (1 - t)  (1 - \mathcal{Q}_{a}) \mathcal{K}^{\rm S}_{b}\Big] 
\\ \notag
&& {} \hspace{1.0cm} + \mathrm{i} \, \frac{2}{k r} \mu  (1 - \mu^{2})  \Big[t \mathcal{K}^{\rm D}_{a} - (1 - t) \mathcal{K}^{\rm D*}_{b}\Big] 
\Bigg\} \, k \, \partial_{k} P 
\\ \notag
&& {} +  \frac{ 1}{(k r)^{2}} f  (1 - \mu^{2})  \Bigg\{ 4 t  (1 - t)  f \mu^{2} (1 - \mu^{2})
\\  
&& {} \hspace{3.50cm} + (4 \mu^2-1)  \Big[t^{2}  \mathcal{K}^{\rm S}_{a} + (1 - t)^{2} \mathcal{K}^{\rm S}_{b}\Big]  \Bigg\} \, k^{2} \, \partial_{k}^{2} P \;. \label{P4} 
\end{eqnarray}
\endgroup

Note that in the expressions \eqref{P3} and \eqref{P4}, there are real terms  with a factor of $\mu \partial_{k_z} \mathcal{K}^{\rm S}_a$ or  $\mu \partial_{k_z} \partial_k\mathcal{K}^{\rm S}_a$, which appear to be terms multiplied by an odd power of $\mu$. However the $k_z$ derivative introduces another factor of $\mu=k_z/k$, as can be seen in \eqref{a5} and \eqref{a6}, so that these terms are indeed even in $\mu$.

The multipoles of the power spectra are given by
\begin{equation}
P_{ab}^{{\rm I}\,(\ell)}(k) = \frac{(2\ell + 1)}{2}\int_{-1}^{+1}\ud \mu\,P_{ab}^{\rm I}( k, \mu)\,\mathcal{L}_{\ell}(\mu)~~ \mbox{where I} =  \mbox{S,R,W,RW or R+W+RW}.\label{eMultipole}
\end{equation}
\red{Here $\mathcal{L}_{\ell}$ are Legendre polynomials.
The
detailed expressions for the dominant multipoles $\ell = 0, 1, 2$ are given   in \autoref{AppC}.}

\section{Illustrating the relativistic and wide-angle effects}

\begin{figure}[! h]
\centering
\includegraphics[width=7.0cm]{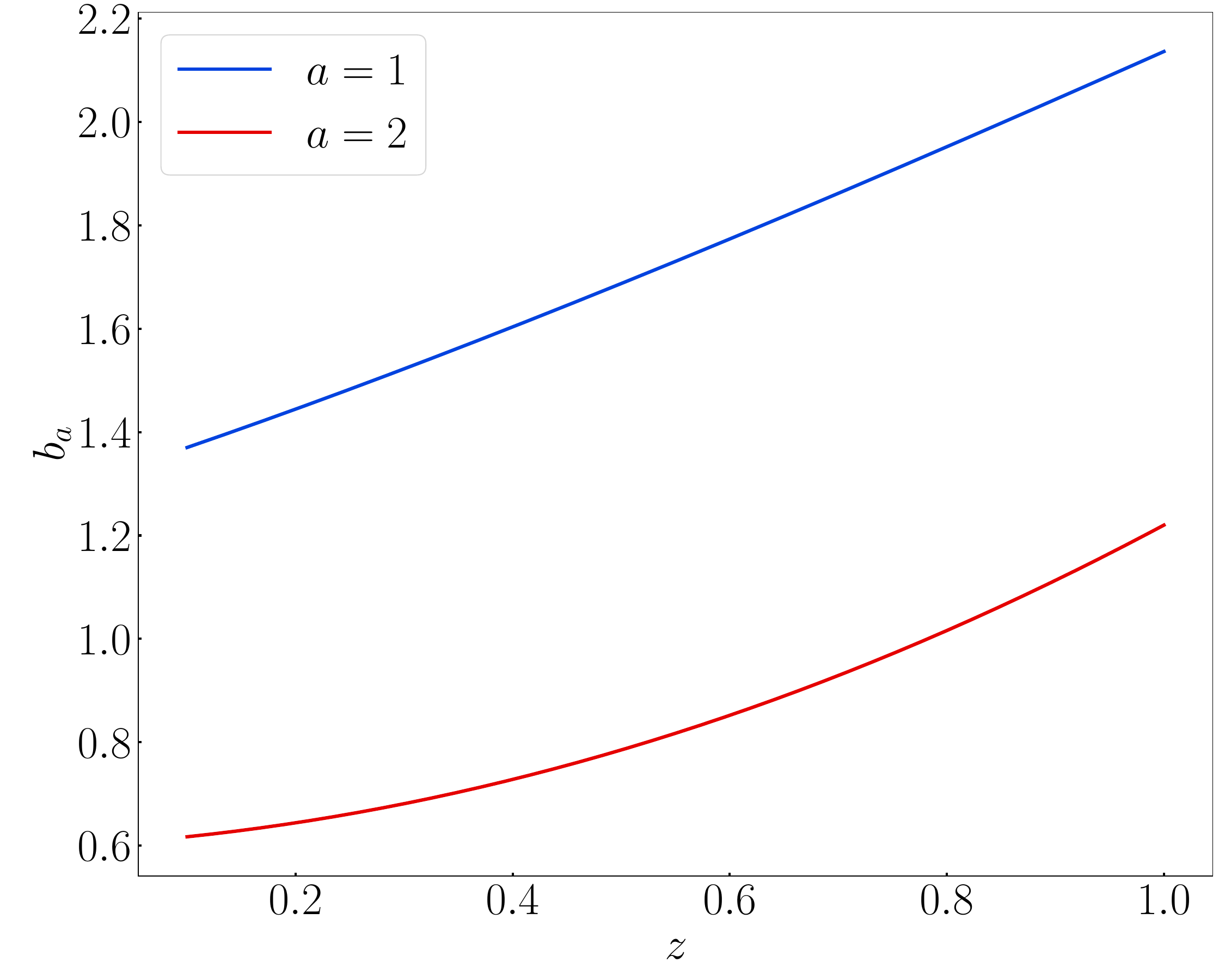} 
\includegraphics[width=7.0cm]{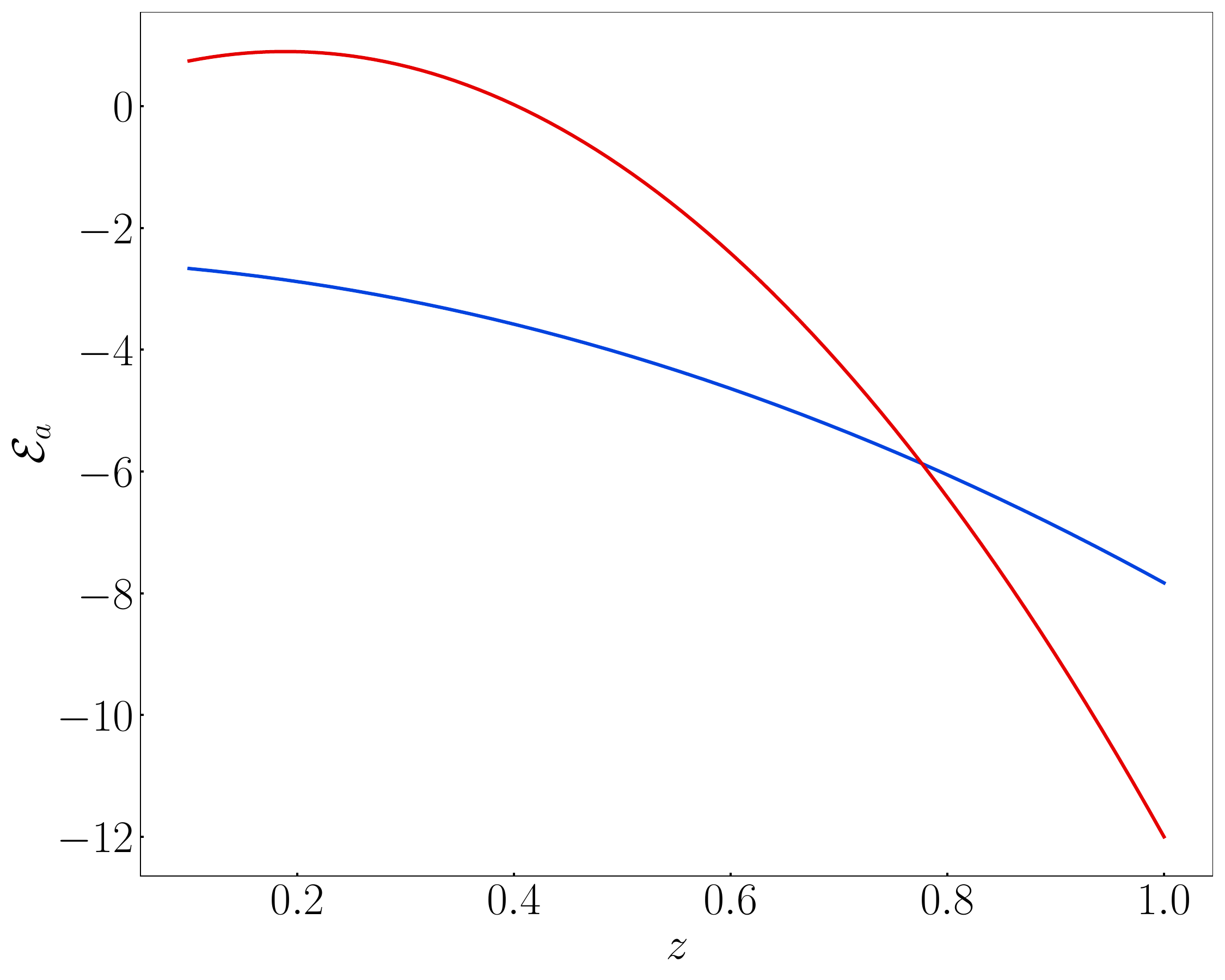} \\
\includegraphics[width=7.0cm]{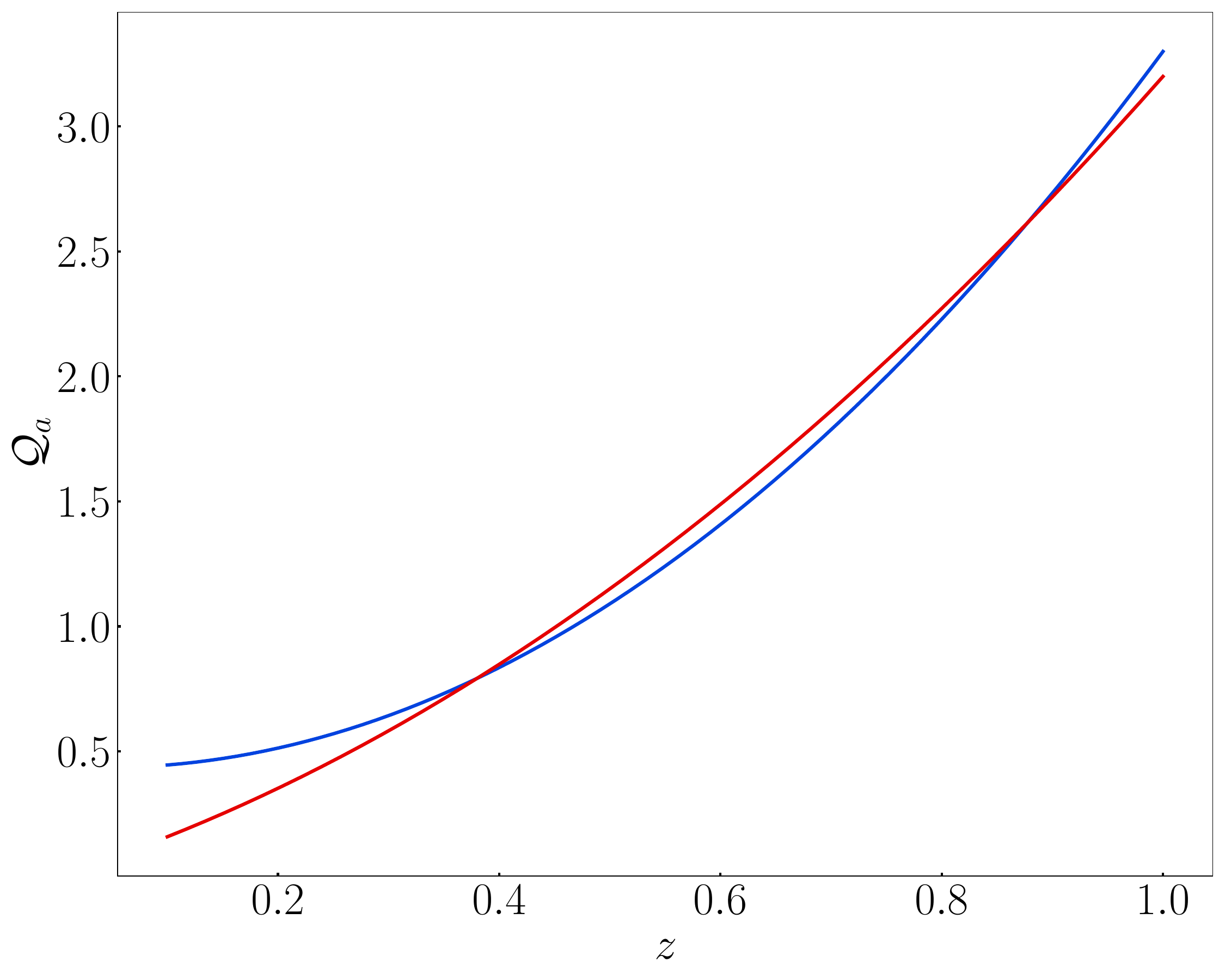}
\includegraphics[width=7.0cm]{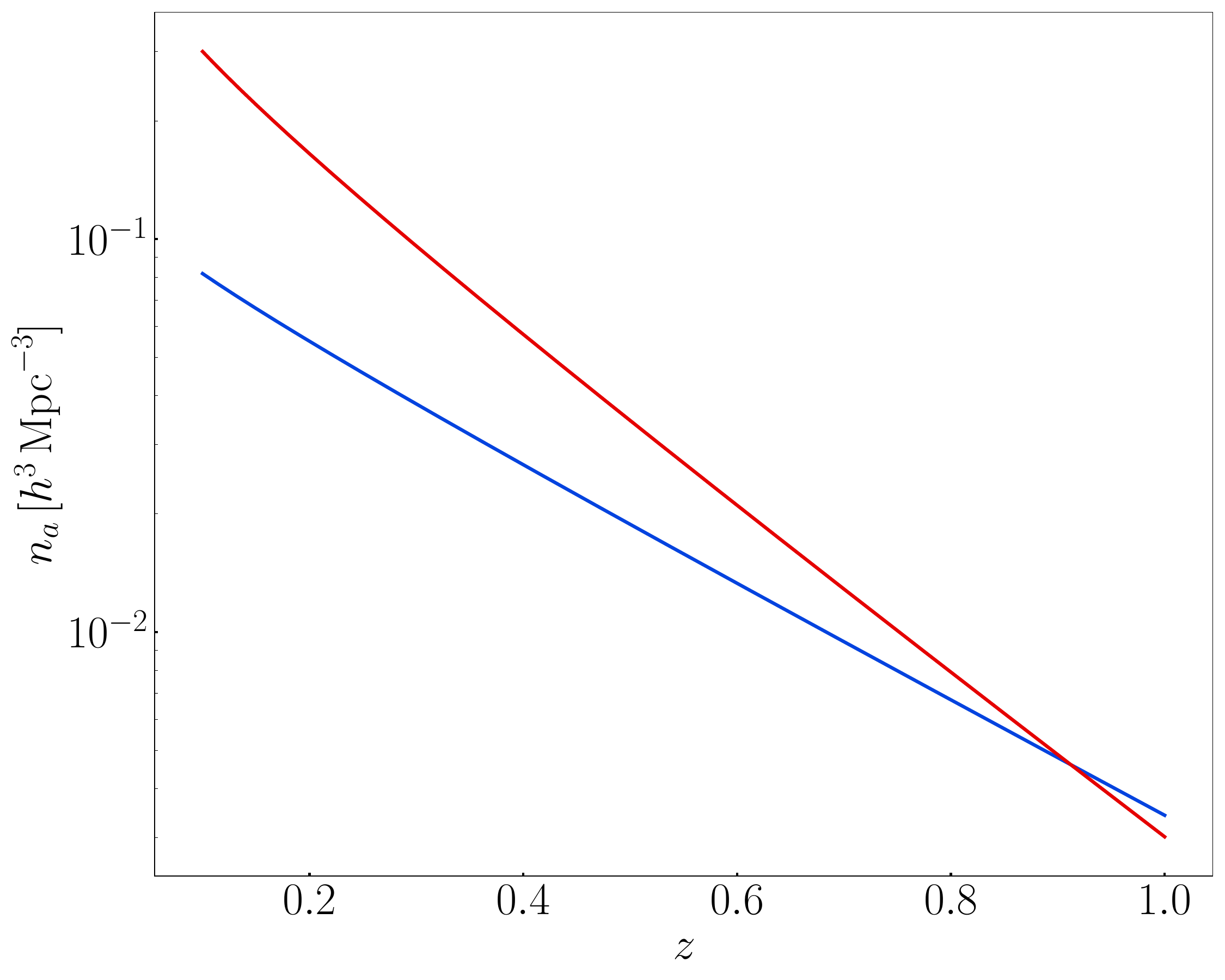}
\caption{ Clustering bias (\emph{top left}), evolution bias (\emph{top right}), magnification bias (\emph{bottom left}) and number density (\emph{bottom right}) for the 2 futuristic galaxy samples, where $a=1$ is `beyond-DESI' and $a=2$ is SKA Phase 2. } \label{fig2}
\end{figure}

In order to illustrate the relativistic and wide-angle effects, we need examples of galaxy samples.
For each sample we require the clustering, evolution and magnification biases, together with the number densities. 
It is important that these quantities are {\em physically self-consistent}. This is achieved by using clustering biases $b_a$ that are based on simulations or a halo model, and by deriving  $\mathcal{E}_a$, $\mathcal{Q}_a$ and $n_a$ from a physically motivated luminosity function. To this end, we adapt the models  described in \cite{Maartens:2021dqy} to the following futuristic mock samples:
\begin{itemize}
\item $a=1$: a futuristic  `beyond-DESI' BGS sample with a fainter magnitude cut, $m_c=22$.
\item $a=2$: a futuristic SKA Phase 2 HI galaxy sample with flux density cut $S_c=5\,\mu$Jy. 
\end{itemize}
For the clustering biases $b_a$ ($a=1,2$) we use \cite{Maartens:2021dqy}:
\begin{align}
 b_{1} &= \frac{1.3}{D(z)}\;, 
\qquad\qquad\quad b_{2} = 0.60 + 0.12 z + 0.50 z^{2}\;.\label{ebg2}   
\end{align}
We apply the futuristic flux cuts, $m_c=22$ (for $a=1$) and $S_c=5\,\mu$Jy (for $a=2$), to the luminosity function models in \cite{Maartens:2021dqy}. Then we find simple fits to the results:
\begin{align}
\mathcal{E}_{1} &=  -2.54 - 0.81 z - 4.48 z^{2} \;, \qquad \quad\quad\qquad \, \mathcal{E}_{ 2} = 0.2 + 7.4z - 19.6z^{2} \;,\label{ebeQ1} \\
\mathcal{Q}_{1} &=  0.44 - 0.26 z + 3.12 z^{2} , \qquad \qquad\quad \qquad \mathcal{Q}_{2} = 1.4z + 1.8z^{2}\;, \label{ebe2Q2} \\
n_{1} &= 0.09 z^{-0.10}\,\exp{(-3.27 z)}~ h^{3}{\rm Mpc}^{-3}, \quad~~~ n_{2} = 0.3z^{-0.2} \, \exp{(-4.6z)}~ h^{3}{\rm Mpc}^{-3}.
\label{ng2}
\end{align}
For the redshift range and sky area, we assume
\begin{align}
0.1 \le z \le 1\,, \qquad  \Omega_{\rm sky} = 15\,000~\deg^{2}\,.
\end{align}
The biases $b_a,\mathcal{E}_a,\mathcal{Q}_a$ and number densities $n_a$ are presented in \autoref{fig2}. 

\begin{figure}[! ht]
\centering
\includegraphics[width=0.45\textwidth]{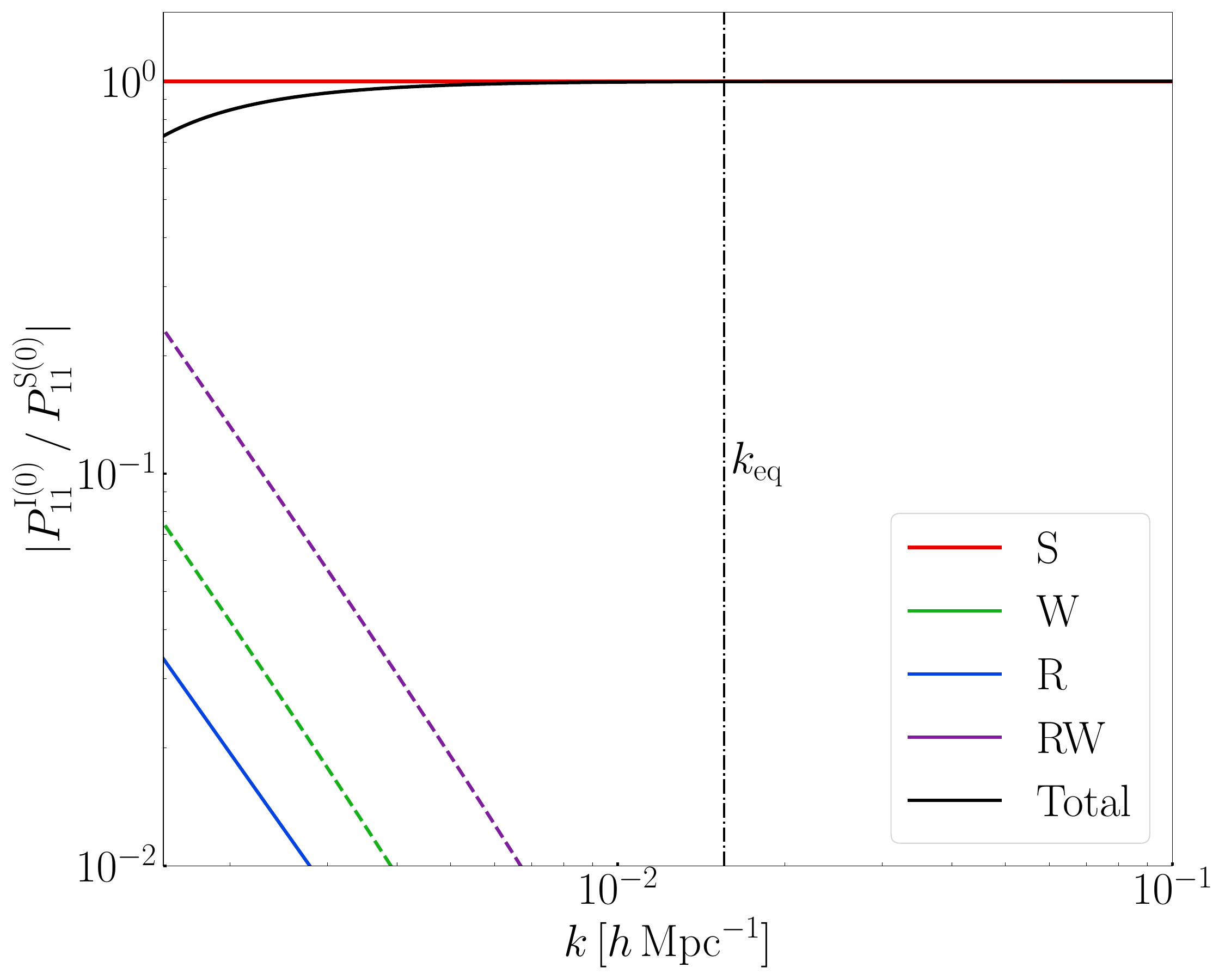} 
\includegraphics[width=0.45\textwidth]{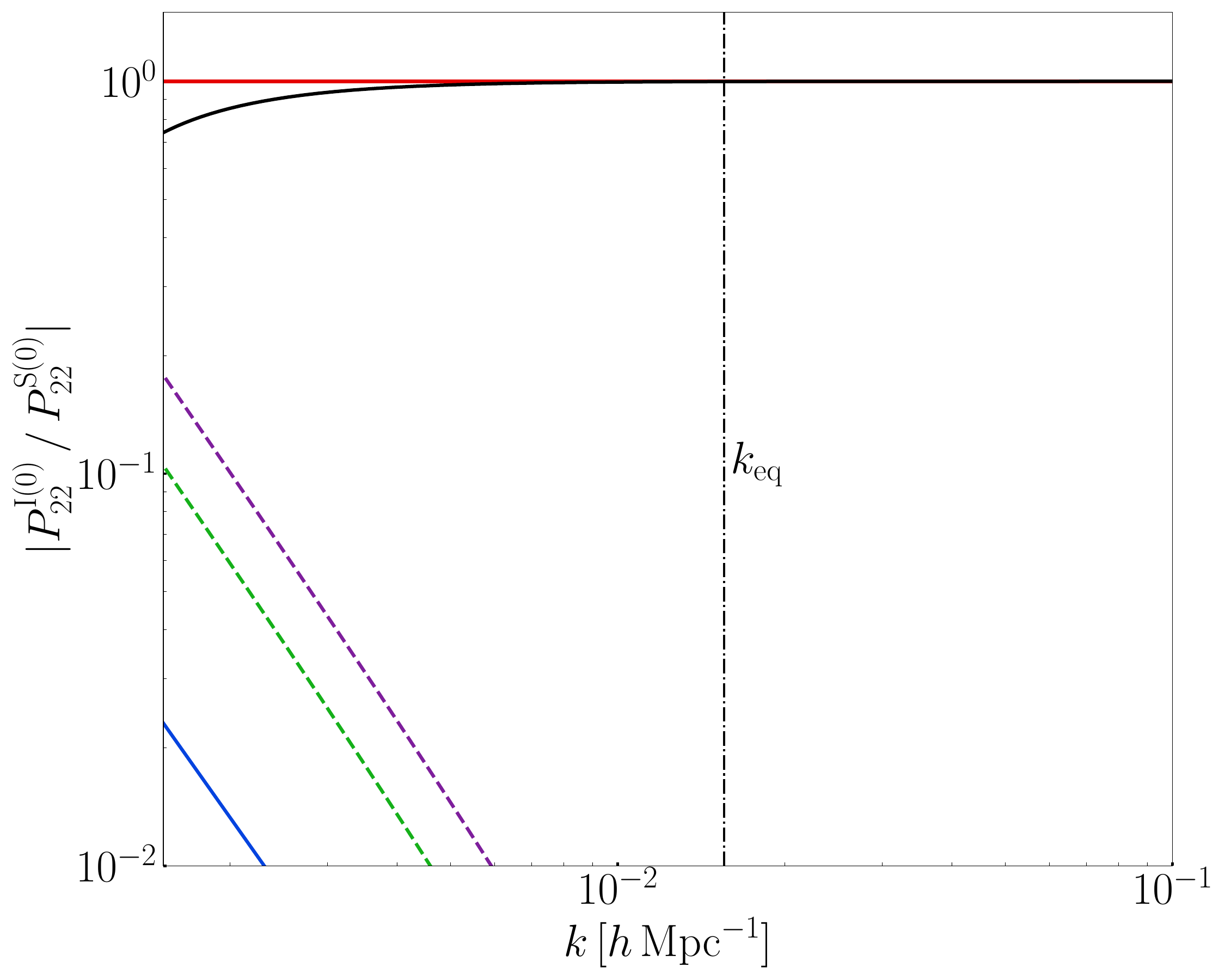} \\
\includegraphics[width=0.45\textwidth]{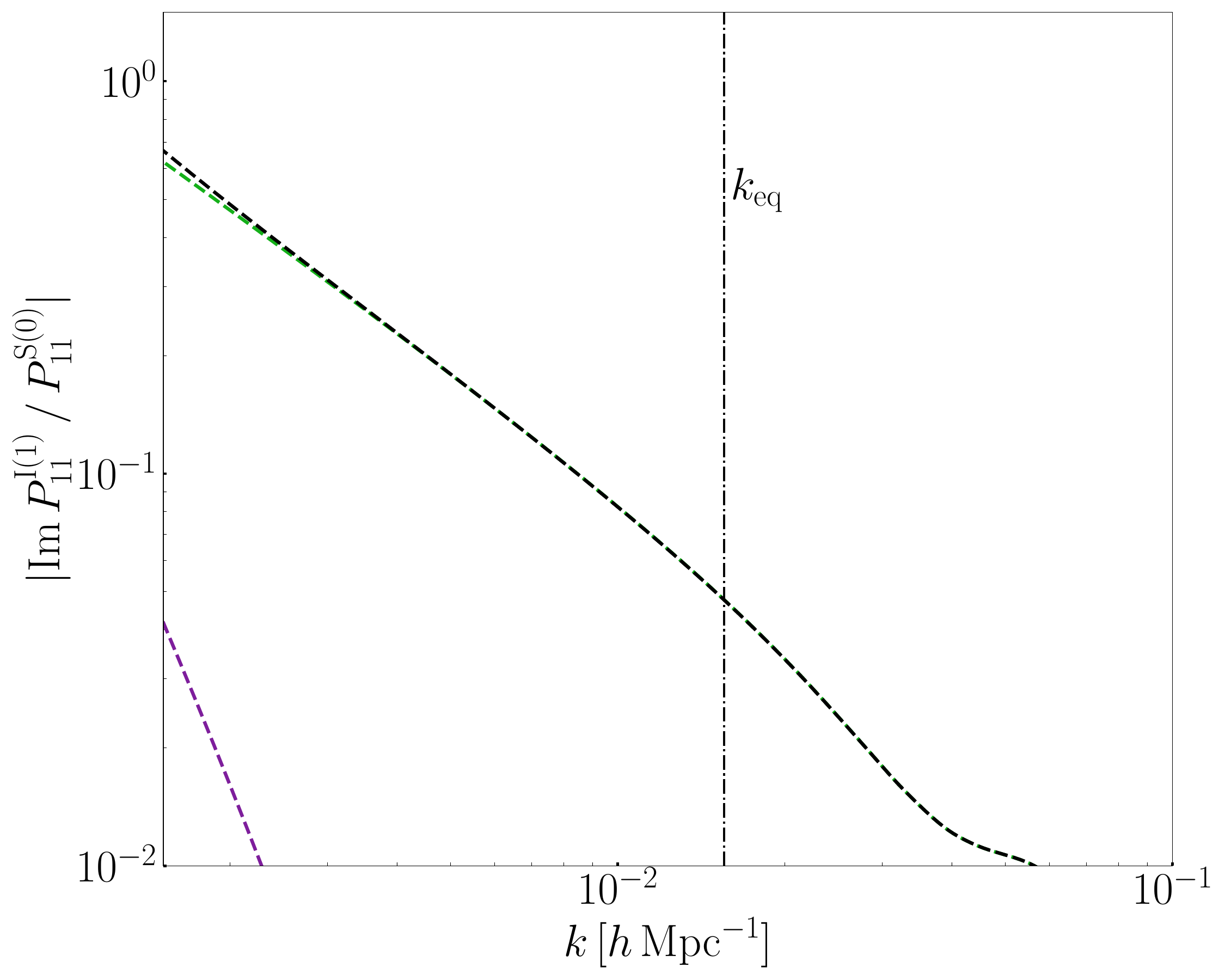} 
\includegraphics[width=0.45\textwidth]{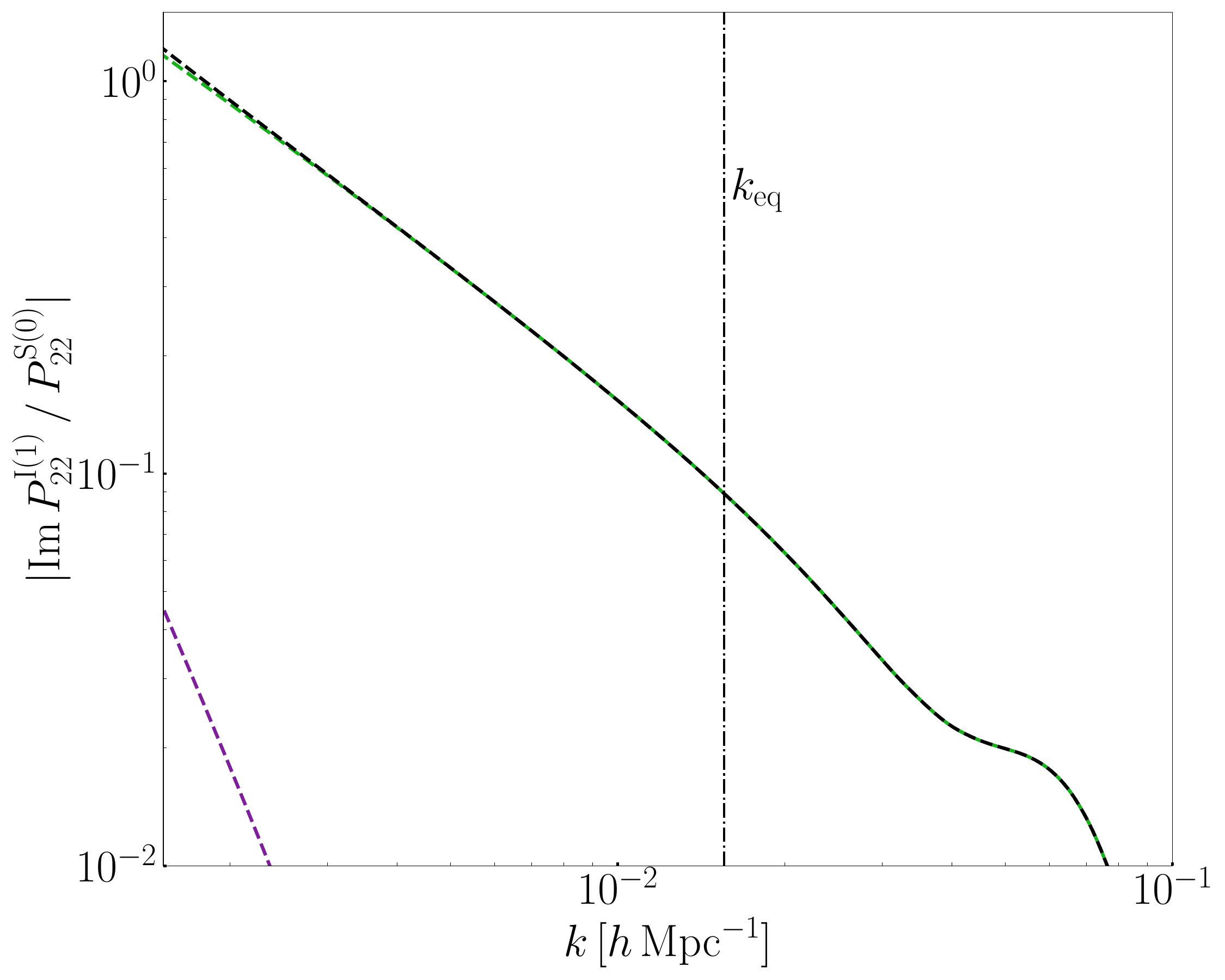} 
\caption{For the auto-power, the
fractional correction to the standard monopole (\emph{top row}) and the dipole  {relative to the standard monopole}  (\emph{bottom row}), at $z=0.5$. Galaxy samples are $a=1$  ({\em left}) and $a=2$ ({\em right}). The line of sight is endpoint, $t=0$.  Dashed lines indicate negative values. 
} \label{fig3a}
\end{figure}

\begin{figure}[! ht]
\centering
\includegraphics[width=7.0cm]{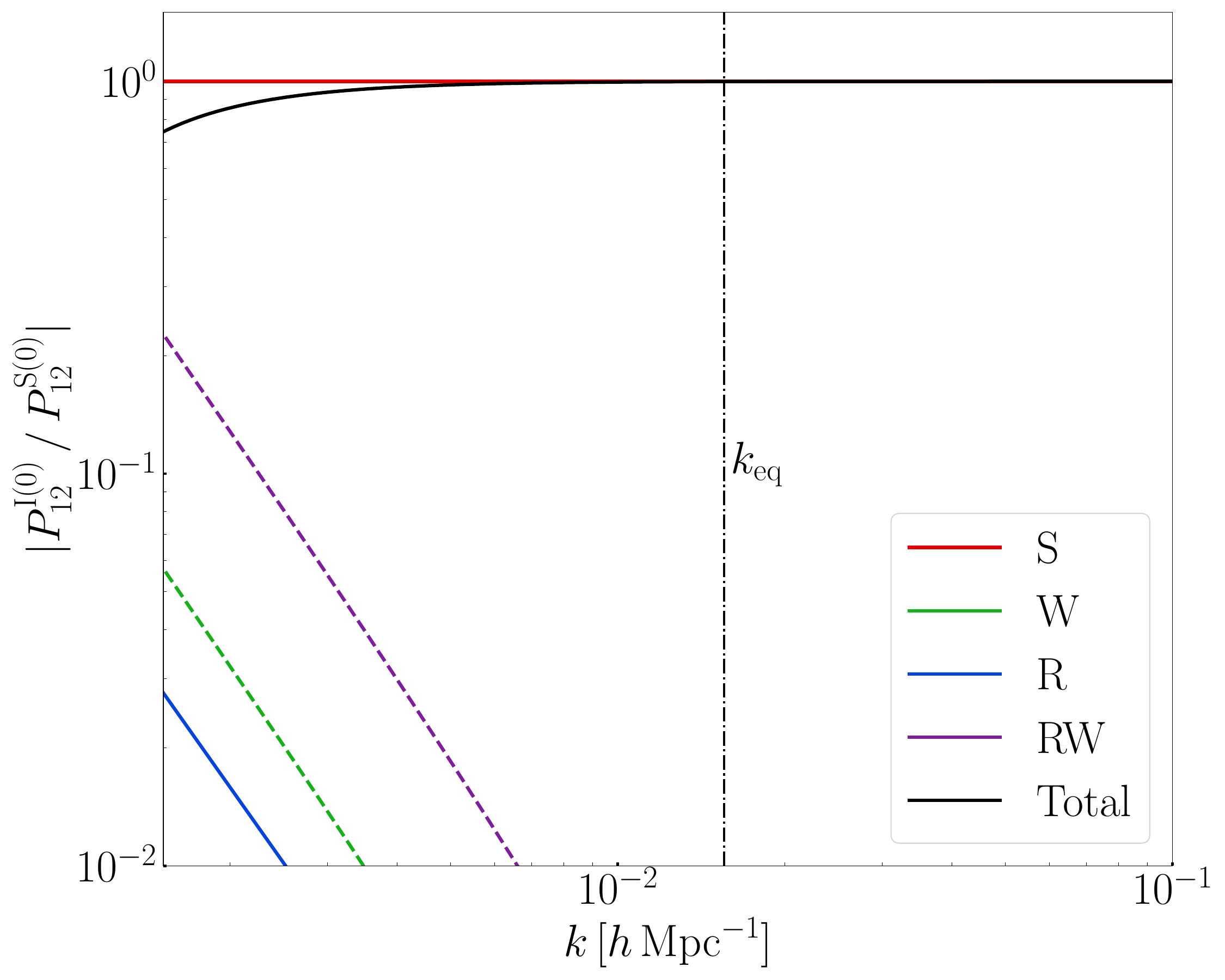} 
\includegraphics[width=7.0cm]{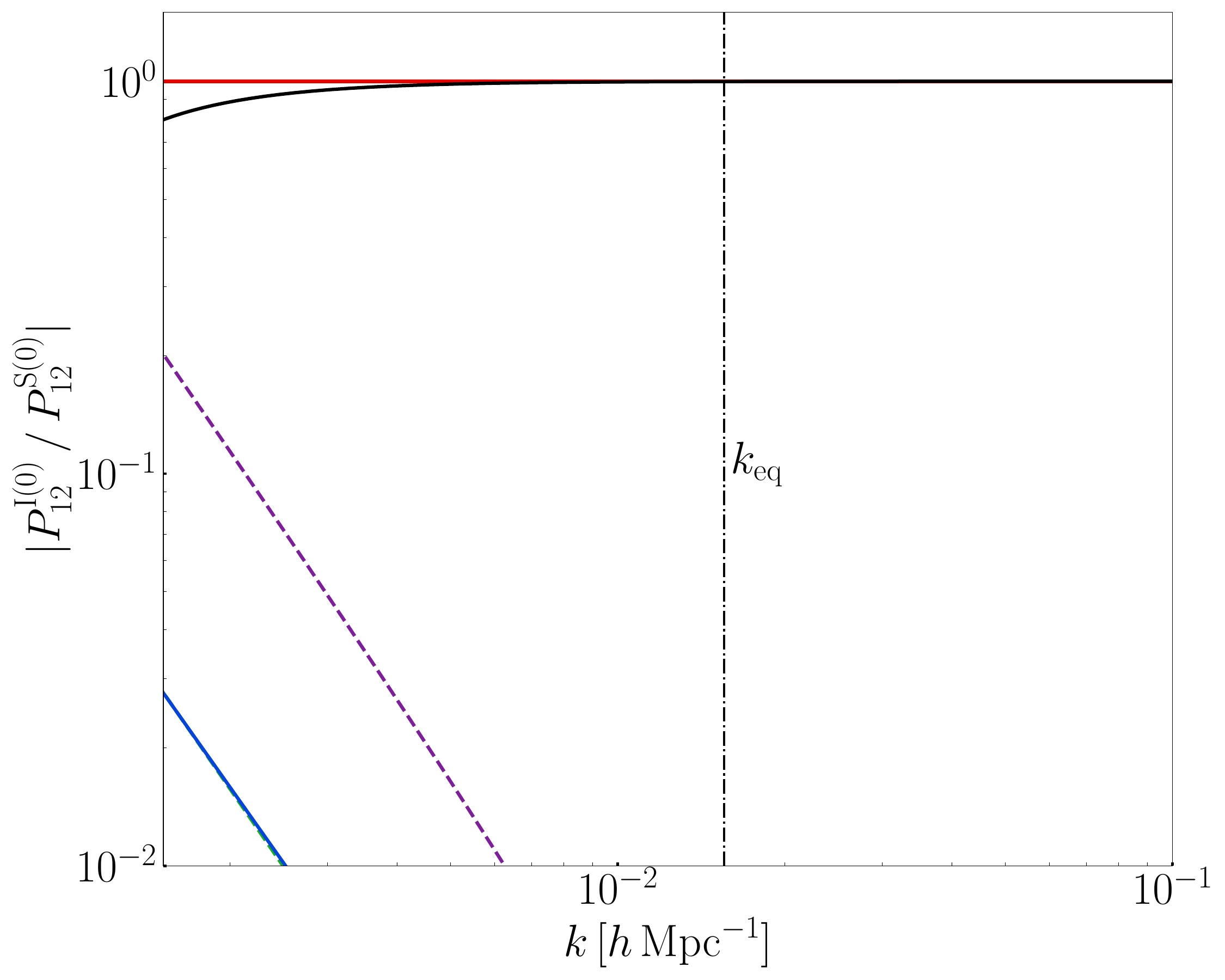} \\
\includegraphics[width=7.0cm]{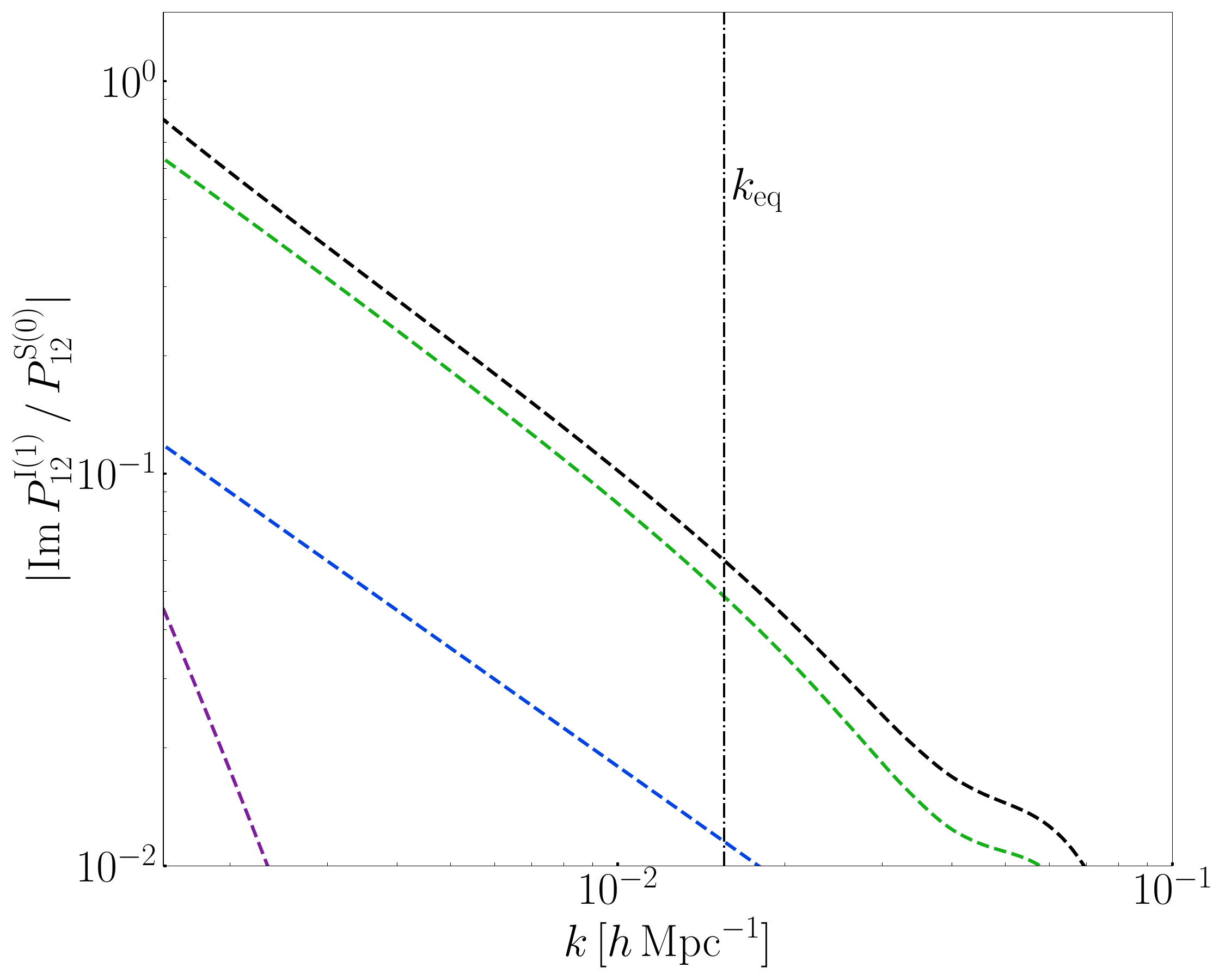} 
\includegraphics[width=7.0cm]{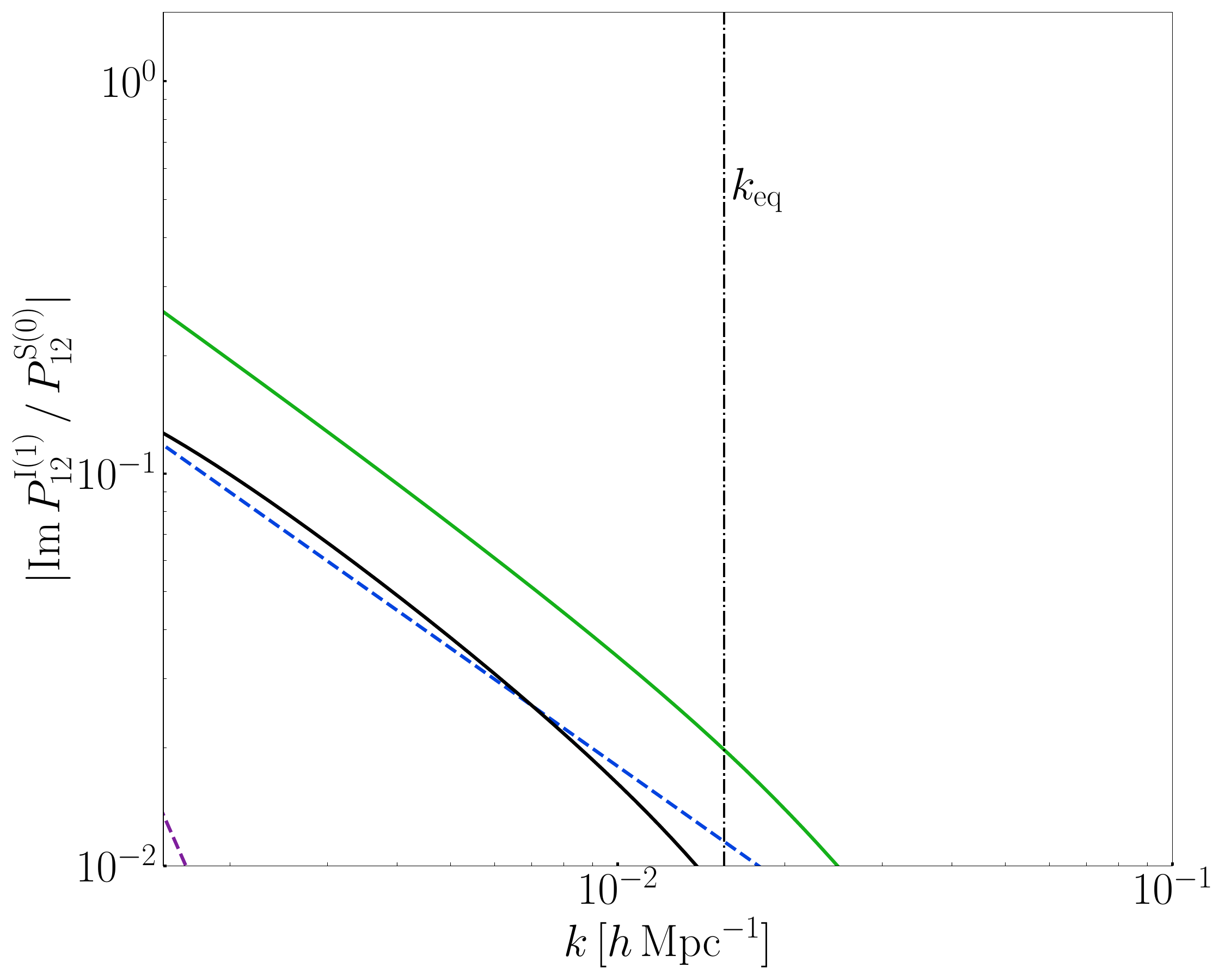} 
\caption{As in \autoref{fig3a}, for the cross-power between samples $a=1$ and 2.
Two configurations are shown: endpoint ($t=0$, \emph{left}) and midpoint ($t=0.5$, \emph{right}).  
} \label{fig3}
\end{figure}

\begin{figure}[! h]
\centering
\includegraphics[width=10.0cm]{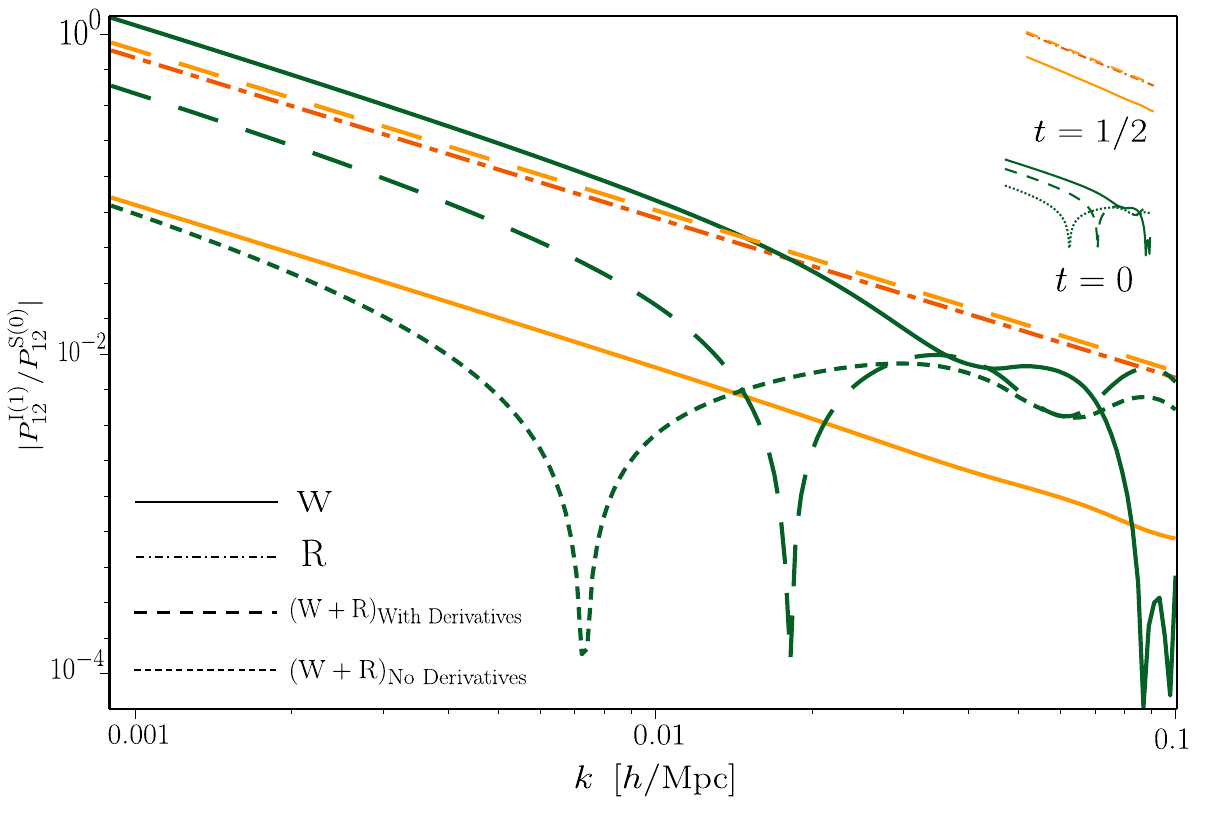} 
\caption{Imaginary part of the  cross-power dipole (which is the main contributor in the imaginary part of the cross-power) as a fraction of the standard  cross-power monopole. Here we use a different pair of tracers: tracer 1 is a SKA-like 21cm intensity mapping survey, while tracer 2 is a Euclid-like spectroscopic H$\alpha$ survey. The redshift is  $z=1$. 
The relativistic (R, red) curve is for both lines of sight, since R is independent of $t$. The long-dashed curves show the wide-angle + relativistic (W+R) corrections,  including the radial derivative corrections, as in \cite{Paul:2022xfx}. For $t=0$, we show also the curves when the 
radial derivatives are neglected (short-dashed curve), as in the main results of this paper.
\red{To avoid clutter, we do not show the $t=1/2$ case with no radial derivatives.}
} \label{fig6x}
\end{figure}

Examples of the monopole and dipole of the power spectra  for these galaxy samples are shown in \autoref{fig3a} and \autoref{fig3}. 
\red{Note that for the midpoint line of sight, $t=1/2$, the dipole  vanishes
in the auto-power case.}
\autoref{fig6x} presents the cross-power dipole for a different pair of tracers, i.e.\ 21cm intensity mapping (SKA-like) and H$\alpha$ galaxies (Euclid-like), at redshift $z=1$. This further illustrates the large variety  in the relativistic and wide-angle effects for different tracers. Note that the leading order relativistic contributions to the dipole do not depend on the line of sight. Furthermore, \autoref{fig6x} shows that the neglect of radial redshift derivatives can lead to non-negligible changes in signal, as shown in the ($\mathrm{W}+\mathrm{R}$) ratio for the $t=0$ case. (See \cite{Paul:2022xfx} for more details.)

\section{Detecting the relativistic and wide-angle effects}

Our aim is to test whether the detection of relativistic and wide-angle effects is feasible in futuristic galaxy surveys. 
Following \cite{Montano:2023zhh}, we
compute the $\chi^{2}$ values between the model $P_{ab}^{\rm S}+P_{ab}^{\rm I}$, which includes the relativistic and/or wide-angle effect I,  and the standard model $P_{ab}^{\rm S}$, which excludes the effect I. We consider all the cases of $P_{ab}^{\rm I}$, where I=W, R, RW or W+R+RW.
Our aim is to estimate the statistical significance of measuring effect I, against the null hypothesis of the absence of such an effect in the data. 

In our case, we use synthetic data generated by taking the theoretical prediction of the full power spectrum \eqref{ePtotal}. Therefore the $\chi^{2}$ for effect I in a redshift bin centred at $z_i$ is simply
\begin{equation}
\chi^{2}(z_i)^{\rm I} = \sum_{k,\mu}\frac{\big|P_{ab}^{\mathrm{I}}(z_i,k,\mu)\big|^{2}}{\mathrm{Var}\big[P_{ab}(z_i,k,\mu)\big]} \;,  \label{eChi}
\end{equation}
where the variance is given by
\begin{align}
\mathrm{Var}\big[P_{aa}\big] = 
\frac{2}{N_{\bm k}}\,  \big|\tilde{P}_{aa}\big|^2 \,,\quad
\mathrm{Var}\big[P_{12}\big] = 
\frac{1}{N_{\bm k}}\, \left[\,\big|\tilde P_{12}\big|^{2} + \big|\tilde{P}_{11}\big| \big|\tilde{P}_{22}\big| \,\right] \;.\label{eVar}
\end{align}
Here 
\begin{align}
    \tilde P_{ab} = P_{ab}+ 
    \red{P_{ab}^{\rm shot}}
    %\frac{\delta^\mathrm{Kronecker}_{ab}}{n_{a}}
    \qquad \mbox{and} \qquad P_{ab}=
    P_{ab}^{\rm S}+P_{ab}^{\rm R+W+RW}\,.
\end{align}
\red{The shot noise is given by $P_{aa}^{\rm shot}=n_a^{-1}$ and $P_{12}^{\rm shot}=0$.}
\begin{figure}[! ht]
\centering
\includegraphics[width=0.75\textwidth]{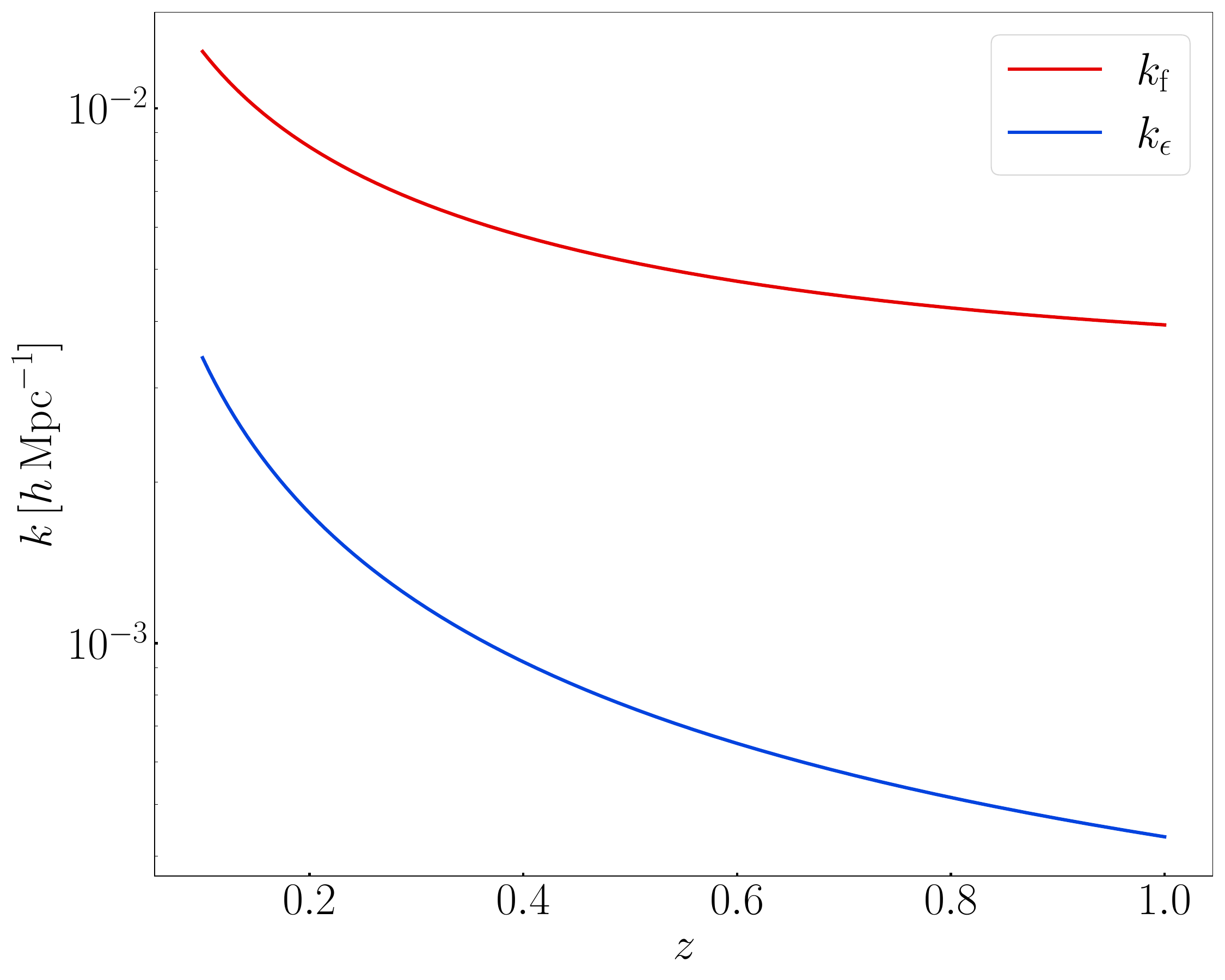}
\caption{Longest wavelength modes in \eqref{kmin}.} \label{figk}
\end{figure} 
Note that $P_{ab}$ in the variance in \eqref{eChi}--\eqref{eVar} is given by \eqref{ePtotal} -- i.e.\ it includes {\em all} relativistic and wide-angle effects.
The number of Fourier modes \red{in a redshift bin centred on $z_i$}  is 
\begin{equation}
N_{\bm k}\red{(z_i)} 
= \frac{2\,\pi\,k^{2}\,\Delta k\,\Delta \mu}{k_{\mathrm{f}}\red{(z_i)}^{3}} \;.\label{Nmodes}
\end{equation}
Here the fundamental mode, $k_{\mathrm{f}}\red{(z_i)}$, corresponds to the longest wavelength that can be measured in the redshift bin \red{centred on $z_i$. The Fourier modes in each bin  satisfy $k_{\rm min}(z_i)\leq k \leq k_{\rm max}(z_i)$, where the minimum and maximum modes are given in \eqref{eBins} and \eqref{kmin}}. 

The $\chi^{2}$ cumulative over redshift bins is
\begin{equation}
\chi^{2}(\le z_i)^{\rm I} = \sum_{j=1}^{i}\, \chi^{2}(z_j)^{\rm I} \;.\label{cumChi2} 
\end{equation}
The detection significance for effect I can be estimated as
\begin{align}
    \mathcal{S}(z_i)^{\rm I} =\Big[\chi^{2}(z_i)^{\rm I} \Big]^{1/2},
\end{align}
and similarly for the cumulative $\mathcal{S}(\leq z_i)^{\rm I}$.

For the numerical calculation
we choose
\begin{equation}
\Delta z = 0.1\;, \quad \Delta \mu = 0.04\;, \quad \Delta k = 
k_{\mathrm{f}}\;, \label{eBins}
\quad k_{\rm max} = 0.08\left(1+z\right)^{2/(2+n_{s})}\;h\,\mathrm{Mpc}^{-1}\;.
\end{equation} 
The conservative choice of $k_{\rm max}$ in \eqref{eBins} means that linear perturbations remain accurate. The minimum $k$ in each redshift bin, corresponding to the longest wavelength mode, is given by
\begin{align}\label{kmin}
    k_{\mathrm{min}}(z) = {\rm max}\,\big[k_{\rm f}(z) \,,\,k_\epsilon(z) \big]\,,
\end{align}
where $k_\epsilon$ is given in \eqref{keps}. These wavenumbers are shown in \autoref{figk}, confirming that $k_{\mathrm{min}} = k_{\rm f}$  in the case of the two futuristic surveys over redshifts $z\leq 1$ and sky area $15\,000\,\deg^2$.

\begin{figure}[! ht]
\centering
\includegraphics[width=7.0cm]{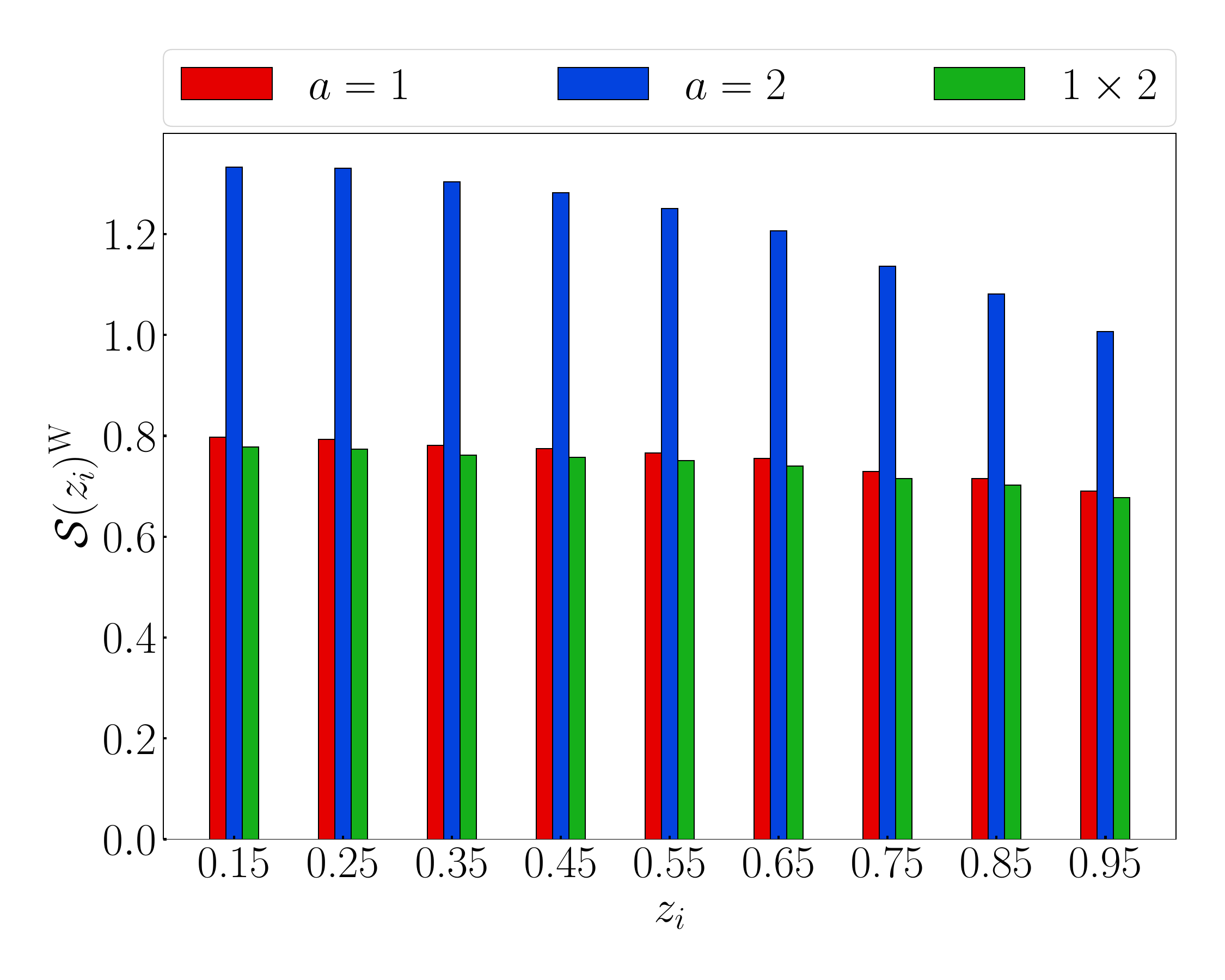} 
\includegraphics[width=7.0cm]{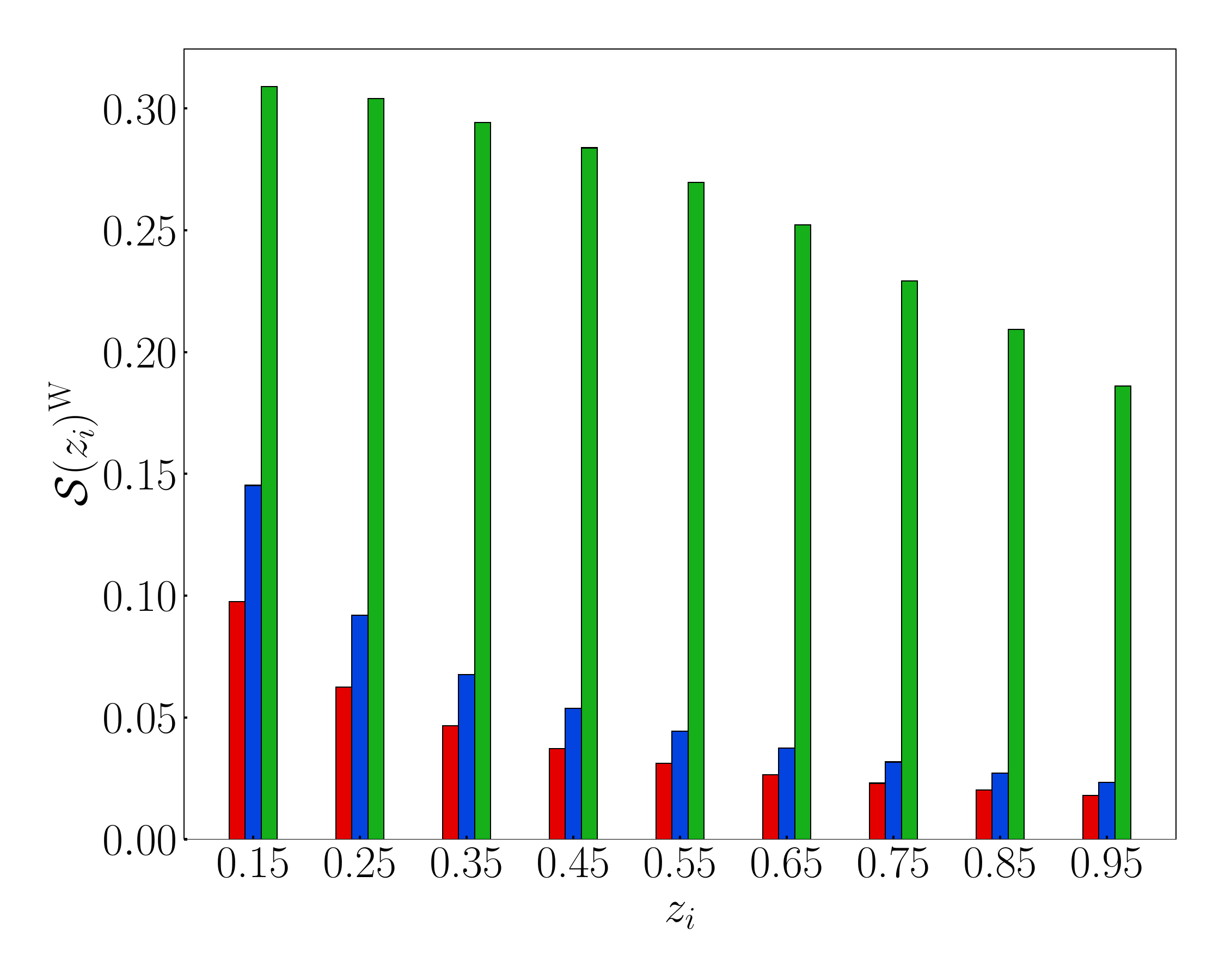} \\ 
\includegraphics[width=7.0cm]{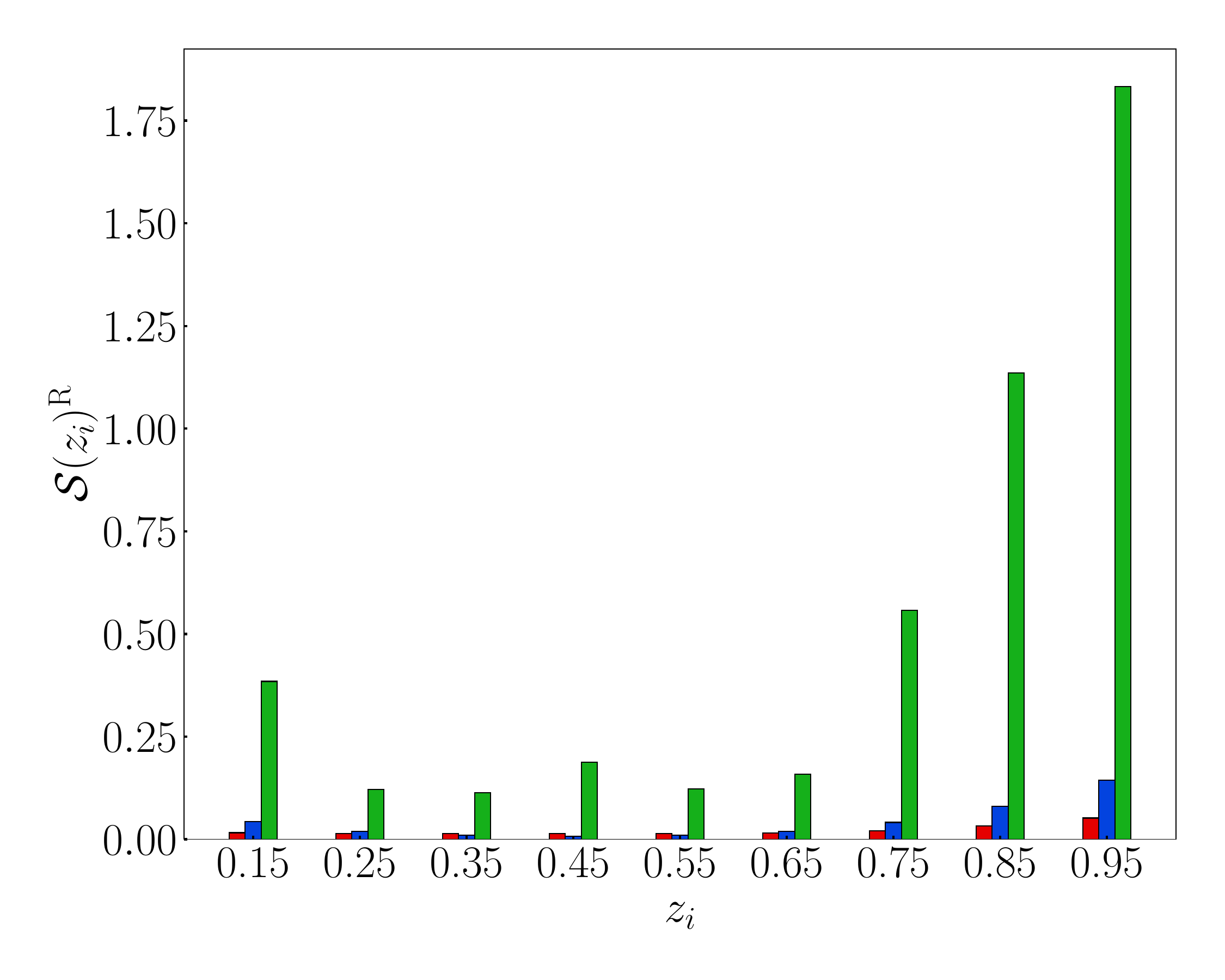} 
\includegraphics[width=7.0cm]{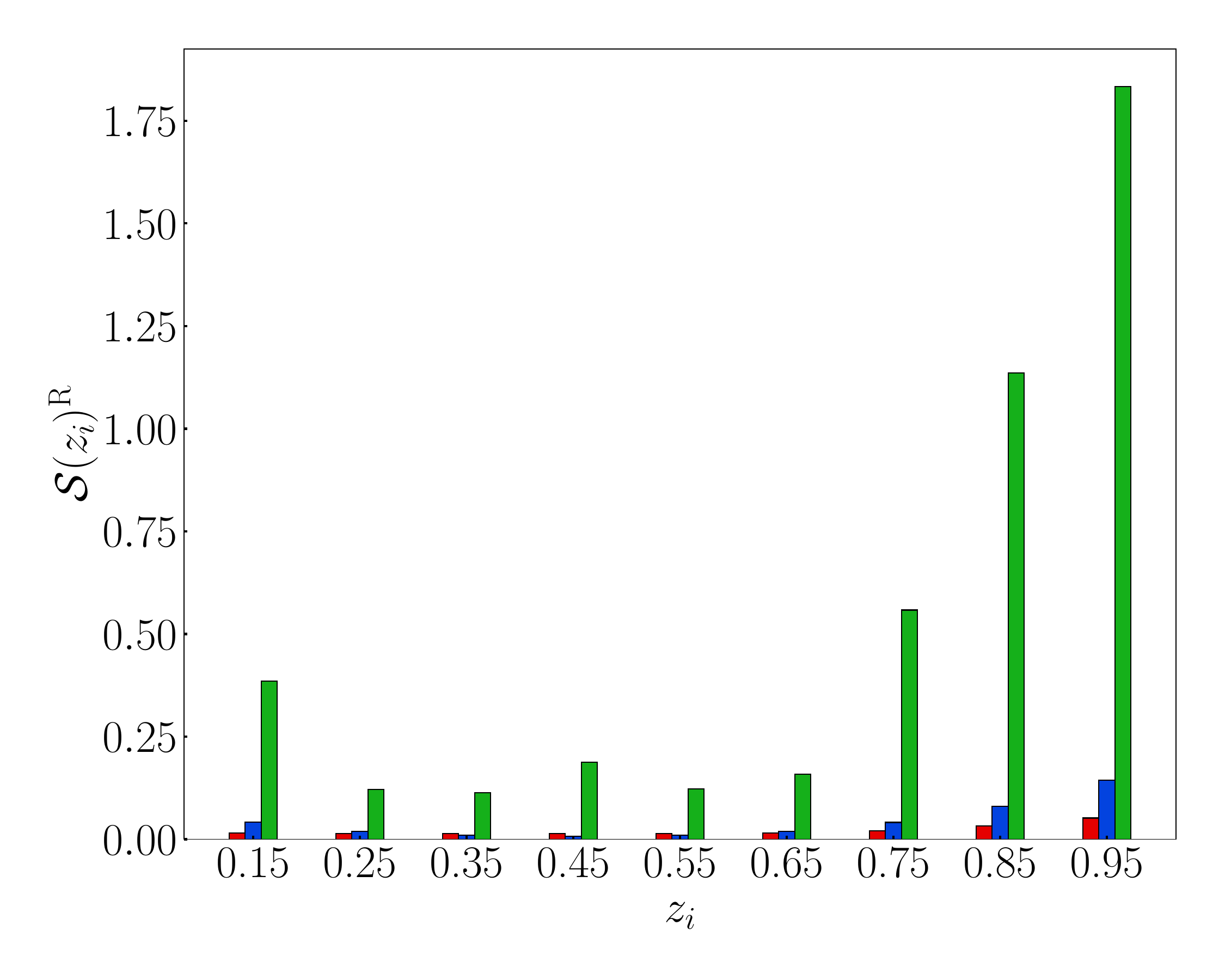} \\
\includegraphics[width=7.0cm]{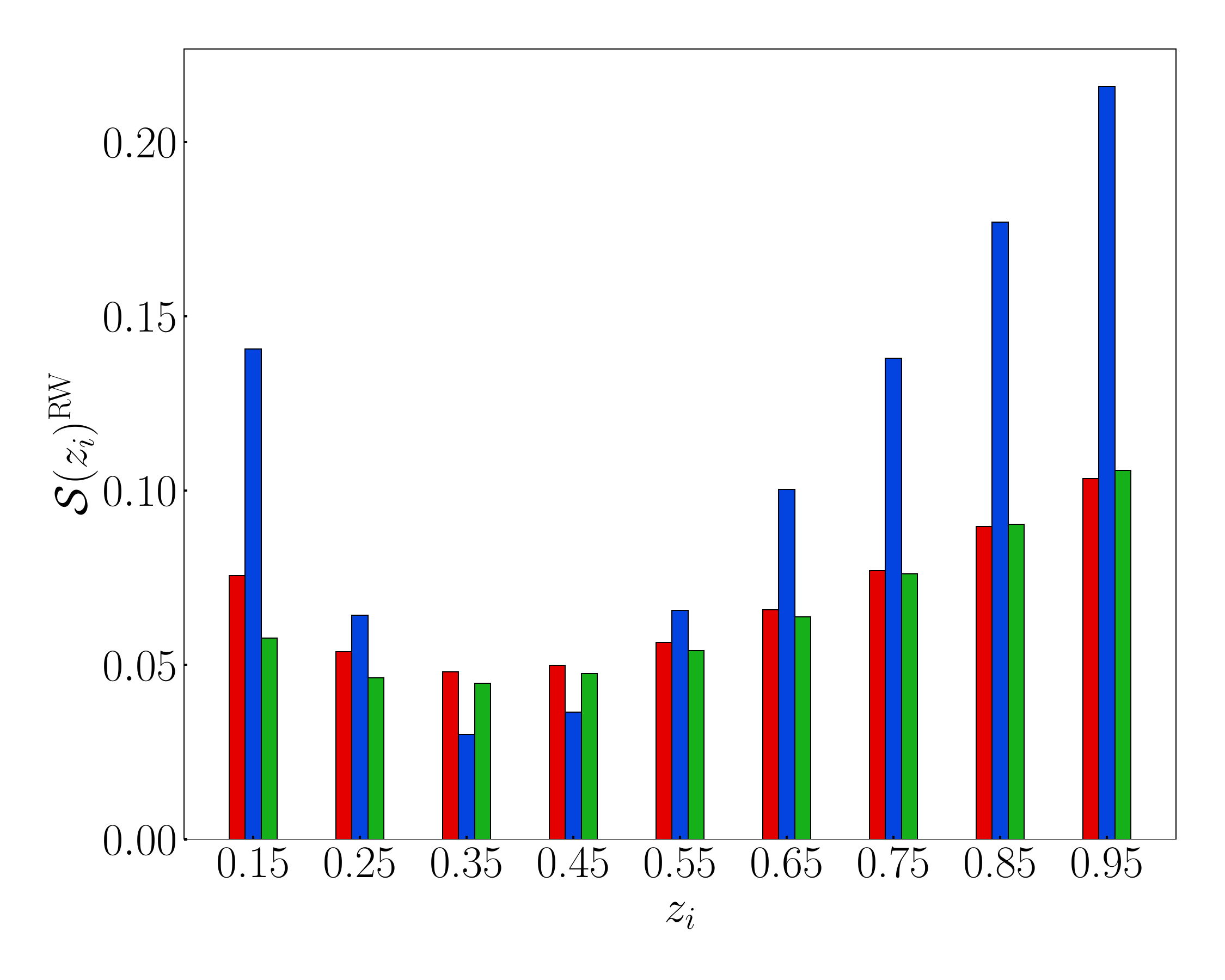} 
\includegraphics[width=7.0cm]{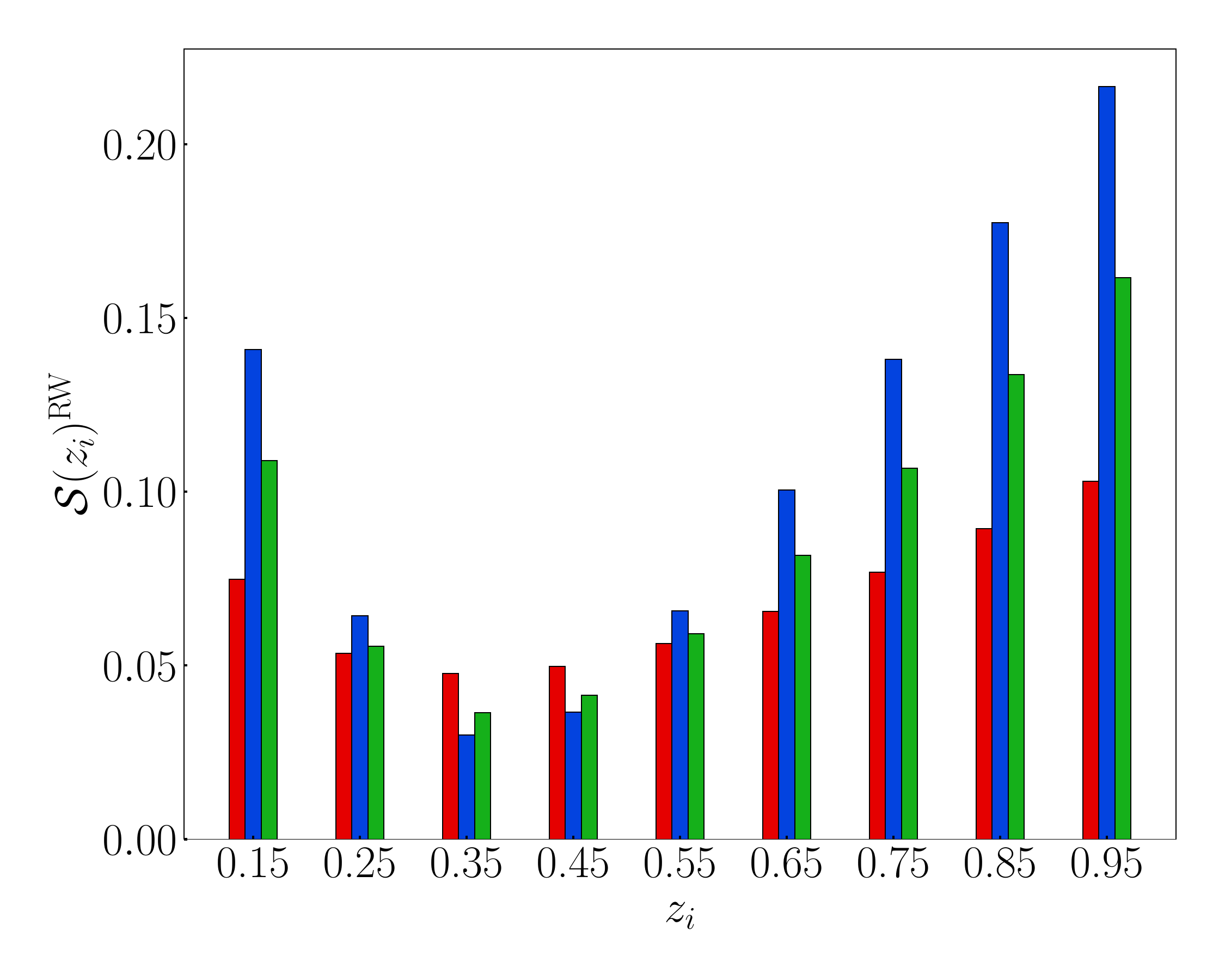} \\
\includegraphics[width=7.0cm]{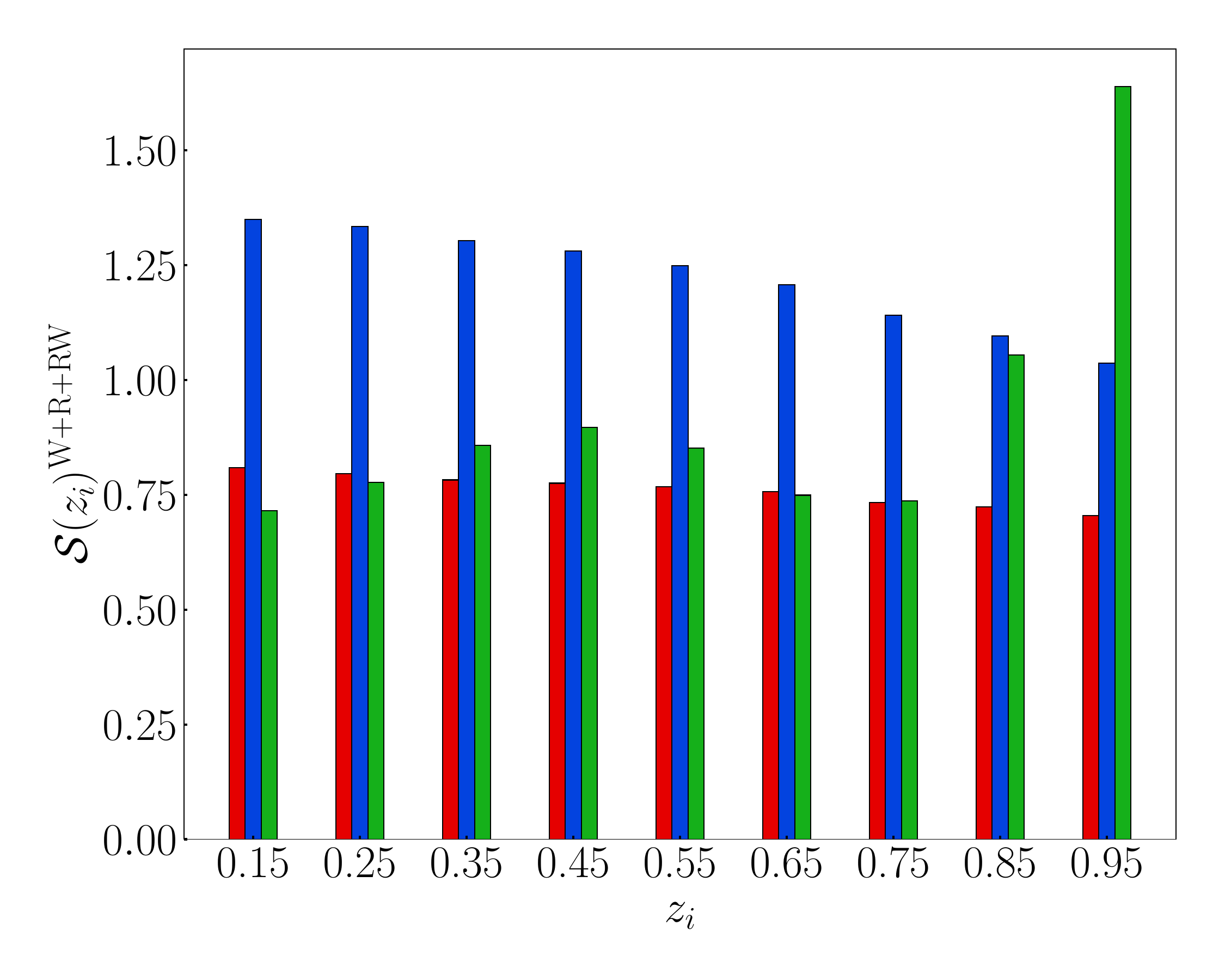} 
\includegraphics[width=7.0cm]{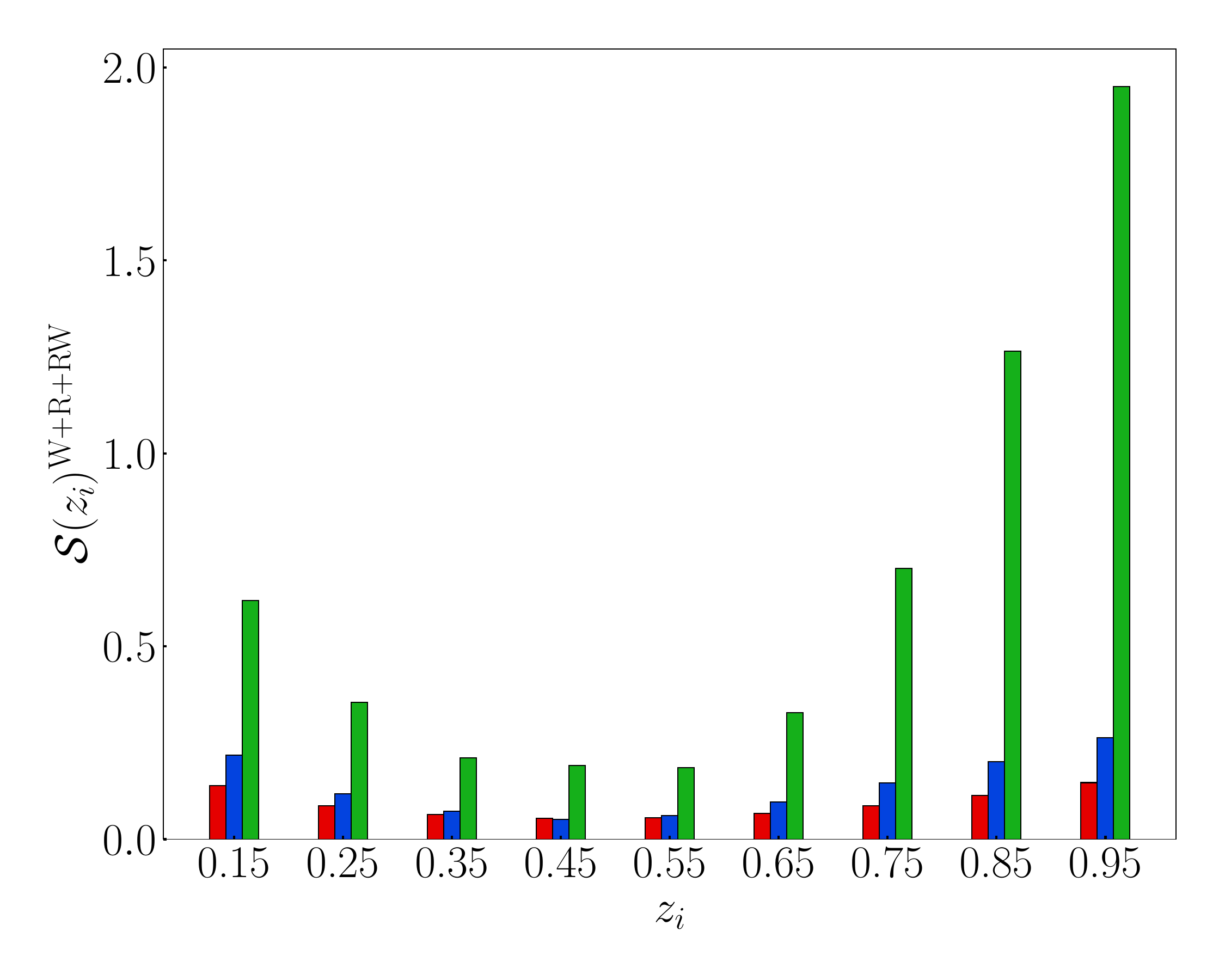}
\caption{For the relativistic wide-angle power $P^{\rm I}_{ab}$, where I=R, W, RW or R+W+RW,
the detection significance $\mathcal{S}^{\rm I}$ is shown
in the auto- and cross-power, in each $z$-bin, for the endpoint  ($t=0$, \emph{left}) and midpoint ($t=0.5$, \emph{right}) lines of sight.
} \label{fig4auto}
\end{figure}

\begin{figure}[! ht]
\centering
\includegraphics[width=7.0cm]{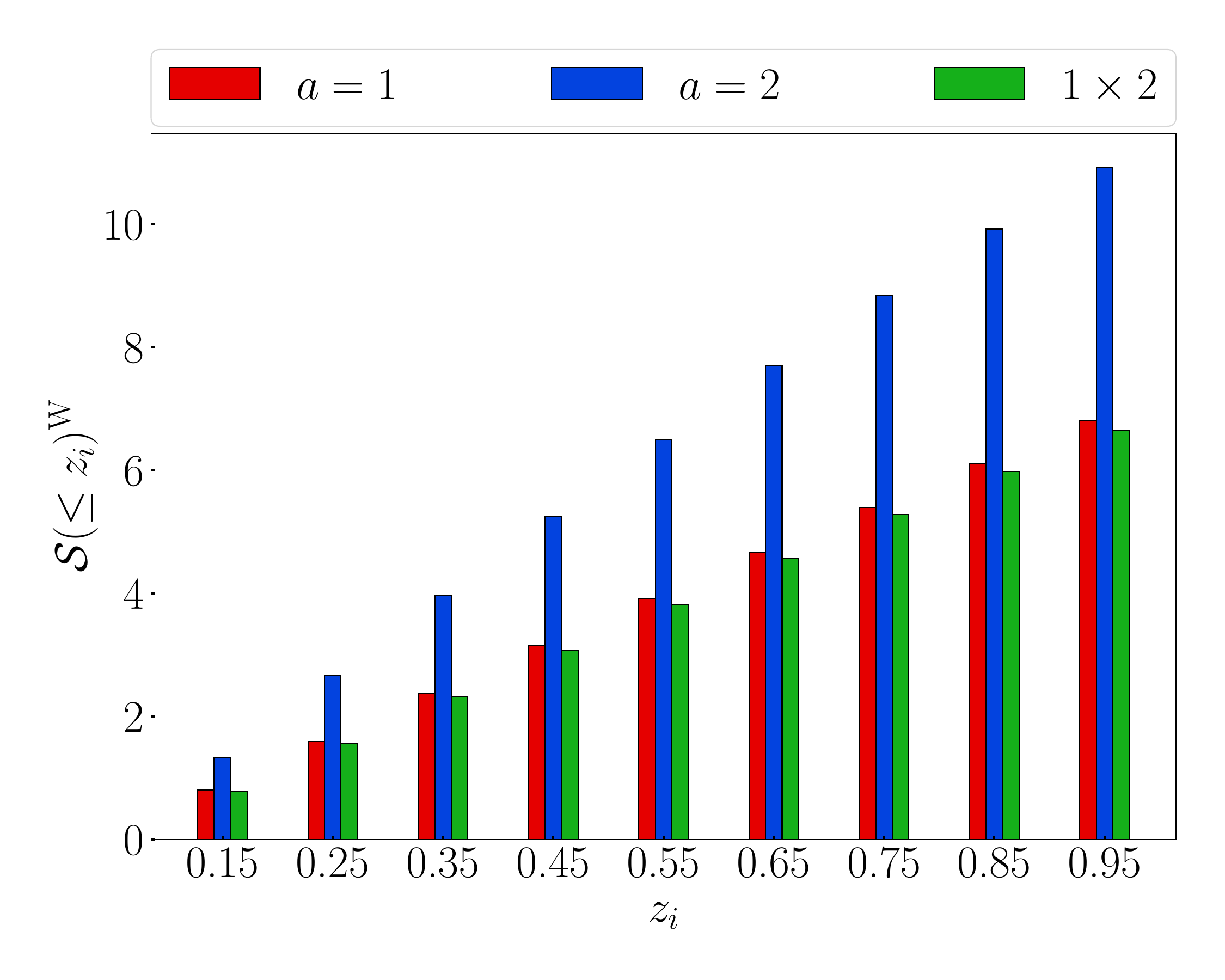} 
\includegraphics[width=7.0cm]{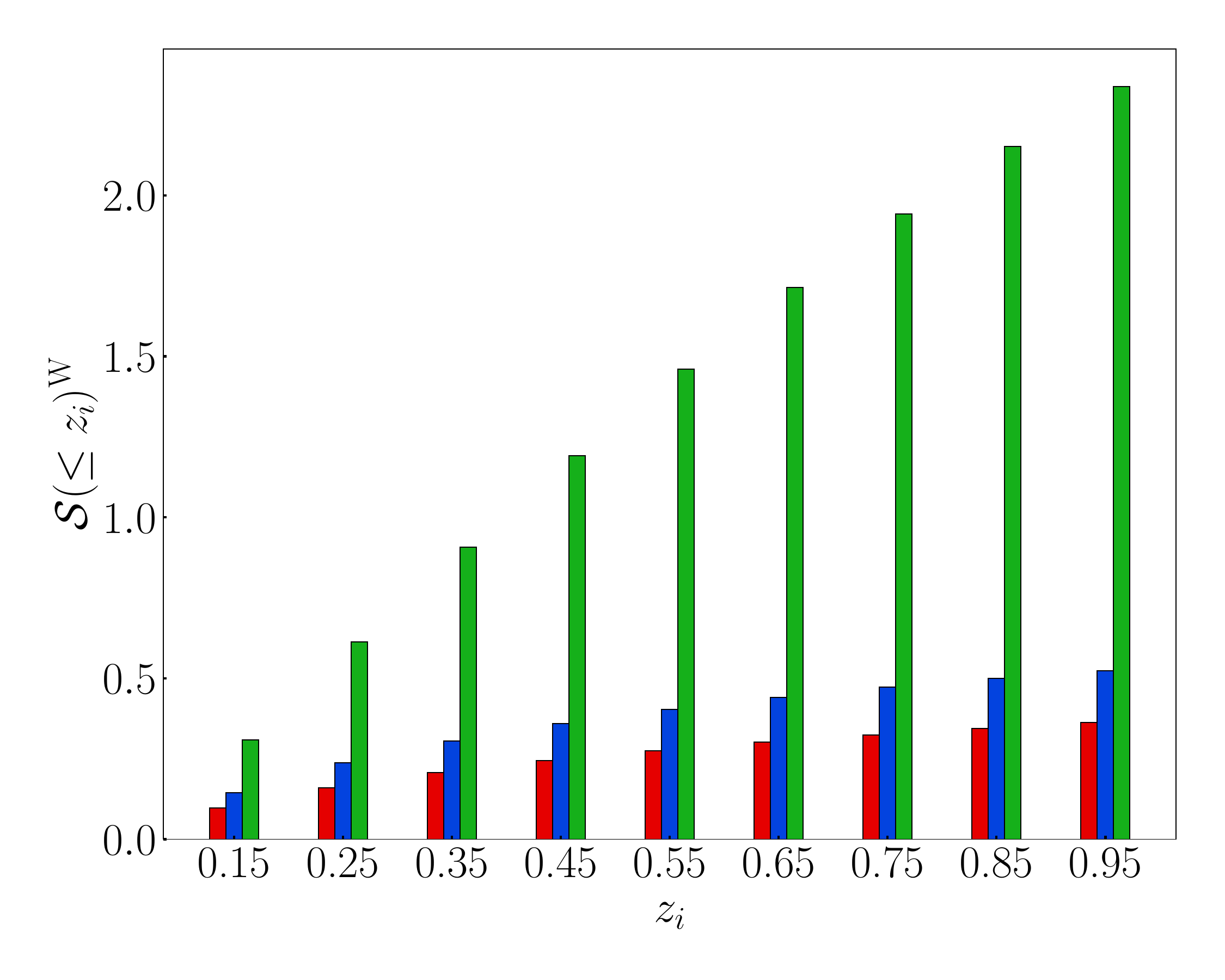} \\ 
\includegraphics[width=7.0cm]{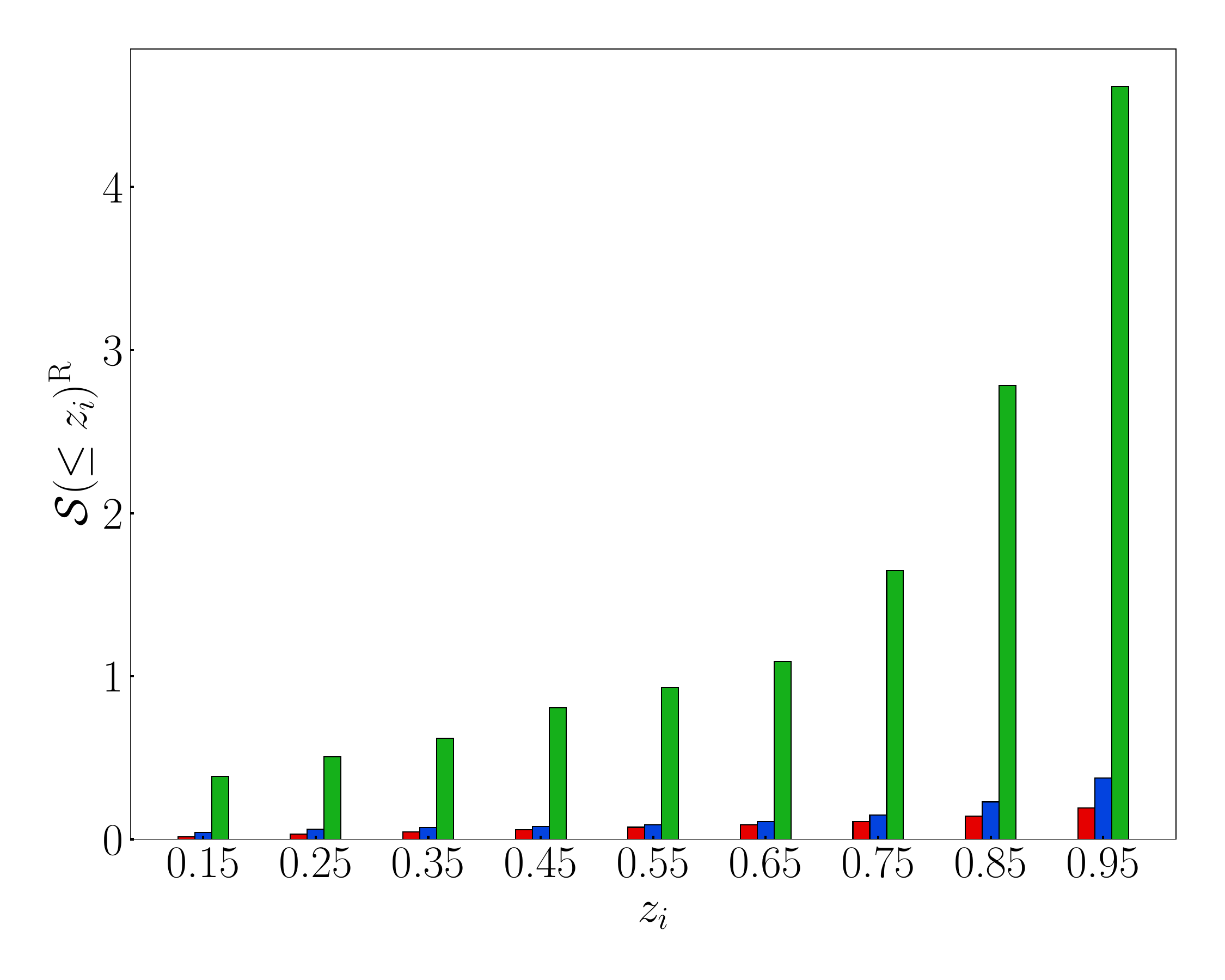} 
\includegraphics[width=7.0cm]{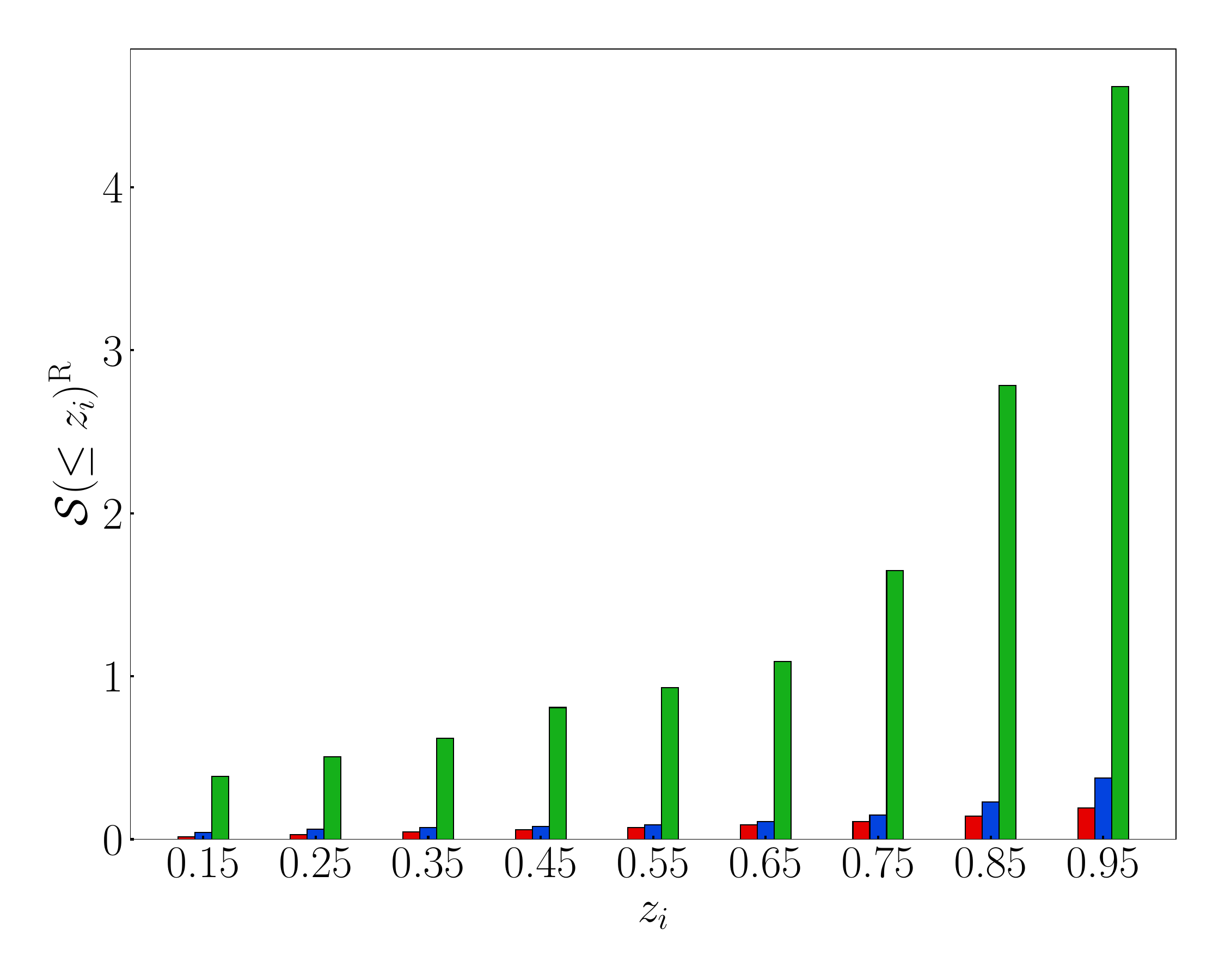} \\
\includegraphics[width=7.0cm]{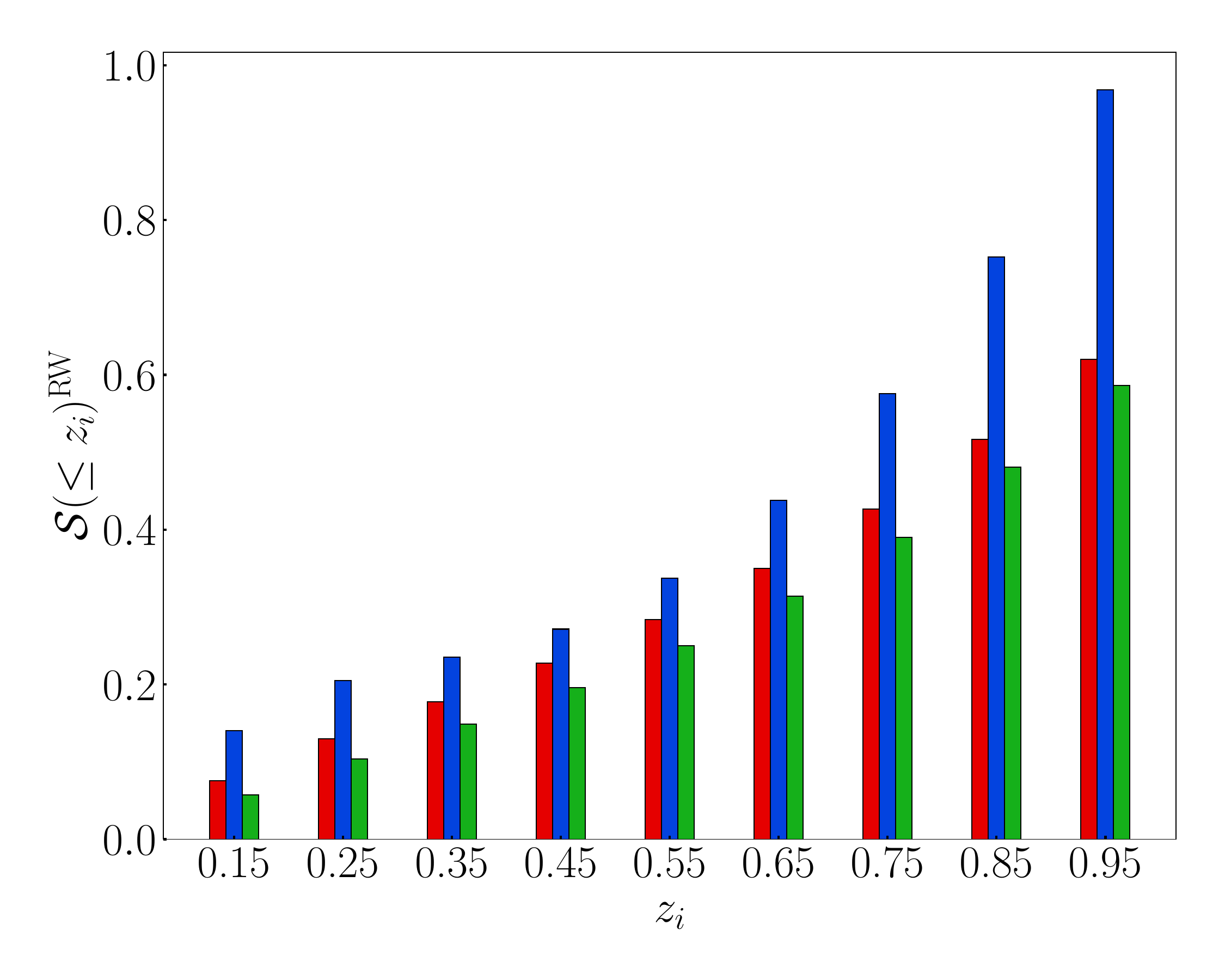} 
\includegraphics[width=7.0cm]{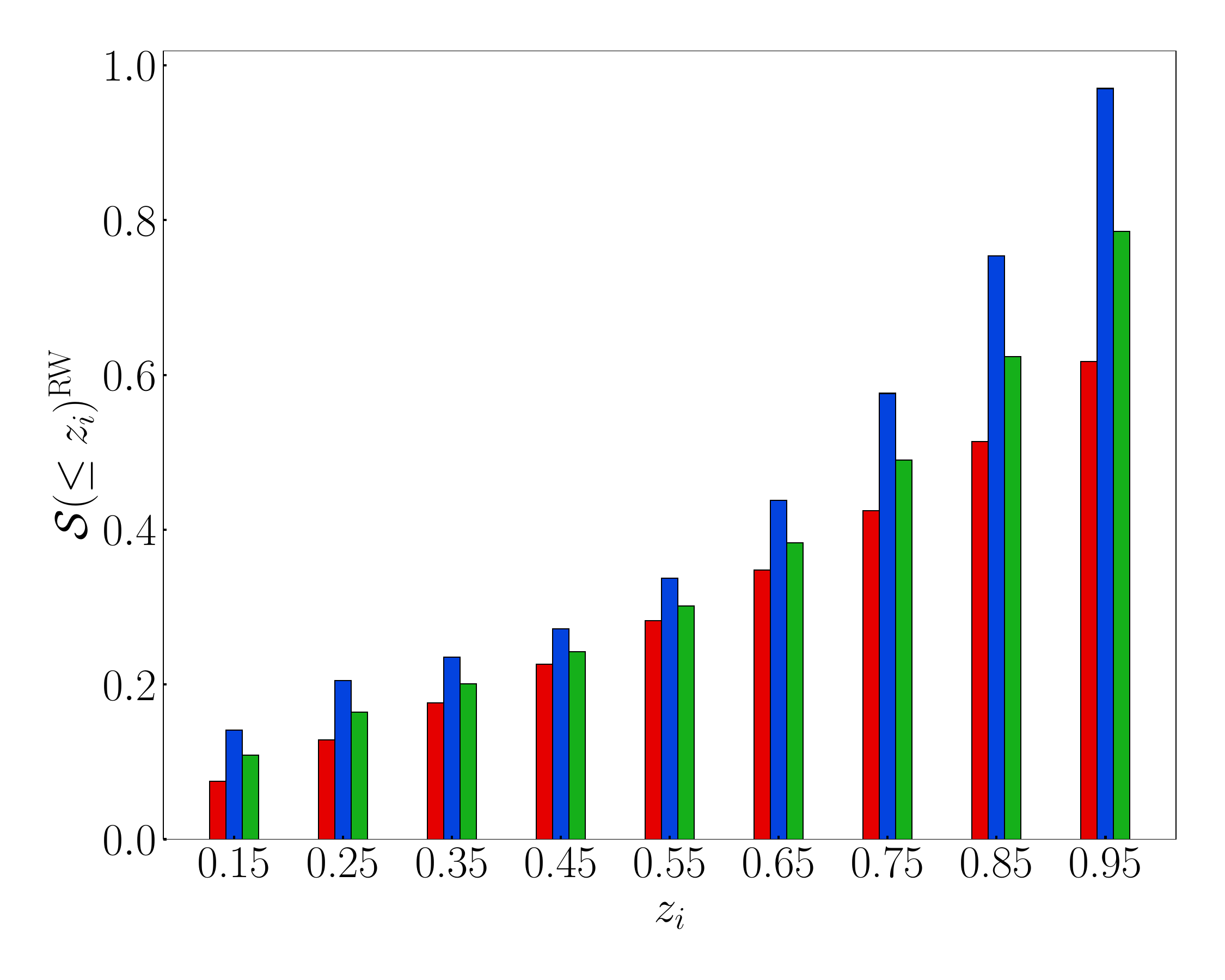} \\
\includegraphics[width=7.0cm]{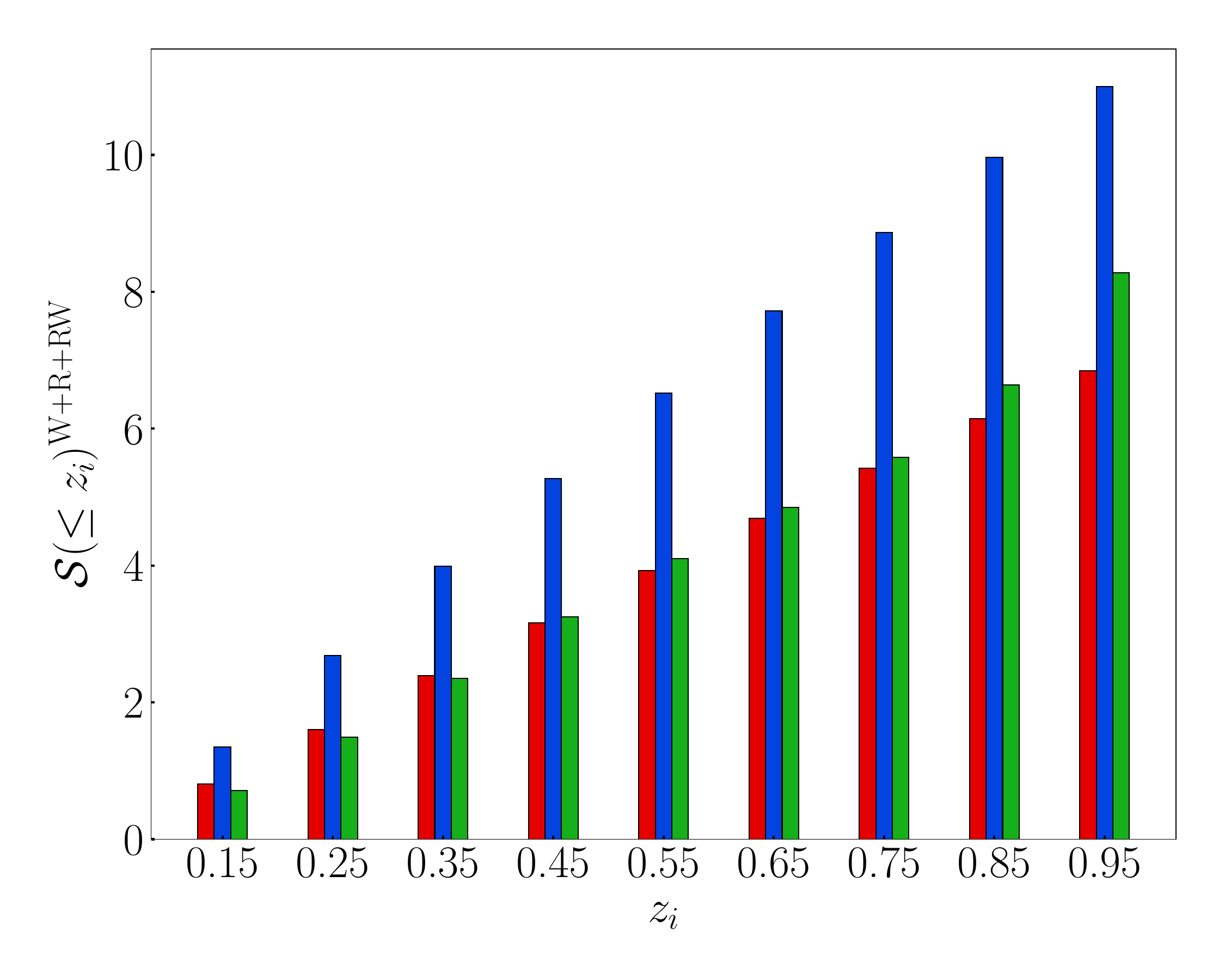} 
\includegraphics[width=7.0cm]{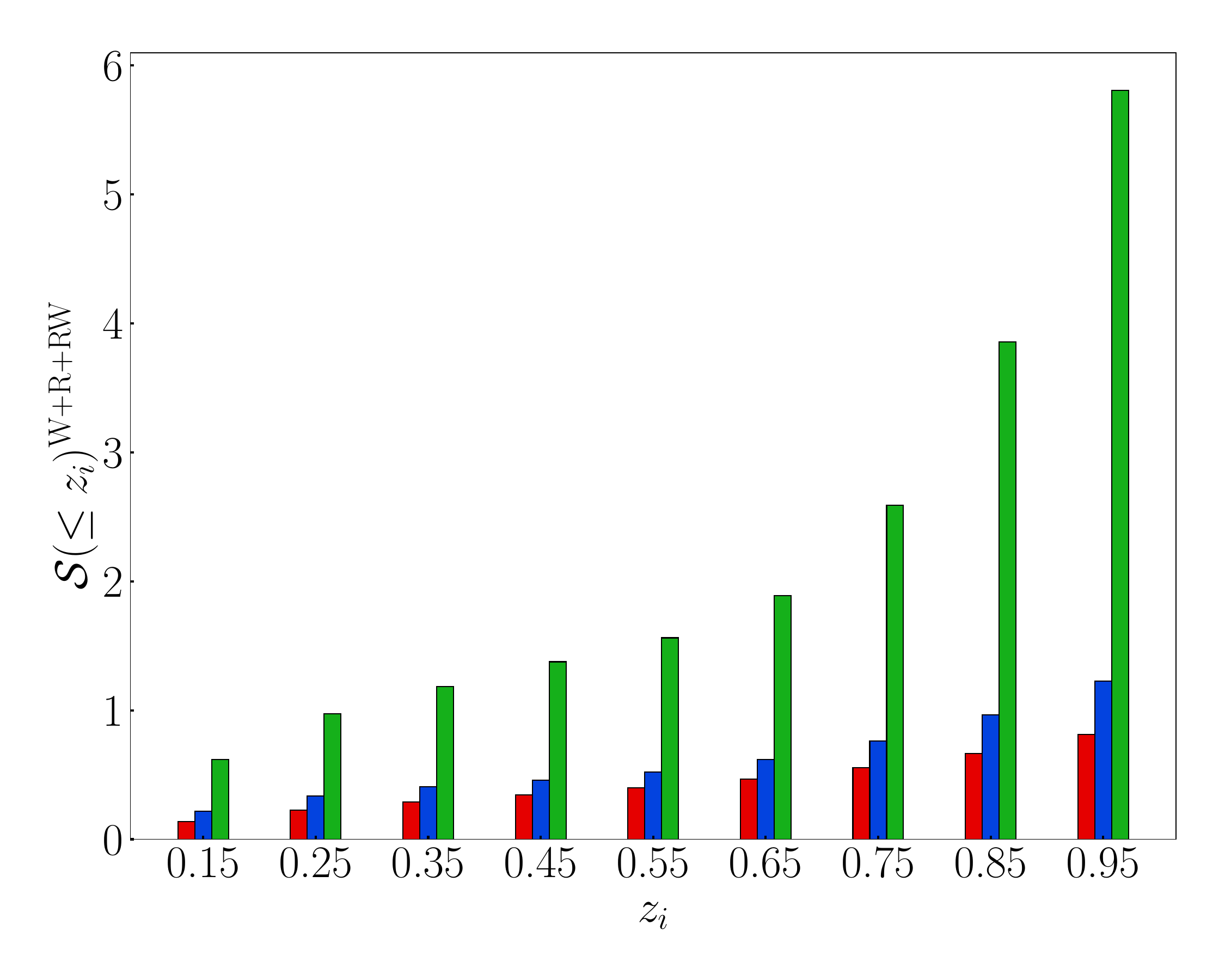}
\caption{As in \autoref{fig4auto}, for the
cumulative detection significance $\mathcal{S}$ of the relativistic and wide-angle corrections in the auto- and cross-power, for the endpoint (\emph{left}) and midpoint (\emph{right}) lines of sight.} \label{fig5auto}
\end{figure}

In the cases $t=0$ (endpoint) and $t=0.5$ (midpoint), the results for the detection significance $\mathcal{S}$ are summarised in \autoref{fig4auto} (per redshift)  and \autoref{fig5auto} (cumulative). The significance is higher for the endpoint configuration, but in both cases, the cumulative significance is
$>5\sigma$. If we combine the signals from auto- and cross-power, the significance for the endpoint $t=0$ line of sight is $>10\sigma$.
We note that for the midpoint line of sight, only the cross-power has a cumulative detection significance $>5\sigma$.

\newpage

\newpage

\clearpage
\section{Conclusions}

\begin{figure}[! hb]
\centering
\includegraphics[width=7.0cm]{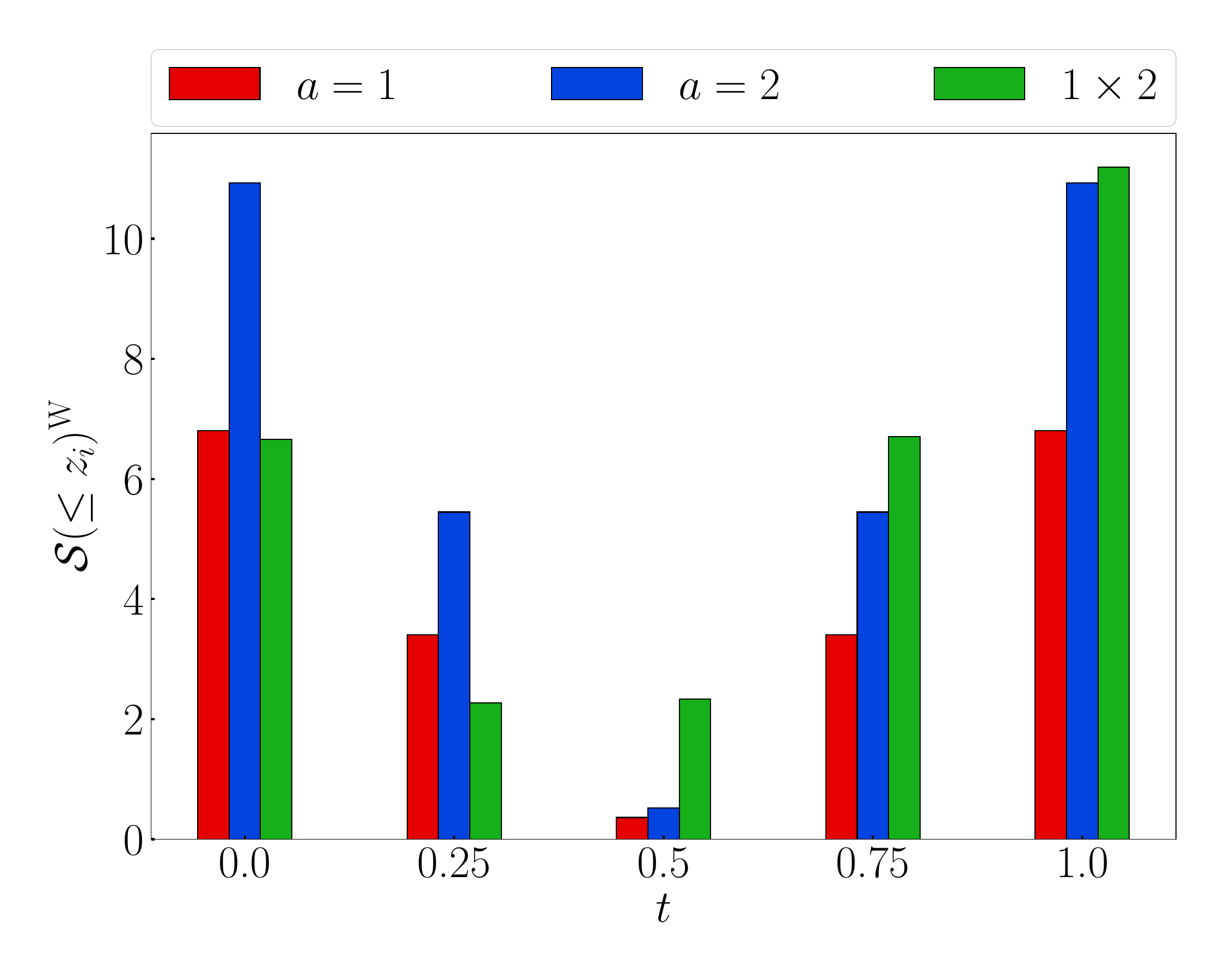} 
\includegraphics[width=7.0cm]{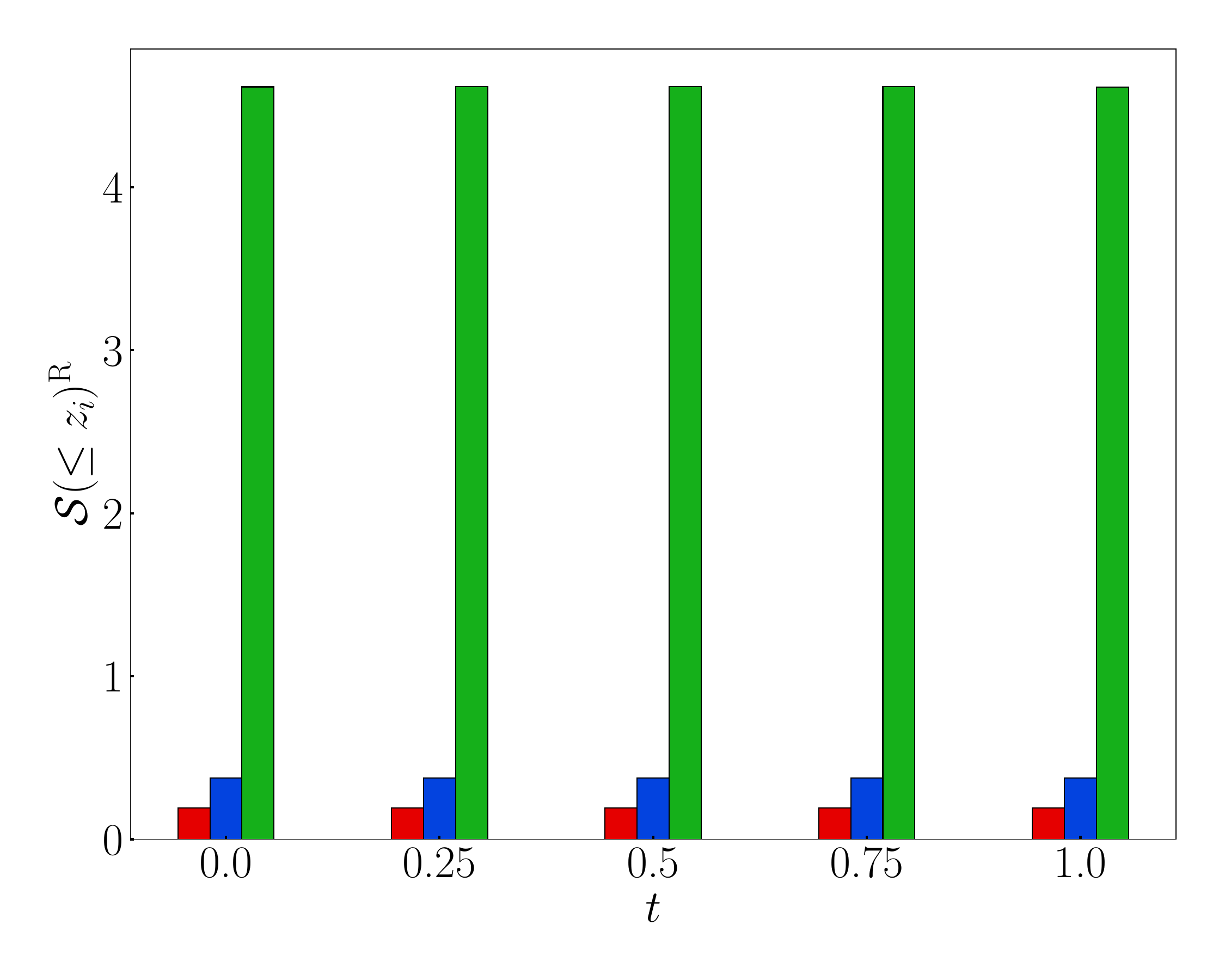} \\
\includegraphics[width=7.0cm]{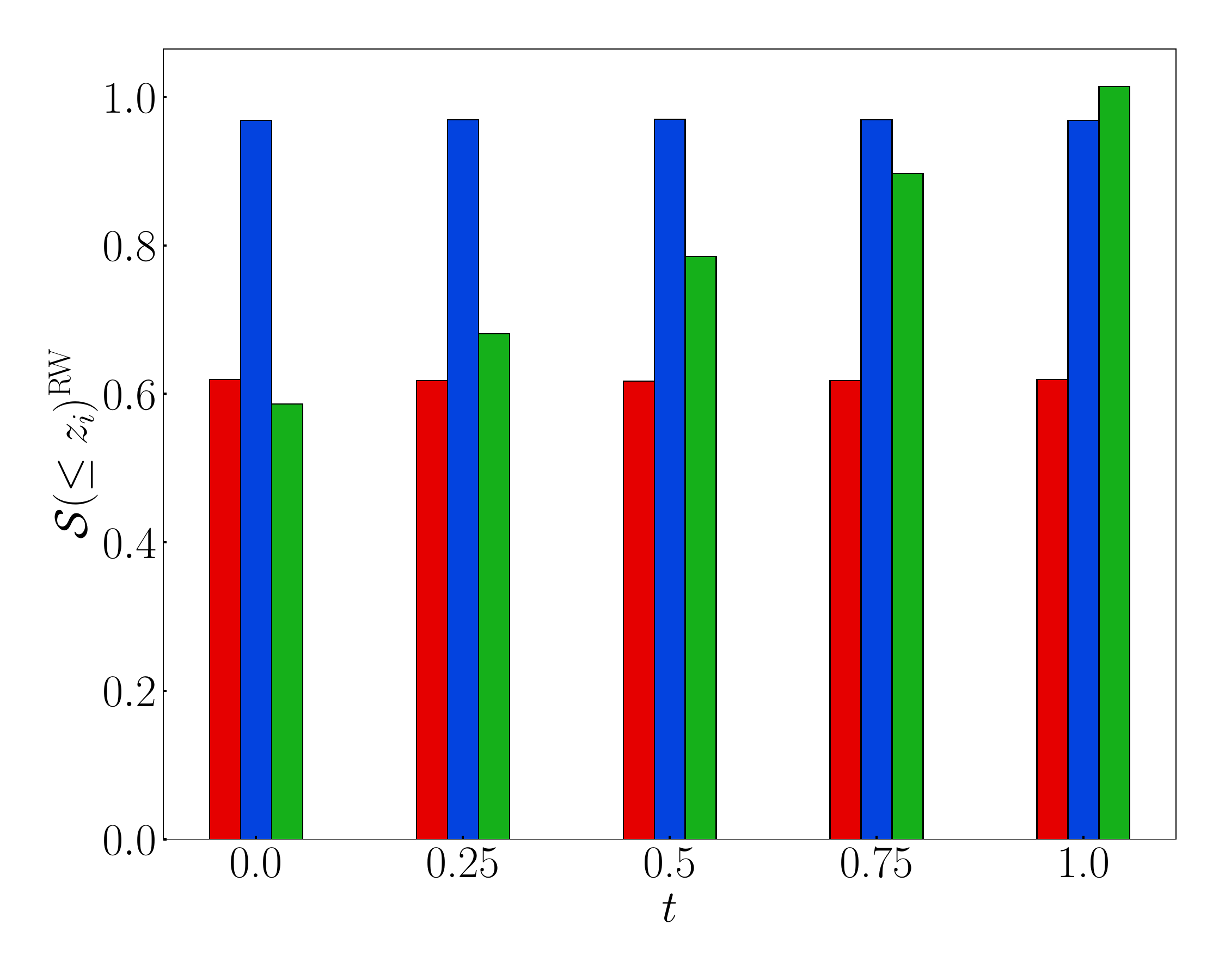} 
\includegraphics[width=7.0cm]{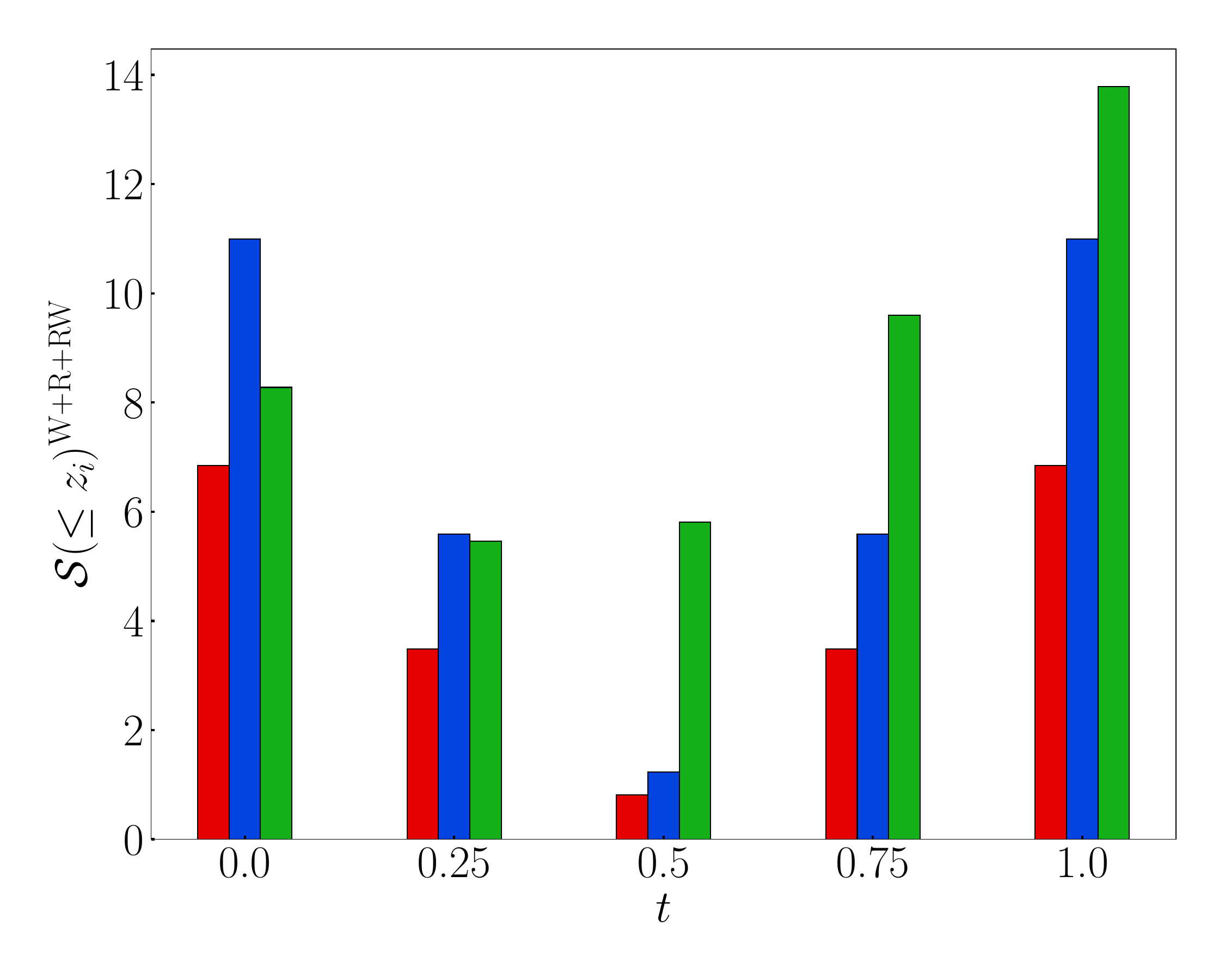} 
\caption{Cumulative detection significance $\mathcal{S}$ of the relativistic and wide-angle corrections in the auto- and cross-power, against the line-of-sight parameter $t$.} \label{fig6}
\end{figure}

This work complements the analysis presented in \cite{Noorikuhani:2022bwc} by generalizing their power spectrum results to two tracers and to arbitrary lines of sight. We omit the integrated relativistic corrections -- mainly the effect of lensing convergence -- which should be reasonable at redshifts $z<1$. The non-integrated relativistic effects in the auto-power  are suppressed by a factor $O[\delta(\mathcal{H}/k)^{2}]$. In the imaginary part of the cross-power by contrast, there are terms $O[\delta(\mathcal{H}/k)]$ \cite{Montano:2023zhh,McDonald:2009ud,Bonvin:2013ogt,Bonvin:2015kuc,Gaztanaga:2015jrs,Irsic:2015nla,Hall:2016bmm,Lepori:2017twd,Bonvin:2018ckp,Lepori:2019cqp}. 
The consequent difference in signal is clearly apparent in the $\mathcal{S}^{\rm R}$ plots in \autoref{fig4auto} and \autoref{fig5auto}.  
The pure relativistic contribution is insensitive to the line-of-sight parameter $t$,  since it is computed in the plane-parallel approximation, as in \eqref{e17_2}. 
On the other hand, the wide-angle effects are obviously sensitive to $t$. 

Both relativistic and wide-angle effects are very sensitive to the different astrophysical properties of the surveys. Nevertheless, \autoref{fig4auto} and \autoref{fig5auto} show that the cumulative R+W+RW signal is 
detectable in the cross-power at $>5\sigma$. For $t=0$, the signal is also detectable in the auto-power. In this case, the combined detection significance from cross- and auto-power is $\sim 15\sigma$. 

What about other values of $t$? \autoref{fig6} shows how the detection significance varies across the full range of line-of-sight parameter $t$.
In the case of the auto-power  $P_{aa}$, there is a symmetry about $t=0.5$, showing that the choice of a particular line of sight $t$ is the mirror of $1-t$. This symmetry is broken in the case of the cross-power $P_{12}$, because the permutation of the tracers leads to different power spectra, as seen in the detailed forms \eqref{P3} and \eqref{P4}. The results for the futuristic surveys indicate significant detection potential $\sim10\sigma$ for the cross-power $P_{12}$ with $t=1$, and for the auto-power $P_{22}$ with $t=0,1$. The highest detection significance is in the $t=1$ configuration, where the combined detection significance is $>15\sigma$.

Our results show that future galaxy surveys, and even more their multi-tracer combination,   should be able to detect the combined relativistic and wide-angle contributions to the power spectra, assuming that other surveys can measure the clustering, evolution and magnification biases with reasonable uncertainties -- and also assuming that ultra-large scale 
systematics can be dealt with. This detectability is dependent on the choice of the line-of-sight parameter $t$, with the strongest detectability for $t=1$ (see \autoref{fig6}).
The choice of $t$ has significant implications for computation, so it is useful that the signal is detectable in all cases  we considered.  
Detectability means that constraints on cosmological parameters could be affected by omitting these corrections to the standard power spectra. In particular, we expect that measurements of
the local primordial non-Gaussianity parameter $f_{\mathrm{NL}}$ could be affected by these corrections. Future work will look at Fisher forecasts on $f_{\mathrm{NL}}$ and other cosmological parameters  \cite{Guedezounme:2024}.

\vfill
\noindent{\bf Acknowledgements}\\
SJ is supported by the Stellenbosch University Astrophysics Research Group fund. SG and RM are supported by the South African Radio Astronomy Observatory and National Research Foundation (grant no. 75415).
SC acknowledges support from the Italian Ministry of University and Research, PRIN 2022 `EXSKALIBUR -- Euclid-Cross-SKA: Likelihood Inference Building for Universe Research', from the Italian Ministry of Foreign Affairs and International
Cooperation (grant no.\ ZA23GR03), and from the European Union -- Next Generation EU.
\red{We used the Python code Sympy for analytical calculations.}

\clearpage
\appendix

\section{Fourier kernels} \label{AppA}

\red{Recall that throughout the paper we work to $\mathcal{O}\big[\epsilon^2,\delta(\cH/k)^2 \big]$.}

\begin{eqnarray}
\mathcal{K}_a^{\rm W} &=& \frac{f}{k^{2}} \Big\{ (t - \sigma_a)^{2} \Big[ (k_{x}^{2} - k_{z}^{2}) \epsilon_{x}^{2} +  (k_{y}^{2} -  k_{z}^{2}) \epsilon_{y}^{2} \notag 
\\ \notag
&& {} \hspace{2.5cm} + 2 (k_{x} k_{y} \epsilon_{x} \epsilon_{y}  - k_{x} k_{z}  \epsilon_{x} \epsilon_{z}  -  k_{y} k_{z} \epsilon_{y} \epsilon_{z})\Big] 
\\ 
&& {}~~~~ + 2  (t - \sigma_a)  \Big[k_{x} k_{z} \epsilon_{x}  + k_{y} k_{z} \epsilon_{y}\Big] \Big\}, 
%+ \mathcal{O}({ \epsilon}^3),
\\  \notag
\mathcal{K}_a^{\rm RW} &=& \mathrm{i} \, \frac{1}{k} \frac{\mathcal{H}}{k} f \Big\{\frac{2 (1 - \mathcal{Q}_{a})}{r \mathcal{H}} \Big[(t - \sigma_a)  k_{z} \epsilon_{z} 
 \\ \notag
 && {} \hspace{2.50cm} + \frac{1}{2}   (t - \sigma_a)^2 \Big( k_{z} (\epsilon_{x}^{2} + \epsilon_{y}^{2} - 2 \epsilon_{z}^{2}) + 2 (k_{x} \epsilon_{x} +  k_{y} \epsilon_{y}) \epsilon_{z} \Big) \Big] 
\\ \notag
&& {} \hspace{1.50cm} + \Big[ 2 \mathcal{Q}_{a} - \mathcal{E}_a + \frac{2 (1- \mathcal{Q}_{a})}{r \mathcal{H}} + \frac{\mathcal{H}^{'}}{\mathcal{H}^2}\Big] \Big[ -  (t - \sigma_a)  (k_{x} \epsilon_{x} + k_{y}  \epsilon_{y})
\\ \notag
&& {} \hspace{1.50cm} + \frac{1}{2}   (t - \sigma_a)^2 \Big( k_{z} (\epsilon_{x}^{2} + \epsilon_{y}^{2}) + 2 (k_{x} \epsilon_{x} +  k_{y} \epsilon_{y}) \epsilon_{z}\Big) \Big] 
 \Big\} 
\\ 
&& {} +  \frac{3}{2} \Omega_{m} \frac{\mathcal{H}^2}{k^2} \frac{(1 - \mathcal{Q}_{a})}{r \mathcal{H}} \Big[ 2   (t - \sigma_a)   \epsilon_{z} +  (t - \sigma_a)^{2} (\epsilon_{x}^{2} + \epsilon_{y}^{2} - 2 \epsilon_{z}^2) \Big],  %+ \mathcal{O}({\bm \epsilon}^3), 
\qquad
\end{eqnarray}
where $\sigma_a = (0, 1)$. 

Derivative terms that appear in power spectrum expressions are
\begin{eqnarray}
\partial_k \mathcal{K}^{\rm S}_{a} &=&  - b_{a\phi} \frac{f_{\rm NL}}{\mathcal{M}} \frac{\partial_k \mathcal{M}}{\mathcal{M}} - \frac{2}{k} f \left(\frac{k_{z}}{k}\right)^{2} ,
\\ 
\partial_k^2 \mathcal{K}^{\rm S}_{a} &=&  2 b_{a\phi} \frac{f_{\rm NL}}{\mathcal{M}} \frac{\big(\partial_k \mathcal{M}\big)^2}{\mathcal{M}^2} - b_{a\phi} \frac{f_{\rm NL}}{\mathcal{M}} \frac{ \partial_k^2 \mathcal{M}}{\mathcal{M}} + \frac{6}{k^{2}} f  \left(\frac{k_{z}}{k}\right)^{2} ,
\\ 
\partial_{k_z} \mathcal{K}^{\rm S}_{I} &=&  \frac{2}{k} f \frac{k_{z}}{k},  \label{a5}
\\  
\partial_{k_z} \, \partial_k  \mathcal{K}^{\rm S}_{a} &=& - \frac{4}{k^2} f \frac{k_{z}}{k},
\label{a6}
\\ 
\partial_k  \mathcal{K}^{\rm D}_{a} &=& - \frac{2}{k} \, \mathcal{K}^{\rm D}_{a} ,
\end{eqnarray}
where $f_{\rm NL}$ is the local primordial non-Gaussianity parameter, $b_{a\phi}$ is the primordial non-Gaussian bias and
\begin{align}
{\cal M}(k,z)=\frac{2D(z)\,T(k)\, k^2}{3\Omega_{m0} H_0^2\, g_{\rm dec}}\,.     
\end{align}
Here $D$ is the linear growth factor, normalised to 1 at $z=0$, $T(k)$ is the  matter transfer function, normalised to 1 at $k=0$, and  $g_{\rm dec}$ is the metric potential growth factor at decoupling.

\section{Coefficients $C_{lmn}^{(ab)}$} \label{AppB}

\begin{eqnarray}
\mathcal{C}_{000}^{(ab)} &=& \mathcal{K}^{\mathrm{S}}_{a}\mathcal{K}^{\mathrm{S}}_{b} + \mathcal{K}^{\mathrm{D}}_{a}\mathcal{K}^{\mathrm{D}*}_{b} + \mathcal{K}^{\mathrm{S}}_{a}\mathcal{K}^{\mathrm{D}*}_{b} + \mathcal{K}^{\mathrm{S}}_{b}\mathcal{K}^{\mathrm{D}}_{a} + \mathcal{K}^{\mathrm{S}}_{a}\mathcal{K}^{\Phi}_{b} + \mathcal{K}^{\mathrm{S}}_{b}\mathcal{K}^{\Phi}_{a} \label{c000} \\ 
\mathcal{C}_{001}^{(ab)} &=& \frac{1}{(kr)}\bigg\{3\Omega_{m}\left(\frac{\mathcal{H}}{k}\right)\Big[t(1-\mathcal{Q}_{b})\mathcal{K}^{\mathrm{S}}_{a} - (1-t)(1-\mathcal{Q}_{a})\mathcal{K}^{\mathrm{S}}_{b}\Big] \nonumber \\
&& \qquad \quad -\mathrm{i}\,2f\frac{k_{z}}{k}\Big[(1-t)(1-\mathcal{Q}_{a})\Big(\mathcal{K}^{\mathrm{S}}_{b}+\mathcal{K}^{\mathrm{D}*}_{b}\Big) + t(1-\mathcal{Q}_{b})\Big(\mathcal{K}^{\mathrm{S}}_{a}+\mathcal{K}^{\mathrm{D}}_{a}\Big)\bigg\} \label{c001}  \\ 
\mathcal{C}_{010}^{(ab)} &=& 2f\frac{k_{y}}{k}\frac{k_{z}}{k}\Big[t\Big(\mathcal{K}^{\mathrm{S}}_{a}+\mathcal{K}^{\mathrm{D}}_{a}+\mathcal{K}^{\Phi}_{a}\Big)-(1-t)\Big(\mathcal{K}^{\mathrm{S}}_{b}+\mathcal{K}^{\mathrm{D}*}_{b}+\mathcal{K}^{\Phi}_{b}\Big)\Big] \\
&&{} + \mathrm{i}\,f\frac{k_{y}}{k}\bigg\{\frac{2}{(kr)}\Big[t(1-\mathcal{Q}_{b})\Big(\mathcal{K}^{\mathrm{S}}_{a}+\mathcal{K}^{\mathrm{D}}_{a}\Big) + (1-t)(1-\mathcal{Q}_{a})\Big(\mathcal{K}^{\mathrm{S}}_{b}+\mathcal{K}^{\mathrm{D}*}_{b}\Big)\Big] \nonumber \\
&&\qquad \qquad~ + \left(\frac{\mathcal{H}}{k}\right)\bigg[(1-t)\bigg(-\mathcal{E}_{ a}+2\mathcal{Q}_{a}+\frac{\mathcal{H}^{\prime}}{\mathcal{H}^{2}}\bigg)\Big(\mathcal{K}^{\mathrm{S}}_{b}+\mathcal{K}^{\mathrm{D}*}_{b}\Big) \nonumber \\
&& \qquad \qquad \qquad \qquad \quad + t\bigg(-\mathcal{E}_{ b}+2\mathcal{Q}_{b}+\frac{\mathcal{H}^{\prime}}{\mathcal{H}^{2}}\bigg)\Big(\mathcal{K}^{\mathrm{S}}_{a}+\mathcal{K}^{\mathrm{D}}_{a}\Big)\bigg]\bigg\} \label{c010} \\ 
C_{100}^{(ab)} &=& C_{010}^{(ab)}\big|_{k_{y}\rightarrow k_{x}} \label{c100} \\  
C_{011}^{(ab)} &=& -2f\frac{k_{y}}{k}\frac{k_{z}}{k}\bigg\{(1-t)^{2}\Big(\mathcal{K}^{\mathrm{S}}_{b}+\mathcal{K}^{\mathrm{D}*}_{b}+\mathcal{K}^{\Phi}_{b}\Big) + t^{2}\Big(\mathcal{K}^{\mathrm{S}}_{a}+\mathcal{K}^{\mathrm{D}}_{a}+\mathcal{K}^{\Phi}_{a}\Big)\nonumber \\
&& \qquad \qquad \quad -\frac{1}{(kr)}\left(\frac{\mathcal{H}}{k}\right)t(1-t)\bigg[-3\Omega_{m} (2-\mathcal{Q}_{a}-\mathcal{Q}_{b}) \nonumber \\
&& \qquad \qquad \qquad \qquad \qquad \qquad \qquad ~~ +f\Big[\bigg(-\mathcal{E}_{ b}+2\mathcal{Q}_{b}+\frac{\mathcal{H}^{\prime}}{\mathcal{H}^{2}}\bigg) (1-\mathcal{Q}_{a}) \nonumber \\
&&  \qquad \qquad \qquad \qquad \qquad \qquad \qquad \qquad ~~~ + \bigg(-\mathcal{E}_{ a}+2\mathcal{Q}_{a}+\frac{\mathcal{H}^{\prime}}{\mathcal{H}^{2}}\bigg) (1-\mathcal{Q}_{b})\Big]\bigg] \nonumber \\
&& \qquad \qquad \quad -\frac{1}{(kr)^{2}}4f\,t(1-t)(1-\mathcal{Q}_{a})(1-\mathcal{Q}_{b})\bigg\} \nonumber \\
&& +\mathrm{i}\,f\frac{k_{y}}{k}\bigg\{\frac{4}{(kr)}\Big[f\left(\frac{k_{z}}{k}\right)^{2}t(1-t)(\mathcal{Q}_{a}-\mathcal{Q}_{b}) + (1-t)^{2}(1-\mathcal{Q}_{a})\Big(\mathcal{K}^{\mathrm{S}}_{b}+\mathcal{K}^{\mathrm{D}*}_{b}\Big) \nonumber \\
&& \qquad \qquad \quad ~~~-t^{2}(1-\mathcal{Q}_{b})\Big(\mathcal{K}^{\mathrm{S}}_{a}+\mathcal{K}^{\mathrm{D}}_{a}\Big)\Big]  \nonumber \\
&& \qquad \quad ~~+ \left(\frac{\mathcal{H}}{k}\right)\bigg[(1-t)^{2}\bigg(-\mathcal{E}_{ a}+2\mathcal{Q}_{a}+\frac{\mathcal{H}^{\prime}}{\mathcal{H}^{2}}\bigg)\Big(\mathcal{K}^{\mathrm{S}}_{b}+\mathcal{K}^{\mathrm{D}*}_{b}\Big) \nonumber \\
&& \qquad \qquad \qquad \qquad ~-t^{2}\bigg(-\mathcal{E}_{ b}+2\mathcal{Q}_{b}+\frac{\mathcal{H}^{\prime}}{\mathcal{H}^{2}}\bigg)\Big(\mathcal{K}^{\mathrm{S}}_{a}+\mathcal{K}^{\mathrm{D}}_{a}\Big)\bigg]\bigg\} \label{c011} \\ 
C_{101}^{(ab)} &=& C_{011}^{(ab)}\big|_{k_{y}\rightarrow k_{x}} \label{c101} 
\end{eqnarray}
\newpage
\begin{eqnarray}
C_{110}^{(ab)} &=& 2f\frac{k_{x}}{k}\frac{k_{y}}{k}\bigg\{(1-t)^{2}\Big(\mathcal{K}^{\mathrm{S}}_{b}+\mathcal{K}^{\mathrm{D}*}_{b}+\mathcal{K}^{\Phi}_{b}\Big) + t^{2}\Big(\mathcal{K}^{\mathrm{S}}_{a}+\mathcal{K}^{\mathrm{D}}_{a}+\mathcal{K}^{\Phi}_{a}\Big) \nonumber \\
&& \qquad \qquad~ -f\,t(1-t)\bigg[4\left(\frac{k_{z}}{k}\right)^{2} + \left(\frac{\mathcal{H}}{k}\right)^{2}\Big[(2\mathcal{Q}_{b}-\mathcal{E}_{ b})(2\mathcal{Q}_{a}-\mathcal{E}_{ a}) \nonumber \\
&& \qquad \qquad \qquad \qquad \qquad \qquad \qquad \qquad \qquad ~~~ +\frac{\mathcal{H}^{\prime}}{\mathcal{H}^{2}}\Big(-\mathcal{E}_{ a}-\mathcal{E}_{ b}+2(\mathcal{Q}_{a}+\mathcal{Q}_{b})+\frac{\mathcal{H}^{\prime}}{\mathcal{H}^{2}}\Big)\Big] \nonumber \\
&& \qquad \qquad \qquad \qquad \qquad ~ +\frac{2}{(kr)}\left(\frac{\mathcal{H}}{k}\right)\bigg[\bigg(-\mathcal{E}_{ b}+2\mathcal{Q}_{b}+\frac{\mathcal{H}^{\prime}}{\mathcal{H}^{2}}\bigg) (1-\mathcal{Q}_{a}) 
\nonumber \\
&& \qquad \qquad \qquad \qquad \qquad \qquad \qquad \qquad ~~~ + \bigg(-\mathcal{E}_{ a}+2\mathcal{Q}_{a}+ \frac{\mathcal{H}^{\prime}}{\mathcal{H}^{2}}\bigg) (1-\mathcal{Q}_{b})\bigg] \nonumber \\
&& \qquad \qquad \qquad \qquad \qquad ~+\frac{4}{(kr)^{2}}(1-\mathcal{Q}_{b})(1-\mathcal{Q}_{a})\bigg] \nonumber \\
&& \qquad \qquad~ +\mathrm{i}\,2f\frac{k_{z}}{k}t(1-t)\bigg[\left(\frac{\mathcal{H}}{k}\right)\big[ \mathcal{E}_{ b} -\mathcal{E}_{ a} - 2(\mathcal{Q}_{b}-\mathcal{Q}_{a})\big] + \frac{2}{(kr)}(\mathcal{Q}_{b}-\mathcal{Q}_{a})\bigg]\bigg\} \label{c110}  \\ 
C_{002}^{(ab)} &=& -\frac{3}{(kr)}\left(\frac{\mathcal{H}}{k}\right)\Omega_{m}\Big[(1-t)^{2}(1-\mathcal{Q}_{a})\mathcal{K}^{\mathrm{S}}_{b} + t^{2}(1-\mathcal{Q}_{b})\mathcal{K}^{\mathrm{S}}_{a}\Big] \nonumber \\
&& -\frac{4}{(kr)^{2}}f^{2}\left(\frac{k_{z}}{k}\right)^{2}t(1-t)(1-\mathcal{Q}_{a})(1-\mathcal{Q}_{b}) \nonumber \\
&& +\mathrm{i}\,\frac{2}{(kr)}f\frac{k_{z}}{k}\Big[t^{2}(1-\mathcal{Q}_{b})\Big(\mathcal{K}^{\mathrm{S}}_{a}+\mathcal{K}^{\mathrm{D}}_{a}\Big) - (1-t)^{2}(1-\mathcal{Q}_{a})\Big(\mathcal{K}^{\mathrm{S}}_{b}+\mathcal{K}^{\mathrm{D}*}_{b}\Big)\Big]\label{c002} \\ \nonumber \\
C_{020}^{(ab)} &=&f\bigg[\left(\frac{k_{y}}{k}\right)^{2}-\left(\frac{k_{z}}{k}\right)^{2}\bigg]\Big[t^{2}\Big(\mathcal{K}^{\mathrm{S}}_{a}+\mathcal{K}^{\mathrm{D}}_{a}+\mathcal{K}^{\Phi}_{a}\Big) + (1-t)^{2}\Big(\mathcal{K}^{\mathrm{S}}_{b}+\mathcal{K}^{\mathrm{D}*}_{b}+\mathcal{K}^{\Phi}_{b}\Big)\Big] \nonumber \\
&& + \frac{1}{(kr)}\left(\frac{\mathcal{H}}{k}\right)\frac{3}{2}\Omega_{m}\Big[t^{2}(1-\mathcal{Q}_{b})\mathcal{K}^{\mathrm{S}}_{a} + (1-t)^{2}(1-\mathcal{Q}_{a})\mathcal{K}^{\mathrm{S}}_{b} \Big] \nonumber \\
&& -f^{2}t(1-t)\left(\frac{k_{y}}{k}\right)^{2}\bigg\{4\left(\frac{k_{z}}{k}\right)^{2}+ \left(\frac{\mathcal{H}}{k}\right)^{2}\Big[(2\mathcal{Q}_{b}-\mathcal{E}_{ b})(2\mathcal{Q}_{a}-\mathcal{E}_{ a}) \nonumber \\
&& \qquad \qquad \qquad \qquad \qquad \qquad \qquad \qquad \quad ~~~ +\frac{\mathcal{H}^{\prime}}{\mathcal{H}^{2}}\Big(-\mathcal{E}_{ a}-\mathcal{E}_{ b}+2(\mathcal{Q}_{a}+\mathcal{Q}_{b})+\frac{\mathcal{H}^{\prime}}{\mathcal{H}^{2}}\Big)\Big] \nonumber \\
&& \qquad \qquad \qquad \qquad \qquad ~ +\frac{2}{(kr)}\left(\frac{\mathcal{H}}{k}\right)\bigg[\bigg(-\mathcal{E}_{ b}+2\mathcal{Q}_{b}+\frac{\mathcal{H}^{\prime}}{\mathcal{H}^{2}}\bigg) (1-\mathcal{Q}_{a}) \nonumber \\
&& \qquad \qquad \qquad \qquad \qquad \qquad \qquad \qquad ~~~ + \bigg(-\mathcal{E}_{ a}+2\mathcal{Q}_{a}+\frac{\mathcal{H}^{\prime}}{\mathcal{H}^{2}}\bigg) (1-\mathcal{Q}_{b})\bigg] \nonumber \\
&& \qquad \qquad \qquad \qquad \qquad ~+\frac{4}{(kr)^{2}}(1-\mathcal{Q}_{b})(1-\mathcal{Q}_{a})\bigg] \bigg\} \nonumber \\
&& + \mathrm{i}\,f\left(\frac{k_{z}}{k}\right)\bigg\{2f\left(\frac{k_{y}}{k}\right)^{2}t(1-t)\bigg[\left(\frac{\mathcal{H}}{k}\right)\big[ \mathcal{E}_{ b} -\mathcal{E}_{ a} - 2(\mathcal{Q}_{b}-\mathcal{Q}_{a})\big] + \frac{2}{(kr)}(\mathcal{Q}_{b}-\mathcal{Q}_{a})\bigg] \nonumber \\
&& \qquad \qquad \quad ~~+\frac{1}{2}\left(\frac{\mathcal{H}}{k}\right)\bigg[(1-t)^{2}\bigg(-\mathcal{E}_{ a}+2\mathcal{Q}_{a}+\frac{\mathcal{H}^{\prime}}{\mathcal{H}^{2}}\bigg)\Big(\mathcal{K}^{\mathrm{S}}_{b}+\mathcal{K}^{\mathrm{D}*}_{b}\Big) \nonumber \\
&& \qquad \qquad \qquad \qquad \qquad \quad ~-t^{2}\bigg(-\mathcal{E}_{ b}+2\mathcal{Q}_{b}+\frac{\mathcal{H}^{\prime}}{\mathcal{H}^{2}}\bigg)\Big(\mathcal{K}^{\mathrm{S}}_{a}+\mathcal{K}^{\mathrm{D}}_{a}\Big)\bigg] \nonumber \\
&& \qquad \qquad \quad ~~+ \frac{2}{(kr)}\Big[(1-t)^{2}(1-\mathcal{Q}_{a})\Big(\mathcal{K}^{\mathrm{S}}_{b}+\mathcal{K}^{\mathrm{D}*}_{b}\Big) - t^{2}(1-\mathcal{Q}_{b})\Big(\mathcal{K}^{\mathrm{S}}_{a}+\mathcal{K}^{\mathrm{D}}_{a}\Big)\Big]\bigg\} \label{e020} \\
C_{200}^{(ab)} &=& C_{020}^{(ab)}\big|_{k_{y}\rightarrow k_{x}} \label{c200}
\end{eqnarray}

\section{Multipoles of the power spectra} \label{AppC}

{\allowdisplaybreaks
The monopoles $(\ell = 0)$ for the standard, relativistic, wide-angle and cross-terms from wide-angle and relativistic contributions are
\begin{align}
P^{\mathrm{S}(0)}_{ab} &= \left[\hat{b}_{a} \hat{b}_{b} + \frac{f}{3} \left(\hat{b}_{a} +  \hat{b}_{b}\right) + \frac{f^{2}}{5} \right] P \;,\label{ExpP0S}
\\
P^{\mathrm{R}(0)}_{ab} &= \Bigg[\frac{1}{3} \frac{1}{k^{2}}  \gamma^{\rm D}_{a} \gamma^{\rm D}_{b} + \left(\frac{f}{3} +  \hat{b}_{b}\right)  \frac{1}{k^{2}} \gamma^{\Phi}_{a}  + \left(\frac{f}{3} + \hat{b}_{a}\right)  \frac{1}{k^{2}}  \gamma^{\Phi}_{b}\Bigg] P \;,\label{ExpP0R}
\\
\notag
P^{\mathrm{W}(0)}_{ab} &= \frac{2}{105} \frac{f}{(k r)^{2}} \Bigg\{ \Big\{ f [1 +  18 t (1 -  t)] - 21 \Big[(1 - t)^{2} \hat{b}_{b}   + t^{2} \hat{b}_{a}\Big]
\\
\notag
& \hspace{3.0cm} - 35 \Big[(1 - t)^{2} k \, \partial_{k} \hat{b}_{b} + t^{2} k \, \partial_{k} \hat{b}_{a}\Big] 
\\
\notag
&  \hspace{4.0cm} - 7 \Big[(1 - t)^{2} k^{2} \, \partial_{k}^{2} \hat{b}_{b} + t^{2} k^{2} \, \partial_{k}^{2} \hat{b}_{a}\Big] \Big\} P
\\
\notag
& {} +  \Big\{ f [11 + 30 t (1 -  t)]  - 35 \Big[(1 - t)^{2} \hat{b}_{b} +  t^{2} \hat{b}_{a}\Big]
\\
\notag
&  \hspace{2.50cm}  - 14 \Big[(1 - t)^{2} k \, \partial_{k} \hat{b}_{b} +  t^{2} k \, \partial_{k} \hat{b}_{a}\Big] \Big\} k \, \partial_{k} P
\\
& {} +  \Big\{ f [5 + 6 t (1 -  t)] - 7 \Big[ (1 - t)^{2} \hat{b}_{b} +  t^{2} \hat{b}_{a} \Big] \Big\} k^{2} \, \partial_{k}^{2} P \Bigg\} \;,\label{ExpP0W}
\\
\notag
P^{\mathrm{RW}(0)}_{ab}(k) &= \frac{2}{15} f \Bigg\{-\frac{2}{k r}  \Big[(1 - t) \frac{1}{k} \gamma^{\rm D}_{b} + t \frac{1}{k} \gamma^{\rm D}_{a}\Big] 
\\
\notag
&  \hspace{1.50cm} + \frac{1}{k r} \frac{\mathcal{H}}{k}  \Bigg[(1 - t) (f + 5 \hat{b}_{b})  \left(\mathcal{E}_{a} - 2 \mathcal{Q}_{a} - \frac{{\mathcal{H}'}}{\mathcal{H}^2}\right) 
\\
\notag
&  \hspace{3.50cm}  +  t (f  + 5 \hat{b}_{a}) \left(\mathcal{E}_{b} - 2 \mathcal{Q}_{b} - \frac{{\mathcal{H}'}}{\mathcal{H}^2}\right) \Bigg]
\\
\notag
&  \hspace{1.50cm}  + \frac{5}{k r} \frac{\mathcal{H}}{k}  \Bigg[ (1 - t) \left(\mathcal{E}_{a} - 2 \mathcal{Q}_{a} - \frac{{\mathcal{H}'}}{\mathcal{H}^2}\right)  k \, \partial_{k} \hat{b}_{b} 
\\
\notag
&  \hspace{3.50cm} + t \left(\mathcal{E}_{b} - 2 \mathcal{Q}_{b} - \frac{{\mathcal{H}'}}{\mathcal{H}^2}\right) k \, \partial_{k} \hat{b}_{a}\Bigg] 
\\
\notag
&  \hspace{1.50cm} +  \frac{1}{(k r)^{2}} \Big[(1 - t) (1 - \mathcal{Q}_{a}) (f - 5 \hat{b}_{b}) +  t (1 - \mathcal{Q}_{b}) (f - 5 \hat{b}_{a})\Big] 
\\
\notag
&  \hspace{1.50cm} -  \frac{5}{(k r)^{2}} \Big[ (1 - t) (1  - \mathcal{Q}_{a})  k \, \partial_{k} \hat{b}_{b} +  t (1 -  \mathcal{Q}_{b}) k \, \partial_{k} \hat{b}_{a}\Big] \Bigg\} P
\\
\notag
& {} + \frac{2}{15} f \Bigg\{-\frac{2}{k r} \Big[(1 - t) \frac{1}{k}  \gamma^{\rm D}_{b} + t \frac{1}{k} \gamma^{\rm D}_{a}\Big] 
\\
\notag
& \hspace{1.50cm} + \frac{1}{k r}  \frac{\mathcal{H}}{k} \Bigg[(1 - t)  (f + 5  \hat{b}_{b}) \left(\mathcal{E}_{a} - 2 \mathcal{Q}_{a} - \frac{{\mathcal{H}'}}{\mathcal{H}^2}\right) 
\\
\notag
& \hspace{3.50cm} + t (f + 5 \hat{b}_{a}) \left(\mathcal{E}_{b} - 2 \mathcal{Q}_{b} - \frac{{\mathcal{H}'}}{\mathcal{H}^2}\right)\Bigg] 
\\
\notag
&  \hspace{1.50cm} + \frac{1}{(k r)^{2}} \Big[(1 - t) (1 - \mathcal{Q}_{a}) (f - 5 \hat{b}_{b}) + t  (1 - \mathcal{Q}_{b}) (f - 5 \hat{b}_{a}) \Big] \Bigg\} k \, \partial_{k} P 
\\
\notag
& + \frac{1}{15} \frac{f}{(k r)^{2}} \Bigg\{ -2 \Big[(1 - t)^{2} \frac{1}{k^{2}} \gamma^{\Phi}_{b} + t^{2} \frac{1}{k^{2}} \gamma^{\Phi}_{a}\Big] 
\\
\notag
& \hspace{2.50cm} + 3  \frac{\mathcal{H}}{k}  \Bigg[(1 - t)^{2} \left(\mathcal{E}_{a} - 2 \mathcal{Q}_{a} -  \frac{{\mathcal{H}'}}{\mathcal{H}^{2}}\right)  \frac{1}{k} \gamma^{\rm D}_{b}  
\\
& \hspace{4.0cm} + t^{2} \left(\mathcal{E}_{b} - 2 \mathcal{Q}_{b} -  \frac{{\mathcal{H}'}}{\mathcal{H}^{2}}\right) \frac{1}{k} \gamma^{\rm D}_{a} \Bigg] 
\\ \notag
& \hspace{2.50cm} + 8 (1 - t) t \frac{\mathcal{H}^{2}}{k^{2}} f \left(\mathcal{E}_{b} - 2 \mathcal{Q}_{b} - \frac{{\mathcal{H}'}}{\mathcal{H}^{2}}\right) \left(\mathcal{E}_{a} - 2 \mathcal{Q}_{a} - \frac{{\mathcal{H}'}}{\mathcal{H}^{2}}\right) \Bigg\}  k^{2} \, \partial_{k}^{2} P  \;,\label{ExpP0RW}
\end{align}
where
\begin{align}
    \hat{b}_a=b_a+f_{\rm NL}\,\frac{b_{a\phi}}{\mathcal{M}}\,.
\end{align}

The  dipoles $(\ell = 1)$ for each contribution are
\begin{align}
P^{\mathrm{S}(1)}_{ab} &= 0 \;,\label{ExpP1S}
\\
P^{\mathrm{R}(1)}_{ab} &= \mathrm{i} \, \frac{1}{5} \left[(3 f + 5 \hat{b}_{b}) \frac{1}{k} \gamma^{\rm D}_{a}  - (3 f + 5 \hat{b}_{a}) \frac{1}{k} \gamma^{\rm D}_{b} \right] P \;,\label{ExpP1R}
\\
\notag
P^{\mathrm{W}(1)}_{ab} &= - \mathrm{i} \, \frac{4}{35}  \frac{f}{k r}  \Bigg\{\Bigg[9 (1 - 2 t) f  + 21 \Big[(1 - t) \hat{b}_{b} -  t \hat{b}_{a}\Big] 
\\
\notag
& \hspace{3.0cm} + 7 \Big[(1 - t) k \, \partial_{k} \hat{b}_{b} - t k \, \partial_{k} \hat{b}_{a}\Big] \Bigg] P 
\\
&  \hspace{2.20cm}  + \Bigg[3 (1 - 2 t) f + 7 \Big[(1 - t) \hat{b}_{b} -  t \hat{b}_{a}\Big] \Bigg] k \, \partial_{k} P  \Bigg\}  \;,\label{ExpP1W}
\\
\notag
P^{\mathrm{RW}(1)}_{ab} &= \mathrm{i} \, \frac{1}{5} \frac{1}{(k r)^{2}} \frac{\mathcal{H}}{k} \Bigg\{- 15 \Omega_{m} \Big[(1 - t) (1 -  \mathcal{Q}_{a}) k \, \partial_{k} \hat{b}_{b} - t (1 - \mathcal{Q}_{b}) k \, \partial_{k} \hat{b}_{a} \Big]
\\
\notag
& \hspace{2.7cm} + 3 f \Bigg[(1 - t)^{2} \left(\mathcal{E}_{a} - 2 \mathcal{Q}_{a}  -   \frac{{\mathcal{H}'}}{\mathcal{H}^{2}}\right)  k^{2} \, \partial_{k}^{2}  \hat{b}_{b} 
\\
\notag
& \hspace{4.0cm} - t^{2}  \left(\mathcal{E}_{b}  - 2 \mathcal{Q}_{b}  -  \frac{{\mathcal{H}'}}{\mathcal{H}^{2}} \right) k^{2} \, \partial_{k}^{2} \hat{b}_{a} \Bigg] \Bigg\} P 
\\
\notag
& + \mathrm{i} \, \frac{1}{5} \Bigg\{ - 4 \frac{f}{k r} \Big[(1 - t) \frac{1}{k^{2}} \gamma^{\Phi}_{b} - t \frac{1}{k^{2}} \gamma^{\Phi}_{a} \Big] 
\\
\notag
& \hspace{1.2cm} - 2 \frac{f}{k r} \frac{\mathcal{H}}{k} \Big[(1 - t)  \left(\mathcal{E}_{a} - 2 \mathcal{Q}_{a}  -  \frac{{\mathcal{H}'}}{\mathcal{H}^2} \right)  \frac{1}{k} \gamma^{\rm D}_{b} - t \left( \mathcal{E}_{b} - 2 \mathcal{Q}_{b}  - \frac{{\mathcal{H}'}}{\mathcal{H}^{2}}\right) \frac{1}{k} \gamma^{\rm D}_{a} \Big] 
\\
\notag
& \hspace{1.2cm} + \frac{2}{7} \frac{f}{(k r)^{2}} \frac{\mathcal{H}}{k} \Big[(1 - t) (21 t + 5) f \left(\mathcal{E}_{a} - 2 \mathcal{Q}_{a} -  \frac{{\mathcal{H}'}}{\mathcal{H}^2}\right) 
\\
\notag
& \hspace{3.5cm} + t (21 t - 26) f \left(\mathcal{E}_{b} - 2 \mathcal{Q}_{b}  -  \frac{{\mathcal{H}'}}{\mathcal{H}^2}\right)
\\
\notag
& \hspace{1.2cm} + 21 (1 - t)^{2} \left(\mathcal{E}_{a} - 2 \mathcal{Q}_{a} -  \frac{{\mathcal{H}'}}{\mathcal{H}^2}\right) k \, \partial_{k} \hat{b}_{b} - 21 t^{2}  \left(\mathcal{E}_{b} - 2  \mathcal{Q}_{b} -  \frac{{\mathcal{H}'}}{\mathcal{H}^2}\right) k \, \partial_{k} \hat{b}_{a} \Big]
\\
\notag
& \hspace{1.2cm} - \frac{6}{7}  \frac{f}{(k r)^{2}} \Big[- (1 - t)^{2} \frac{1}{k} \gamma^{\rm D}_{b} + t^{2} \frac{1}{k} \gamma^{\rm D}_{a}\Big]
\\
\notag
& \hspace{1.2cm} - \frac{3}{(k r)^{2}} \frac{\mathcal{H}}{k} \Omega_{m} \Big[(1 - t) (1 - \mathcal{Q}_{a}) (3 f + 5 \hat{b}_{b}) -  t (1 - \mathcal{Q}_{b}) (3 f + 5 \hat{b}_{a}) \Big] 
\\
\notag
& \hspace{1.2cm} - 2 \frac{f}{(k r)^{2}} \Big[(1 - t) (1 - \mathcal{Q}_{a}) \frac{1}{k} \gamma^{\rm D}_{b} - t (1 -  \mathcal{Q}_{b}) \frac{1}{k} \gamma^{\rm D}_{a} \Big]  \Bigg\} k \, \partial_{k} P
\\
\notag
& + \mathrm{i} \, \frac{1}{35}  \frac{f}{(k r)^{2}}  \Bigg\{ \frac{\mathcal{H}}{k} \Big[(1 - t) (7 t + 9) f \left(\mathcal{E}_{a}  - 2 \mathcal{Q}_{a} -  \frac{{\mathcal{H}'}}{\mathcal{H}^2}\right) + t (7 t - 16) f  \left(\mathcal{E}_{b} - 2 \mathcal{Q}_{b} -  \frac{{\mathcal{H}'}}{\mathcal{H}^2}\right)
\\
\notag
& \hspace{2.5cm} + 21 (1 - t)^{2} \left(\mathcal{E}_{a} - 2 \mathcal{Q}_{a} -  \frac{{\mathcal{H}'}}{\mathcal{H}^2}\right) \hat{b}_{b} - 21 t^{2} \left(\mathcal{E}_{b} - 2 \mathcal{Q}_{b} -  \frac{{\mathcal{H}'}}{\mathcal{H}^2}\right) \hat{b}_{a}\Big]
\\
& \hspace{2.5cm} - 10 \Big[(1 - t)^{2} \frac{1}{k} \gamma^{\rm D}_{b} - t^{2}  \frac{1}{k} \gamma^{\rm D}_{a}\Big] \Bigg\} \, k^{2} \, \partial_{k}^{2} P
 \;.\label{ExpP1RW}
\end{align}

The quadrupoles $(\ell = 2)$ are
\begin{align}
P^{\mathrm{S}(2)}_{ab} &= \frac{2}{21} f \Big[6 f + 7 \hat{b}_{a} + 7 \hat{b}_{b}\Big] P \;,\label{ExpP2S}
\\
P^{\mathrm{R}(2)}_{ab} &= \frac{2}{3} \Big[ \frac{1}{k^{2}} \gamma^{\rm D}_{a} \gamma^{\rm D}_{b} + f \frac{1}{k^{2}} \, \gamma^{\Phi}_{a} + f \frac{1}{k^{2}} \, \gamma^{\Phi}_{b}\Big] P 
\;,\label{ExpP2R}
\\
\notag
P^{\mathrm{W}(2)}_{ab} &=  \frac{2}{21} \frac{f}{(k r)^{2}} \Bigg\{ \Big\{2 f [-13 + 54 t (1 -  t)] - 66  \Big[(1 - t)^{2} \hat{b}_{b} + t^{2} \hat{b}_{a}\Big]
\\
\notag
& \hspace{2.7cm} + 22 \Big[(1 - t)^{2} k \, \partial_{k} \hat{b}_{b} + t^{2} k \, \partial_{k} \hat{b}_{a}\Big] 
\\
\notag
& \hspace{2.7cm} + 11 \Big[(1 - t)^{2} k^{2} \, \partial_{k}^{2} \hat{b}_{b} + t^{2} k^{2} \, \partial_{k}^{2} \hat{b}_{a} \Big] \Big\} P 
\\
\notag
& \hspace{1.5cm} + 2 \Big\{11 \Big[ (1 - t)^{2} k \, \partial_{k} \hat{b}_{b} + t^{2} k \, \partial_{k} \hat{b}_{a}\Big] + 2 f [2 + 3 t (1 -  t)]
\\
& \hspace{2.7cm} + 11 \Big[(1 - t)^{2} \hat{b}_{b} + t^{2} \hat{b}_{a}\Big] \Big\} k \, \partial_{k} P
\\
& \hspace{1.5cm} + \Big\{3 f [1 - 2 t (1 - t)] + 11 \Big[(1 - t)^{2} \hat{b}_{b} +  t^{2} \hat{b}_{a}\Big] \Big\} k^{2} \, \partial_{k}^{2} P \Bigg\}
\;,\label{ExpP2W}
\\
\notag
P^{\mathrm{RW}(2)}_{ab}(k) &= \frac{2}{21} f \Bigg\{ \frac{2}{k r} \frac{\mathcal{H}}{k}  \Big[(1 - t) (5 f  + 7 \hat{b}_{b}) \left(\mathcal{E}_{a} - 2 \mathcal{Q}_{a}  - \frac{{\mathcal{H}'}}{\mathcal{H}^{2}}\right) 
\\
\notag
& \hspace{2.7cm} + t (5 f  + 7 \hat{b}_{a}) \left(\mathcal{E}_{b} - 2 \mathcal{Q}_{b}  - \frac{{\mathcal{H}'}}{\mathcal{H}^{2}}\right)\Big]
\\
\notag
& \hspace{1.2cm} - \frac{7}{k r} \frac{\mathcal{H}}{k}  \Big[(1 - t) \left(\mathcal{E}_{a}  - 2 \mathcal{Q}_{a} - \frac{{\mathcal{H}'}}{\mathcal{H}^{2}}\right) k \,  \partial_{k} \hat{b}_{b} 
\\
\notag
& \hspace{2.7cm} + t \left(\mathcal{E}_{b}  - 2 \mathcal{Q}_{b} - \frac{{\mathcal{H}'}}{\mathcal{H}^{2}}\right) k \, \partial_{k} \hat{b}_{a}\Big]
\\
\notag
& \hspace{1.2cm} -  \frac{20}{k r}  \Big[(1 - t) \frac{1}{k} \, \gamma^{\rm D}_{b} + t \frac{1}{k} \, \gamma^{\rm D}_{a}\Big]
\\
\notag
& \hspace{1.2cm} + \frac{28}{(k r)^{2}} \Big[(1 - t) (1 - \mathcal{Q}_{a}) k \, \partial_{k} \hat{b}_{b} + t (1 - \mathcal{Q}_{b}) k \, \partial_{k} \hat{b}_{a} \Big]
\\
\notag
& \hspace{1.2cm} - \frac{2}{(k r)^{2}}  \Big[(1 - t) (1 -  \mathcal{Q}_{a}) (13 f + 28 \hat{b}_{b}) 
\\
\notag
& \hspace{3.0cm} +  t (1 -  \mathcal{Q}_{b})  (13 f + 28 \hat{b}_{a}) \Big]  \Bigg\} P
\\
\notag
& + \frac{2}{21} f \Bigg\{\frac{1}{k r} \frac{\mathcal{H}}{k} \Big[(1 - t)  (f  - 7  \hat{b}_{b}) \left(\mathcal{E}_{a} - 2 \mathcal{Q}_{a} - \frac{{\mathcal{H}'}}{\mathcal{H}^{2}}\right)  
\\
\notag
& \hspace{2.7cm} + t (f  - 7  \hat{b}_{a}) \left(\mathcal{E}_{b}  - 2  \mathcal{Q}_{b} - \frac{{\mathcal{H}'}}{\mathcal{H}^{2}}\right) \Big]
\\
\notag
& \hspace{1.2cm} - \frac{2}{k r}  \Big[(1 - t) \frac{1}{k}  \, \gamma^{\rm D}_{b} + t \frac{1}{k} \, \gamma^{\rm D}_{a}\Big]
\\
\notag
& \hspace{1.2cm} +  \frac{2}{(k r)^{2}}  \Big[(1 - t) (1 - \mathcal{Q}_{a} ) (5 f  + 14  \hat{b}_{b}) 
\\
\notag
& \hspace{3.0cm} + t (1 - \mathcal{Q}_{b}) (5 f + 14 \hat{b}_{a}) \Big] \Bigg\} k \,  \partial_{k} P 
\\
\notag
& + \frac{1}{21} \frac{f}{(k r)^{2}} \Bigg\{ - 16 (1 - t) t \frac{\mathcal{H}^{2}}{k^{2}} f \left(\mathcal{E}_{a} - 2 \mathcal{Q}_{a} - \frac{{\mathcal{H}'}}{\mathcal{H}^{2}}\right)  \left(\mathcal{E}_{b} - 2 \mathcal{Q}_{b} - \frac{{\mathcal{H}'}}{\mathcal{H}^{2}}\right) 
\\
\notag
& \hspace{2.2cm} + 3 \frac{\mathcal{H}}{k} \Big[(1 - t)^{2}  \left(\mathcal{E}_{a} - 2 \mathcal{Q}_{a}  - \frac{{\mathcal{H}'}}{\mathcal{H}^{2}}\right) \frac{1}{k} \, \gamma^{\rm D}_{b} 
\\
\notag
& \hspace{3.5cm} + t^{2} \left(\mathcal{E}_{b}  - 2 \mathcal{Q}_{b}  - \frac{{\mathcal{H}'}}{\mathcal{H}^{2}} \right)  \frac{1}{k} \, \gamma^{\rm D}_{a} \Big]
\\
& \hspace{2.2cm} + 22 \Big[(1 - t)^{2} \frac{1}{k^{2}} \, \gamma^{\Phi}_{b} + t^{2} \frac{1}{k^{2}} \, \gamma^{\Phi}_{a}\Big]  \Bigg\}  k^{2} \, \partial_{k}^{2} P 
\; .
\label{ExpP2RW}
\end{align}
}

\clearpage
\bibliographystyle{JHEP}
\bibliography{reference_library}

\end{document}